\documentclass[a4paper]{cas-dc}
\usepackage{xurl}
\makeatletter
\def\Hy@Warning#1{}
\providecommand{\@LN}[2]{}
\makeatother

\usepackage[authoryear]{natbib}
\newcommand{\mbdim}{kg.m$^{-2}$.s$^{-1}$}
\newcommand{\ptraj}{P(\text{traj})}
\newcommand{\pxz}{P(xz)}
\newcommand{\pxy}{P(xy)}
\newcommand{\pyz}{P(yz)}
\newcommand{\simdot}{\dot{\sigma}_{\rm im}}

\begin{document}

\shorttitle{On the relevance of the exposure age for Saturn's rings.}    

\shortauthors{Ricerchi and Crida}  

\title[mode=title]{Saturn's rings age I.: Reconsideration of the exposure age.}  

\author[1,2]{Gregorio Ricerchi}[orcid=0009-0002-5271-8658]

\cormark[1]

\ead{gregorio.ricerchi@oca.eu}

\credit{}

\affiliation[1]{organization={Université Côte d'Azur, Observatoire de la Côte d'Azur, CNRS, Laboratoire Lagrange, Bd de l'Observatoire, CS 34229, 06304 Nice cedex 4, France}}

\author[1]{Aur\'elien Crida}[orcid=0000-0002-1293-9782]
\cormark[2]
\ead{aurelien.crida@oca.eu}

\credit{}

\affiliation[2]{organization={MAUCA — Master track in Astrophysics, Université Côte d’Azur \& Observatoire de la Côte d’Azur, Parc Valrose, 06100 Nice, France}}
\cortext[1]{Principal corresponding author}
\cortext[2]{Corresponding author}

\begin{abstract}
At the end of the \emph{Cassini} mission, Saturn's rings have been claimed to be spectacularly young compared to the age of the Solar System: their unusual ice-rich composition corresponds to initially pure ice rings polluted by interplanetary dust particles for $100$ to $400$~Myr. Since then, this exposure age has been commonly accepted as the real age of the rings. 

In this paper, we review the processes that are involved in determining the exposure age. We aim to see how the exposure age depends on various parameters and how relevant it is to define the real rings age. First, a new expression for the gravitational focusing onto planar rings, important parameter but crudely defined in the literature, is derived. Then, an analytical formula describing how the dust fraction varies with time in static or viscously evolving rings is provided, including possible vaporisation at impact. Finally, we introduce a cleaning process from space weathering to possibly alter dust and reduce its amount to make rings look younger than they are.

We first found that the gravitational focusing is $5$ times less important than previously thought, which automatically increases the exposure age from $0.5$ to $2$ Gyr. Moreover, depending on the impact properties (vaporisation rate, space weathering efficiency), several billion years can easily be reached. Finally, we find that the dust fraction in the rings converges towards a finite value which, in particular with an efficient space weathering mechanism, can be close to the observed one in the current rings. In this case, neither the age nor the initial composition of the rings can be derived, and the measure of the dust fraction and bombardment rate only constrains the physical parameters of the impacts and the efficiency of the space weathering. As long as the latter parameters are not known, the exposure age argument in favour of young rings is completely undercut.
\end{abstract}

\begin{highlights}
\item Derivation of gravitational focusing for planar rings 
\item Analytical evolution of the fraction of dust in the rings through time
\item Better understandings and outcomes of a cleaning mechanism in the dust fraction evolution
\end{highlights}

\maketitle

\section{Introduction}

Since its first observation by Galilei in 1610, many other scientists (Huygens, Cassini, Maxwell) observed and tried to explain the most famous characteristic of Saturn: its rings. All giant planets of our Solar System have rings but none of them have such complex and rich rings as Saturn. While the ones of Jupiter, Uranus and Neptune are fainter, lighter and darker (i.e. rich in dust) \citep{nicholson_infrared_1984, porco_color_1987,de_pater_rings_2018}, those of Saturn are much more massive and almost completely pure in ice. Their mass has been measured to be $1.54 \pm0.49\ 10^{19}$ kg, which corresponds to $0.41 \pm 0.13$ Mimas masses \citep{iess_measurement_2019} and they are composed of $\geqslant95\%$ water ice \citep{doyle_radiative_1989, grossman_microwave_1990, zhang_exposure_2017, zhang_cassini_2017}.

Many ideas are on the table to explain their origin and fit these observational constraints: remnants of Saturn's circumplanetary disc \citep{pollack_formation_1976}, tidal disruption of a lost Titan-sized moon \citep{canup_origin_2010} or from a comet passing too close to Saturn \citep{dones_recent_1991}, a destabilisation of a lost moon \citep{wisdom_loss_2022, teodoro_recent_2023} and so on. All these scenarios have their strengths and weaknesses and result in recent or primordial rings. Regarding the question of their age, we must define different ages that represent different concepts. The \emph{exposure} age refers to the period during which the rings appear to have been exposed to a bombardment of dust particles. The \emph{structure} age refers to the time structures (gaps, plateaus, edges) have been present in the rings. The \emph{dynamical} age refers to the time the rings density profile has evolved through the evolution of the orbital angular momentum. Lastly, the \emph{erosion} age defines the future of the rings as they lose mass. More details and context on these topics are given in the recent review by \cite{crida_age_2025}. We refer the reader to this paper for a global picture of the problem, while this article is devoted to the exposure age and the evolution of the composition of the rings.

Micrometeoroid bombardment (hereafter MB) is known to darken the rings over time by bringing dust into the particles of the rings \citep{doyle_radiative_1989, cuzzi_compositional_1998}. The near-purity of the rings in ice led to the assumption that they were formed purely from ice and darkened with MB. \cite{cuzzi_compositional_1998} already estimated the exposure age to be around $100$ Myr. Recently, with the end of the \emph{Cassini} mission, \cite{kempf_micrometeoroid_2023} performed a rigorous analysis of the flux of particles getting close to Saturn (in particular, the total flux at infinity and the velocity of incoming particles) and deduced an exposure age of $100-400$ Myr old. Since then, this age has been used as the real age of the rings. In that case, a fortunate event must have occurred around Saturn in the recent past to create pure ice rings of mass equivalent to a $\sim 300$~km diameter sphere of density 900~kg/m$^3$. Although some scenarios have been proposed, this is not intuitive in a Solar System thought to be quiet after the giant planets instability \citep{gomes_origin_2005, morbidelli_chaotic_2005, tsiganis_origin_2005}. Conversely, if the rings formed with the planet, a process is needed to make the exposure age differ from the real age and maintain their amazingly pure composition over the age of the Solar System. Space weathering is a great candidate for a cleaning mechanism that reduces the rings dust fraction over time, and pollution resistance is another possibility. If such a process is strong enough to prevent the rings from becoming too polluted, then the exposure age is biased and can not be extrapolated into a formation age of the rings \citep{crida_are_2019}. 

MB also modifies the structure and dynamics of rings, through mass loading and ballistic transport \citep{ip_ring_1984, lissauer_ballistic_1984, durisen_ballistic_1989, durisen_ballistic_1992, estrada_meteoroid_2018}. These processes are thought to play a role for many local structures and characteristics in the rings. They also open the question of the future of the rings, as they would disappear after a few hundred million years of evolution. These effects are discarded in this paper, where we focus on the composition of the rings. They will be studied in a forthcoming paper which will use the results and constraints on MB from this one.

In this paper, we study the possibility of rings being almost bereft of dust after long timescales. To address this problem, the structure of the work is as follows. Section~\ref{sec:Fg_work} is a complete rework of the gravitational focusing which is critical in the MB process chemically and dynamically speaking. The idea is to derive the correct expression for planar rings taking carefully into account the incoming velocity of the particles, which has not been done before. In Section~\ref{sec:mathematical model}, we develop an analytical model for the time evolution of the average dust fraction in the rings. New physical processes are introduced: space weathering and partial vaporisation of the impactor and the target. This leads to a rich variety of possible outcomes, which are explored in Section~\ref{sec:results}. Finally, our results are summarized and discussed in section~\ref{sec:sdp}.

\section{On the gravitational focusing}\label{sec:Fg_work}

One of the most important parameters relative to the bombardment is the focus caused by Saturn's gravity.
In the literature, the formula of \cite{cuzzi_bombardment_1990} is commonly used:
\begin{equation}\label{eq:fg_CD90}
    F_g(r) = F_g \left( \frac{r}{r_B} \right)^{-0.8}
\end{equation}
where $r$ is the distance to Saturn's centre and $r_B=1.8R_S$, $R_S$ being the radius of Saturn ($60\ 268$ km). $F_g$ is a constant coefficient taken equal to $3$ in the original paper, and to $30$ in \citet{estrada_constraints_2023} due to a change in the estimate of the velocity at infinity of the micrometeorites, from $14.4$~km/s \citep{humes_results_1980} to $4.3$~km/s \citep{kempf_micrometeoroid_2023}.

In this section, we will see how we can derive a more accurate expression of the gravitational focusing for rings, starting from a spherical target (Sec.~\ref{sec:Fg and dFg sphere}) to a plane target (Sec.~\ref{sec:Fg and dFg rings}). Then, we will re-derive the mean velocity of the particles at Saturn's Hill sphere (Sec.~\ref{sec:Fg_vinfty}). Finally, we discuss about the validity of our analysis (which results in a division by about 5 of the gravitational focusing) and compare it with old expressions used (Sec.~\ref{sec:comp_fg_discussion}).

\subsection{Gravitational focusing for a sphere and its local expression}\label{sec:Fg and dFg sphere}

\paragraph{Global gravitational focusing}

The geometrical cross-section of a sphere of radius $r$ is of course $\sigma_{\rm geo}=\pi r^2$. But the gravity of an object of mass $M_p$ deflects the trajectories so that the maximum impact parameter to hit the sphere is larger than $r$ and the actual cross-section is $\sigma_{\rm grav}>\sigma_{\rm geo}$. From the energy and angular momentum conservation equations between infinity and the closest approach to the planet, one finds for particles arriving with a velocity $v_\infty$ at Saturn's Hill sphere \citep[e.g.][]{morfill_consequences_1983}
\begin{equation}\label{eq:fg_sphere full}
    F_g(r) = \frac{\sigma_{\text{grav}}}{\sigma_{\text{geo}}} = 1 + \frac{2GM_p}{rv_{\infty}^2}= 1 + \frac{v^2_{\text{esc}}(r)}{v_{\infty}^2} = 1 + 2\Theta\ \ 
\end{equation}
where $\Theta=v^2_{\text{esc}}(r)/2v_{\infty}^2$ is the Safronov number \citep{safronov_evolution_1972}.

Using $v_{\infty}=14.4$ km/s, Eq.~\ref{eq:fg_sphere full} is well approximated for Saturn by Eq.~\ref{eq:fg_CD90} with a coefficient $F_g=3$. However, with $v_\infty=4.3$ km/s, a power law fit of Eq.~\ref{eq:fg_sphere full} yields $F_g(r) = 39 \left( \frac{r}{r_B} \right)^{-0.97}$, a steeper slope and larger coefficient than what \cite{estrada_constraints_2023} use (see solid blue and dotted green curves on Fig.~\ref{fig:new dFg rings new vinfty}).

\paragraph{Local gravitational focusing}
One must remember however that the formulation of Eq.~\ref{eq:fg_sphere full} is to fall \emph{inside} the sphere of radius $r$ and not to fall \emph{at the distance} $r$, which is why we call this $F_g(r)$ 'global'. To find the increase in the bombardment rate that applies \emph{locally} at a radius $r$, let us consider two spheres of radii $r_1, r_2$ with $r_2 > r_1$ and do the ratio of the differences of $\sigma_{\text{grav}}$ to the differences of $\sigma_{\text{geo}}$ of each sphere. This leads to a new $F_g$ that we call $\delta F_g$ such that
\begin{equation}
    \delta F_g = \frac{\pi(r_2^2F_g(r_2) - r_1^2F_g(r_1))}{\pi(r_2^2 - r_1^2)},
\end{equation}
which is the gravitational focusing to have the closest approach between $r_1$ and $r_2$ from the planet.
If $r_1 \rightarrow r_2$, we end up with the derivative with respect to $r^2$ of $r^2$ times the 'global' gravitational focusing $F_g(r)$. We have  
\begin{equation}\label{eq:fg_sphere local}
    \delta F_g \rightarrow \frac{d(r^2F_g(r))}{dr^2} = 1 + \frac{GM_p}{rv_{\infty}^2} = 1+\Theta = F_g^*(r)\ \ 
\end{equation}
and we call this new quantity $F_g^*(r)$ the \emph{local gravitational focusing} from now on. It is the one that should be used to estimate the bombardment rate at a given distance $r$ from Saturn. Using $v_\infty=4.3$~km/s, one gets $F_g^*(r_B)=20$. This is the same formula as the one \cite{morfill_consequences_1983} obtained when deriving the flux at Saturn's rings $F(\text{rings})$ from the flux at infinity $F_0$: 
\begin{equation}
    F(\text{rings}) = F_0\left(1 + \frac{GM_p}{rv_{\infty}^2}\right) = F_0\cdot F_g^*(r),
\end{equation}
which has been used in many papers \citep[e.g.][]{charnoz_did_2009, colwell_origins_1992, colwell_disruption_1994}. However, this agreement is fortuitous since our formula applies to a spherical target, while one should consider planar rings, which implies a different way of defining the gravitational cross-section and thus the $F_g^*(r)$ for such geometry. 

\subsection{Derivation of \texorpdfstring{$F_g(r)$ and $F_g^*(r)$}{Fg(r) and Fg*(r)} for rings}\label{sec:Fg and dFg rings}
The effective cross section of a planar disc of radius $r$ under isotropic bombardment is neither straightforward nor intuitive. For example, while the effective cross section for projectiles following a straight trajectory is always $\pi r^2$ for a sphere, it is $\pi r^2 \sin\theta$ for a planar disc where $\theta$ is the angle between the incoming trajectory and the disc plane. Averaged over an isotropic distribution of the directions, this yields $0.5 \pi r^2$.

\begin{figure}
    \centering
    \includegraphics[width=\columnwidth]{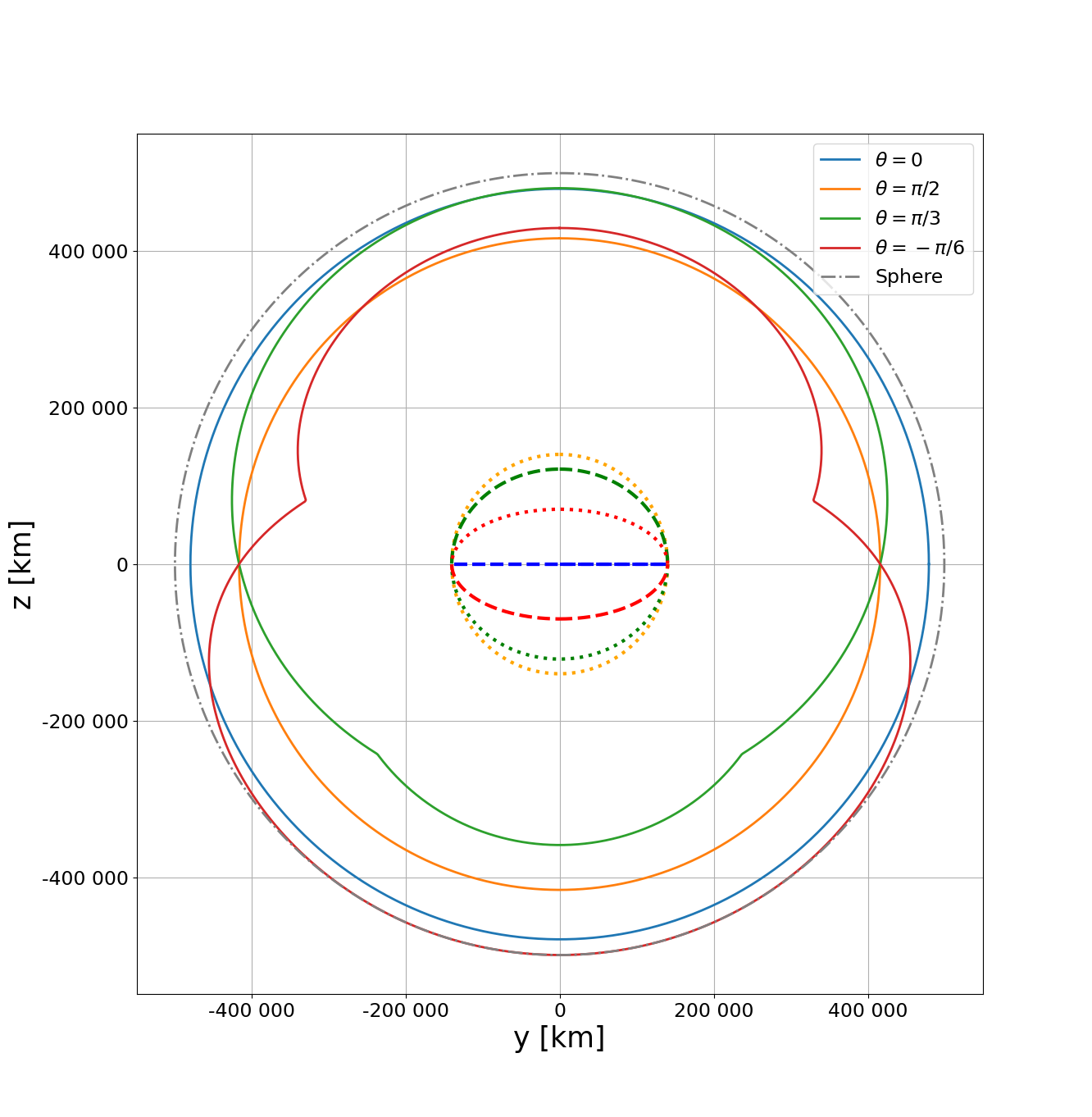}
    \caption{Plain lines: Effective cross sections to hit the rings at distance $r\leqslant r_{\rm Roche}=140\ 000$ km, for different values of the angle $\theta$ between the incoming trajectory and the rings plane and with a velocity of $v_\infty=6.8$ km/s. The 'mushroom' shapes show when we change the face of the rings the particles strike (first constraint of Appendix~\ref{sec: final exp dFg rings}). The geometrical cross section of the rings inclined by an angle $\theta$ is represented by the different ellipses and circle, with the same colour code for the effective cross sections. The segment of the rings in the $x>0$ is dashed while the $x\leqslant0$ segment is dotted. Finally, the gray dash-dotted line is the cross section to hit a sphere of radius $r=140\ 000$ km. }
    \label{fig:mushroom_cross_sections}
\end{figure}

Taking into account the bending of the trajectories by a central mass requires a more complex analysis presented in Appendix~\ref{appendix: derivation Fgstar}. As an example, Fig.~\ref{fig:mushroom_cross_sections}  shows in the $y-z$ plane the cross-sections to hit the rings with an initial velocity $v_\infty = 6.8$~km/s parallel to the $x$-axis with various inclinations of the rings. The blue circle corresponds to the case where the ring plane is the $x-y$ plane 
(that is: the rings are seen edge-on\footnote{Obtaining a circle may be counter-intuitive in this $\theta=0$ case. But since all the bent trajectories cross the $x$-axis on the other side of the planet, this is where they all meet the ring plane and thus the calculation of the maximum impact parameter is independent of the azimuth in the $y-z$ plane.}), 
and the other curves correspond to rings rotated along the $y$ axis by an angle $\theta$. In the end, averaging over $\theta$ for an isotropic distribution, we find that the local $F_g^*(r)$ that applies to planar rings reads:
\begin{equation}\label{eq:fg_generalised rings}
    F_g^*(r) = s(r/R_p)\cdot\underbrace{\left( 0.5 + 0.82 \Theta \right)}_{\widetilde{F_g^*}}
\end{equation}
where $\Theta$ has been defined above.
The term $\widetilde{F_g^*}$ represents the \emph{unshielded} local gravitational focusing to fall at a radius $r$ on a planar ring. It differs from the same expression for a sphere given by Eq.~\ref{eq:fg_sphere local} in the coefficients (notably $0.5$ instead of $1$ as explained above), but not the functional form. The $s(r/R_p)$ factor represents the shielding caused by the planet of radius $R_p$ which intercepts some trajectories; it actually depends slightly on $v_\infty$, in a non simple way (see Appendix~\ref{appendix: derivation Fgstar}). We will see in the next subsection what expression will be needed in our work.

\subsection{New estimation of the velocity $v_\infty$}
\label{sec:Fg_vinfty}
The value of $4.3$ km/s for $v_\infty$ is the modal value in the histogram shown in figure 2 of \cite{kempf_micrometeoroid_2023}. This does not make it a representative value of $v_\infty$ at Saturn's Hill sphere. Indeed, the $v_\infty$ distribution observed deep inside Saturn's potential well is precisely biased by the gravitational focusing\,: it is easier to deflect low-velocity particles than those with a greater velocity. Hence, the \emph{Cassini} sample necessarily suffers from an over-representation of low $v_\infty$ particles. In order to "defocus" the distribution, each particle $i$ should be given a coefficient $1/F_g(i)$ where $F_g(i)$ is computed using Eq.~\ref{eq:fg_sphere full} at the distance of detection (the particle enters the sphere of this radius), using the $v_\infty$ of the corresponding particle.
\begin{figure}
    \centering
    \includegraphics[width=\columnwidth]{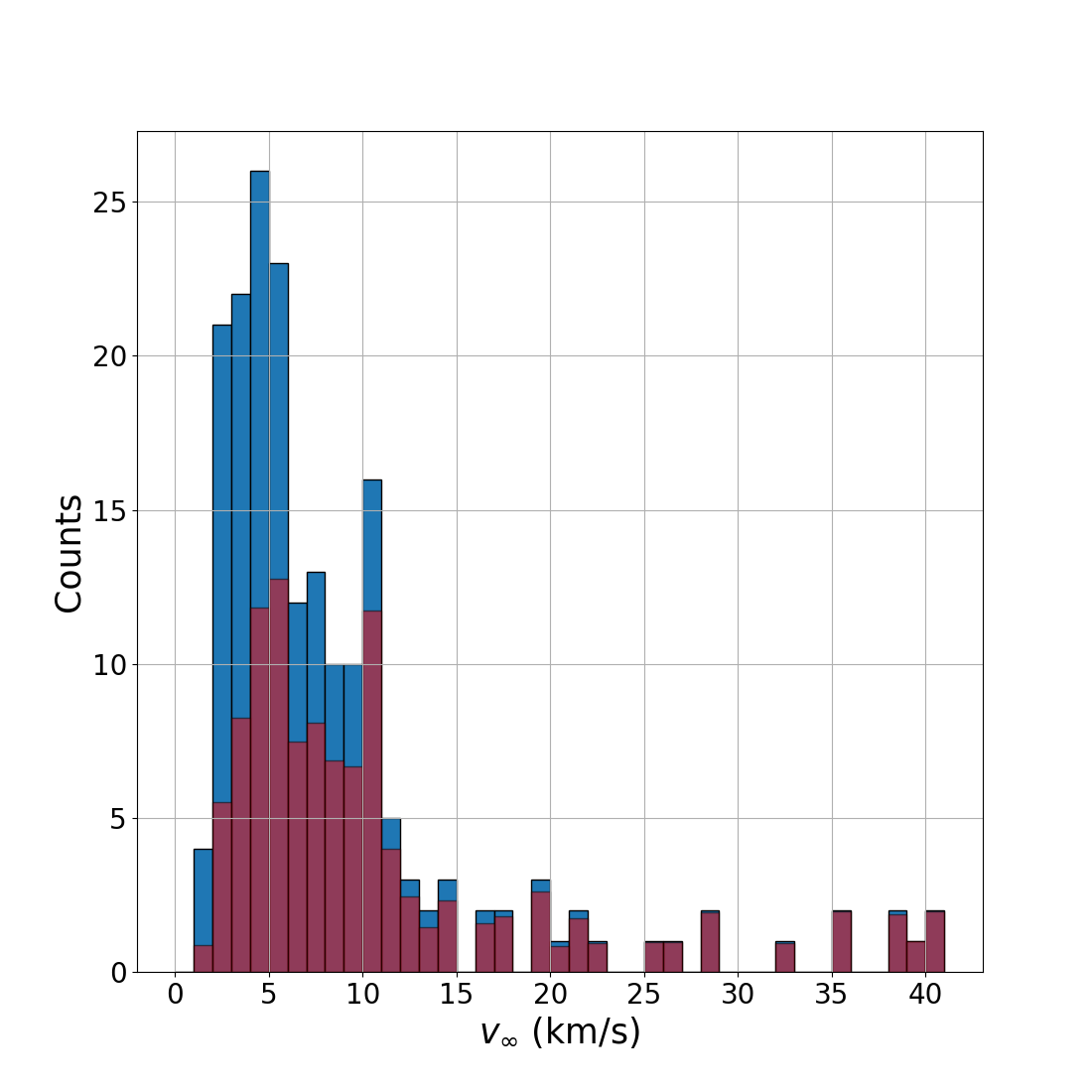}
    \caption{Blue: Distribution of the velocity at infinity of the dust particles detected by \emph{Cassini}, where $v_\infty$ has been obtained with our approximation of Eq.~\ref{eq:vrel at saturn} and with the data of \cite{kempf_micrometeoroid_2023}. The histogram is similar to theirs (see main text for details). Red: Debiased distribution of the velocity at infinity of the particles as they enter Saturn's Hill sphere. Since they are more focussed by Saturn's gravity, slow particles represent a larger fraction of the \emph{Cassini} sample than of the unfocussed distribution.}
    \label{fig:hist vinfty}
\end{figure}
From the data in the supplementary information of \citet[][Table S2]{kempf_micrometeoroid_2023}, we can first find the distance $r_i$ of \emph{Cassini} to Saturn when it detected particles with $r_S,z_S$ (in polar coordinates with $z_S$ the height from Saturn's symmetry plane). Furthermore, we can estimate the relative velocity $v_{\rm rel}$ of the particle at Saturn (which is $v_\infty$ in Saturn's frame) with the orbital parameters. From $a,e,i$, the relative velocity at Saturn's orbit is given by
\begin{equation}\label{eq:vrel at saturn}
    v_\infty = \sqrt{v_r^2 + (v_\phi\ \text{cos}(i) - v_{\phi,S})^2 + (v_\phi \text{sin}(i))^2}
\end{equation}
where $v_r, v_\phi$ are the radial and azimuthal velocity of the particle on its orbit inclined by an angle $i$ with respect to Saturn's orbit, and $v_{\phi,S}$ is the orbital velocity of Saturn at 9.5 AU (the inclination and eccentricity of Saturn are neglected). Our derived histogram of $v_\infty$ is given in Fig.~\ref{fig:hist vinfty} (blue bars). It is not exactly the same as the histogram from the aforementioned article for two reasons. First, the computation of $v_\infty$ is different from that of the article but should yield a similar result (unfortunately, $v_\infty$ is not provided directly in the Table). Second, we do not have the same number of trajectories. Indeed, we could not use the data with a semi-major axis $a$ equal to 0 or with missing data in $a,e,i$. Among the $163$ different particles, only $125$ particles were kept and a total of $193$ different trajectories \footnote{The \emph{Cosmic Dust Analyser} instrument only had 2 inclined grids resulting in determining two components of the particles velocity. The third component could be inferred with at most four different possibilities, leading sometimes to different trajectories for a same particle.}.
Interestingly, since the escape velocity from the solar system at Saturn's orbital radius is $13.6$~km/s, particles with a $v_\infty$ relative to Saturn (whose orbital velocity is $9.6$~km/s) larger than $\sim 23$~km/s are necessarily on a hyperbolic trajectory around the Sun.

Nonetheless, our histogram and theirs are very similar, which validates our method: the maximum is for the bin $4-5$~km/s and the mean value of $1/v_\infty^2$ yields $4.3$~km/s. 

If we now attribute to each particle $i$ a weight $1/F_g(i)$ instead of $1$ when counting the particles in each velocity bin, the histogram becomes the red one. This one represents the distribution of $v_\infty$ at Saturn's Hill sphere which yields the observed (blue) distribution by \emph{Cassini}. Therefore, the relevant value for $v_\infty$ is $v_m$ such that:
\begin{equation}
    {v_{\rm m}}^{-2} = \frac{\sum\limits_{i}\ (v_{\infty,i})^{-2}/F_g(r_i, v_{\infty,i})}{\sum\limits_{i}1/F_g(r_i, v_{\infty,i})}.
\end{equation}
We find $v_m=6.8$ km/s. From now on, we will use this new value as the value of $v_\infty$ in our work. Note that this decreases the value of $\Theta$ by a factor $2.5$ compared to the previous value of $4.3$~km/s. 
With this new value of $v_\infty$, a numerical fit of the shielding gives $s(r/R_p)=1-0.34\exp\left(- \frac{(r/R_p-1)}{0.21} \right)$. 

Hence, for our work, the local gravitational focusing for the rings is
\begin{equation}\label{eq:fgstar with 6.8 km/s for v_infty}
    F_g^*(r)=\left(1-0.34\cdot \exp{\left(- \frac{(r/R_p-1)}{0.21} \right)}\right)\cdot\big(0.5 + 0.82\Theta\big)\ \
\end{equation}
where $\Theta$ is computed using $v_\infty=6.8$~km/s. If one kept $v_\infty=4.3$~km/s, the fit of the shielding function $s$ would be $s(r/R_p)=1-0.33\exp\left(-\frac{r/R_p-1}{0.20}\right)$.

\begin{figure}
    \centering
    \includegraphics[width=\columnwidth]{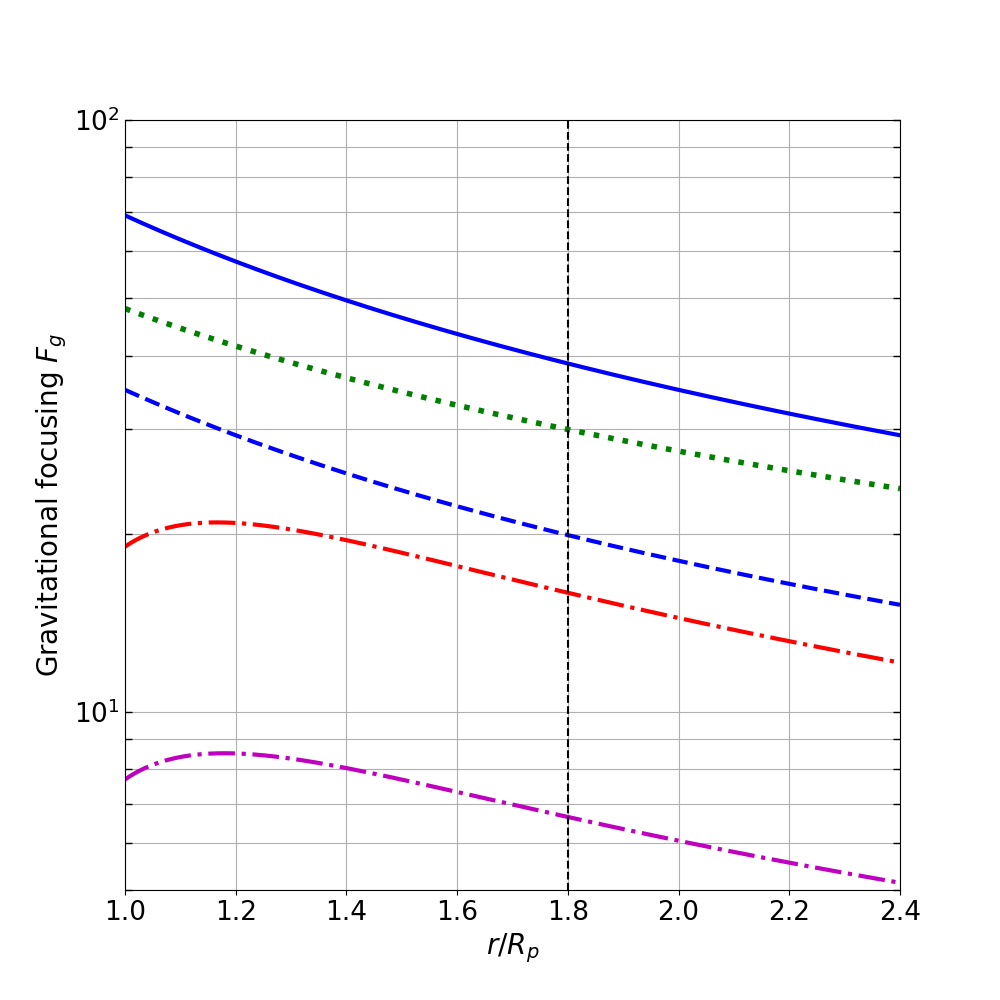}
    \caption{Semi-log plot of different gravitational focusing expressions. Solid and dashed blue: Eq.~\ref{eq:fg_sphere full} and \ref{eq:fg_sphere local} (respectively $F_g$ and $F_g^*$ for a sphere), with $v_\infty=4.3$ km/s. Dotted green: \cite{estrada_constraints_2023}. Dash-dotted red: local gravitational focusing for rings with $v_\infty=4.3$ km/s (Eq.~\ref{eq:fg_generalised rings}). Bottom magenta curve: same as the red one with $v_\infty=6.8$ km/s. The vertical dashed line corresponds to the distance $r_B=1.8$ $R_S$.}
    \label{fig:new dFg rings new vinfty}
\end{figure}

\subsection{Comparisons with previous works}\label{sec:comp_fg_discussion} 

Figure~\ref{fig:new dFg rings new vinfty} shows the gravitational focusing for the different cases explained previously. The top blue curve is the classic \emph{global} gravitational focusing for a sphere (Eq.~\ref{eq:fg_sphere full}) with $v_\infty=4.3$~km/s, while the dashed blue curve shows its \emph{local} version (Eq.~\ref{eq:fg_sphere local}), about two times smaller. The dash-dotted red curve is the \emph{local} gravitational focusing for a planar disc instead of a sphere; it is slightly below the previous one because a disc is harder to hit than a sphere of same radius. Finally, the dashed magenta curve is like the red one but with $v_\infty=6.8$ instead of $4.3$~km/s, which causes a drop of a factor of about 2.5. In the end, we find that the rings are bombarded $\sim 5$ times less than what is used for instance in \citet{estrada_constraints_2023} (green dotted curve).
\begin{figure}
    \centering
    \includegraphics[width=\columnwidth]{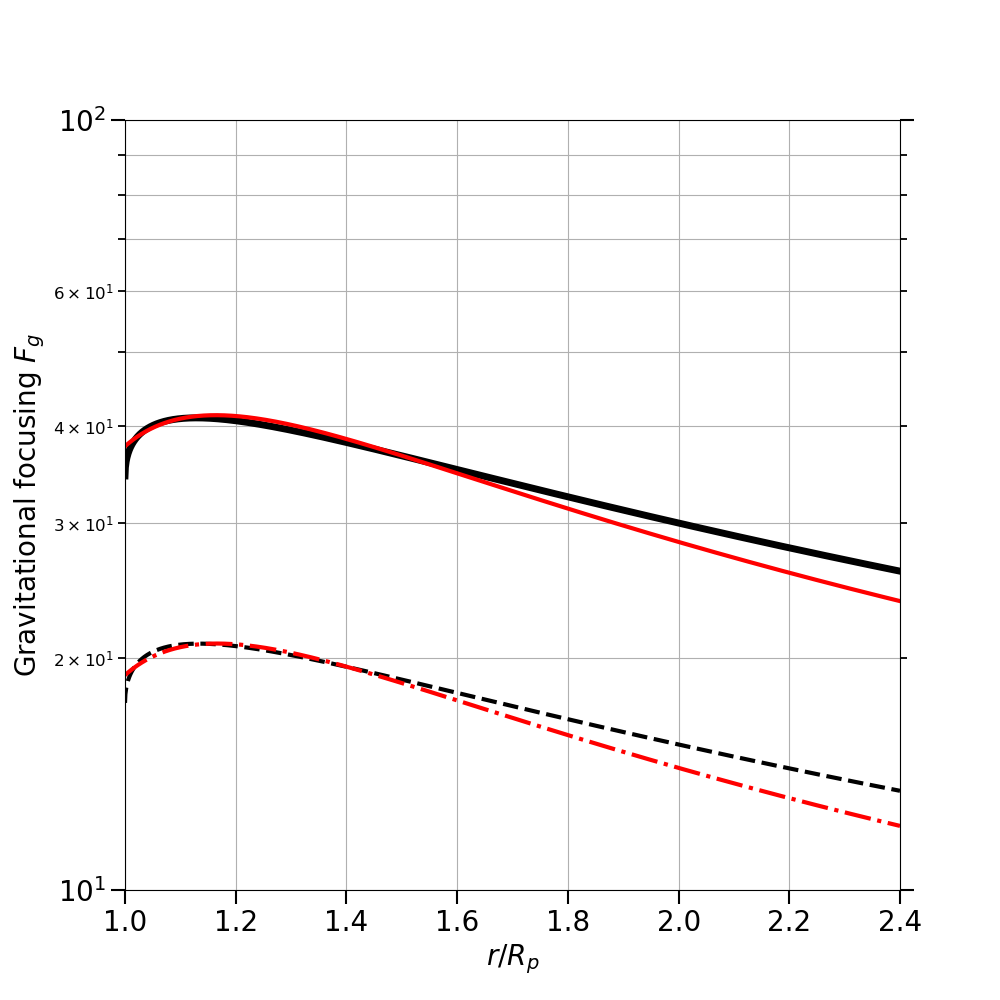}
    \caption{Semi-log plot comparing the expression of \cite{kempf_micrometeoroid_2023} (solid thick black curve), the global (solid red curve) and local gravitational focusing (dash-dotted red curve, same as in Fig.~\ref{fig:new dFg rings new vinfty}) we found for the rings, using $v_\infty=4.3$~km/s. The dashed black curve accounts for the gravitational focusing of the mentioned article when removing the factor $2$ in their equation 12 of their supplementary material.}
    \label{fig:comp_fgKempf}
\end{figure}

In the supplementary material of \cite{kempf_micrometeoroid_2023}, they gave a different expression of $F_g(r)$ (their Eqs. 11 to 13) based on conservation of energy and number density \citep{colombo_earths_1966, spahn_e_2006}. In Fig.~\ref{fig:comp_fgKempf}, both our global $F_g(r)$ (with $v_\infty=4.3$ km/s) and the one from the mentioned articles are extremely similar although we had two different ways of deriving the same physical process (solid black and red lines). However, their $F_g(r)$ is still the global one (because for the number density equations they calculated what is inside $r' \leq r$). Removing the factor $2$ in the term $2\mu/rv_{\infty}^2$ in their Eq. 12 in order to compute $d(r^2F_g(r))/d(r^2)$, the black curve falls at the same level as the dash-dotted red curve, i.e. our local $F_g^*(r)$.

From these few points and a small discussion in Appendix~\ref{app:discuss Fg}, we conclude that Eq.~\ref{eq:fg_generalised rings} correctly describes the focus needed to estimate the local bombardment rate at distance $r$. We apply it and use the magenta curve (Eq.~\ref{eq:fgstar with 6.8 km/s for v_infty} precisely) of Fig.~\ref{fig:new dFg rings new vinfty} in our analysis of the evolution of the rings below --\,directly yielding five-fold larger exposure ages. Note that the expression of $\widetilde{F_g^*}$, which gives the bombardment rate at a given radius, could also be applied to satellites on orbits of radius $r$. Since these bodies have a mass and size much smaller than the planet, they can be considered as a particle of an extended ring of radius $r$. However, the shielding factor will be different for a satellite \citep[see Eq. 19 of][]{lissauer_bombardment_1988} than the one we derived for the rings (see Appendix~\ref{sec: final exp dFg rings}).

\section{Rings evolution description}
\label{sec:mathematical model}
We consider the problem to be axisymmetric so that the description can be done over the $r$ axis. The rings' particles evolve through their angular momentum exchanges such that the rings behave like a sheared viscous fluid. They also evolve from the impacts of dusty particles, with different sizes and injected velocities in Saturn's Hill sphere. These impacts will lead to additional source terms in the mass and angular momentum continuity equations but also changes in the chemical composition.

\subsection{Basic equations}\label{sec:basic equations}
\subsubsection{Global evolution of the rings under viscosity and bombardment}
The rings can be seen as a pressureless and viscous disc that has a Keplerian rotation facing MB. We define $\Sigma(r,t)$ the surface density of the rings at a distance $r$ and time $t$. From \cite{pringle_accretion_1981}, the first equation ruling the rings evolution is the mass continuity equation which reads
\begin{equation}
\label{eq:mass_cons}    
\frac{\partial\Sigma}{\partial t} + \frac{1}{r}\frac{\partial}{\partial r}(\Sigma v_r r) = \simdot(r), 
\end{equation}
where $v_r$ is the radial velocity of the particles and $\dot{\sigma}_{\text{im}}(r)$ is the addition of mass due to micrometeoroid bombardment at a distance $r$. It is decomposed into \citep{cuzzi_compositional_1998}:
\begin{equation} \label{eq:sigma_im_dot}
    \simdot(r) = 2 A(\tau) F_{\infty} F_g^*(r)
\end{equation} 
where $F_g^*(r)$ is the new local gravitational focusing (Eq.~\ref{eq:fg_generalised rings}), $F_\infty$ is the one-sided flat plate flux entering Saturn's Hill sphere and $A(\tau)$ is the impact probability as a function of the normal optical depth $\tau$ defined as $\tau=3\Sigma/4\rho_p r_p$ with $\rho_p, r_p$ the particles density and radius. \cite{cuzzi_compositional_1998} use $A(\tau) = \Big[ 1 - e^{(-\tau/\tau_s)^p} \Big]^{1/p}$, where the parameters $\tau_s$ and $p$ are fitted to $0.515$ and $1.0335$ respectively. In our work, we adopt $\tau_s=0.5$ and $p=1$ for simplification\footnote{The values of $A(\tau)$ computed with our formula and theirs differ at most by $0.013$, near $\tau=0.25$, where $A(\tau)\approx 0.4$.}: 
\begin{equation}\label{eq: simplification A(tau)}
    A(\tau)=1-e^{(-2\tau)}.
\end{equation}

The second equation is the continuity equation of angular momentum with source terms the viscous torque and the momentum carried by the impactors of MB. 
This paper focuses on the compositional variation of the rings and therefore neglects the dynamical effects of the bombardment --\,which is equivalent to assuming that the impactors hit the rings with the same specific orbital angular momentum as the ring particles.
The effects of the radial drift due to bombardment, ballistic transport, and other possible external sources of angular momentum will be treated in a forthcoming work.
By combining the two continuity equations, the temporal evolution of the surface density follows
\begin{equation} \label{eq:main_eq_sd}
    \frac{\partial \Sigma}{\partial t} = \frac{3}{r}\frac{\partial}{\partial r}\Big[\sqrt{r}\frac{\partial}{\partial r}\left(\Sigma\nu\sqrt{r}\right)\Big] + \simdot
\end{equation}
where $\nu$ is the viscosity \citep{daisaka_viscosity_2001}. The first RHS term stems from the radial drift due to the viscous evolution and the second one from the addition of mass.

To model the viscous evolution, we used the algorithm presented in \cite{salmon_long-term_2010}, and were able to recover their figures 3, 4, 10 and 12, with $\dot\sigma_{\rm im}=0$, which validates our code. Note that the viscosity prescription used by \cite{daisaka_viscosity_2001} depends on the optical depth $\tau$ as well; however, this dependence arises from self-gravitating effects, which are dominated by the largest particles. Hence, in agreement with \cite{salmon_long-term_2010}, we use $r_p=1$~m for the evaluation of $\tau$ used in the computation of $\nu$.

\subsubsection{Evolution of the pollution}

In addition to Eq.~\ref{eq:main_eq_sd}, we need to model the evolution of the fraction of non-icy material -- also called pollutant -- inside the rings. We introduce then $\Sigma_{\text{p}}$ the surface density of the pollutant and $f$ the ratio of $\Sigma_{\text{p}}/\Sigma$. Similarly as Eq.~\ref{eq:main_eq_sd}, we have
\begin{equation} \label{eq:sigma_pol simple}
    \frac{\partial\Sigma_{\text{p}}}{\partial t} = \frac{3}{r}\frac{\partial}{\partial r}\Big[f\sqrt{r}\frac{\partial}{\partial r}\left( \Sigma \nu \sqrt{r} \right) \Big]\ + \eta\simdot,
\end{equation}
where the first term of the right-hand side corresponds to the advection of pollutant due to the viscous spreading (internal to the ring system), and $\eta$ is the fraction of pollutant material that keeps its polluting properties after the impact, thus the external source of darkening of the rings from the bombardment. Here, it is assumed that the micrometeoroids are only composed of dust. To take into account a mixture of ice and dust, $\eta$ should be replaced by $\eta f_{\text ext}$ where $f_{\text ext}$ is the non-icy fraction of the impactors. In this paper, we make the conservative assumption that impactors are made of dust only so $f_{\text{ext}}$ = 1.

The value of $\eta$ is not very high and difficult to estimate properly. \cite{cuzzi_compositional_1998} argue for  a range of $\eta$ from 0.05-0.3 to explain some sharp contrasts in different optical regions. They usually take a mean value of 0.1 which will be a reference value.

\subsection{A possible change of paradigm}\label{sec:change of paradigm}
While $\eta$ represented the fraction of accreted material that kept its polluting properties after impact, called retention efficiency in \cite{cuzzi_compositional_1998}, it was originally defined as the fraction of impactor that survived the impact and was accreted unaltered into the rings particles \citep{doyle_radiative_1989}. Hence, what $\eta$ represents is not clear from one work to another. More importantly, depending on the definition given to $\eta$, it must be applied either to Eq.~\ref{eq:main_eq_sd} (surviving fraction definition) or Eq.~\ref{eq:sigma_pol simple} (retention efficiency definition). As a consequence, we suggest to note $\zeta$ \emph{the mass fraction of the impactor that is accreted} onto the ring after impact, while $\eta$ defines the fraction of this accreted mass which counts as pollutant.

\cite{hyodo_pollution_2025} have found that a micrometeoroid hitting non porous ice at 25-35 km/s is fully vaporised together with a certain amount of target material. Only $1$ to $3\%$ of the impactor is subsequently re-accreted onto the rings (more generally, $\zeta \sim 3\%$ of the global vaporised ejecta was re-accreted, which the authors noted $\eta$). Besides, they find that possible vaporisation and loss from the rings particles can happen. For these reasons, a redefinition of Eq.~\ref{eq:mass_cons} to add new source terms is considered: $\dot{\sigma}_{\text{loss}}$ is the loss of mass due to vaporisation upon impact. Again, we neglect here the change of specific angular momentum of the rings due to the escaping steam, which is equivalent to assuming this material evaporates with the same specific orbital angular momentum as the ring particle, before being lost from the rings system as described by \cite{hyodo_pollution_2025}. We end up with  
\begin{equation} \label{eq:new_mass_cons}
    \frac{\partial\Sigma}{\partial t} + \frac{1}{r}\frac{\partial}{\partial r}(\Sigma v_r r) = \simdot + \dot{\sigma}_{\text{loss}},
\end{equation}
where $\dot{\sigma}_{\text{loss}} < 0$ and is decomposed into what is lost from the impactor just after the impact $-(1-\zeta)\simdot$ \citep[consistently with][]{doyle_radiative_1989}, and what is lost from vaporisation of the target material. 
Noting $\mu_p$, $\mu_i$ the ratio of dust and ice vaporised masses on the target to the mass of the impactor, and $\zeta_i$ the accretion efficiency for water vapour, one ends with 
\begin{equation}\label{eq:MSD eq new}
    \frac{\partial\Sigma}{\partial t}=\frac{3}{r}\frac{\partial}{\partial r}\Big[ \sqrt{r}\frac{\partial}{\partial r}\left(\Sigma\nu\sqrt{r}\right)\Big] + (\zeta-A) \simdot
\end{equation}
where $A=\mu_pf(1-\zeta) + \mu_i(1-f)(1-\zeta_i)$ accounts for the vaporisation of the target.
Similarly\,:
\begin{equation}\label{eq:Spol MSD eq new}
\frac{\partial\Sigma_{\text{p}}}{\partial t}=\frac{3}{r}\frac{\partial}{\partial r}\Big[ f\sqrt{r}\frac{\partial}{\partial r}\left(\Sigma\nu\sqrt{r}\right)\Big] + (\eta\zeta - B)\simdot
\end{equation}
where $B=\mu_pf(1-\eta\zeta)$ corresponds to the ring dust vaporised by the impact and lost.
In the end, what really matters for the evolution of the pollution of the rings is neither $\eta$ nor $\zeta$ alone, but the product $\eta\zeta$.
Note that now, the coefficients in front of $\simdot$ can be negative.
With no vaporisation at all ($\zeta=1$, $\mu_p=\mu_i=A=B=0$), we recover the case study of \cite{estrada_constraints_2023}.

\subsection{Analytical time evolution of the fraction of dust $f$}\label{sec:analytical evolution f}
By definition, the temporal evolution of $f$ is 
\begin{equation}\label{eq:df_dt}
 \frac{\partial f}{\partial t}=\frac{\partial}{\partial t}\left(\frac{\Sigma_{\text{p}}}{\Sigma}\right) = \frac{1}{\Sigma}\left(\frac{\partial\Sigma_{\text{p}}}{\partial t}-f\frac{\partial\Sigma}{\partial t}\right).
\end{equation}
By expanding this equation using Eqs.~\ref{eq:MSD eq new} and~\ref{eq:Spol MSD eq new}, the advection terms will not disappear if $f$ is not uniform. Therefore, solving for $f(r,t)$ requires numerical computations.
However, working on the mean value of $f$ called $f_m$ bypasses this problem and allows to find an analytical solution. Eliminating the radial dependences in Eqs. \ref{eq:MSD eq new} and \ref{eq:Spol MSD eq new}, we get:
\begin{equation}
    \frac{\partial f_m}{\partial t} = \frac{\simdot}{\Sigma(t)}\cdot\left(af_m^2 -bf_m+c\right)
\end{equation}
with $a=\mu_p(1-\zeta)-\mu_i(1-\zeta_i)$, $b=\mu_p(1-\eta\zeta)-\mu_i(1-\zeta_i)+\zeta$, $c=\eta\zeta$ and $\simdot=2F_\infty \overline{F^*_g}A(\tau)$ where $\overline{F^*_g}$ is the mean value of $F_g^*(r)$ over a certain distance. 
Since $b>a$, assuming $f_m\ll 1$ (which is the present situation) yields a simple differential equation in the form
\begin{equation}\label{eq: dfm_dt}
    \partial_t f_m =(-b f_m+c)/t_{\rm MB}.
\end{equation}
Note that this equation also applies when we suppose no vaporisation of the target material ($\mu_p=\mu_i=0$), with $b=\zeta$ and still $c=\eta\zeta$, so that $\partial_t f_m =\frac{\zeta}{t_{\rm MB}}(- f_m+\eta)$ in this case.

We introduce $t_{\rm MB}=\Sigma(t)/\simdot$. Depending on the origin of the rings, we have two different evolutions for $\Sigma(t)$. Assuming a \emph{recent formation}, the surface density does not evolve too much in time such that $\Sigma$ is constant. On the contrary, for the case of an \emph{old formation} of the rings that do evolve viscously, we use the asymptotic form of the surface density $\Sigma(t)=\Sigma_0(t/t_0)^{-1/2}$ \citep{crida_complex_2014, crida_age_2025}. Hence, $t_{\rm MB}$ can be written as $t_{\rm MB}=\tau_{\rm MB}(t/t_0)^{-x}$ with $\tau_{\rm MB}=\Sigma_0/\simdot=C^{st}$, $x=1/2\ \rm or\ 0$ in the case with or without viscosity, $t_0=4.5\ \text{Gyr}$ and $\Sigma_0$ represents the current density of the rings if those were a uniform ring, giving $\Sigma_0=365$ kg.m$^{-2}$. In the end, $f_m(t)$ reads
\begin{equation}\label{eq:fm global}
\begin{split}
    f_m(t) &= \frac{c}{b} \cdot \left[ 1 - {\rm exp}\left(-\left(\frac{b\,t}{\tau_{\rm MB}} \right)\left(\frac{4t}{9t_0} \right)^{x}\right)\right] \\ & =\frac{c}{b} \cdot \left[ 1 - {\rm exp}\left(-\left(\frac{t}{\tau_x} \right)^{1+x}\right)\right]
\end{split}
\end{equation}
with $\tau_x=\left( \tau_{\rm MB}/b\right)^{1/(1+x)} \left(9t_0/4 \right)^{x/(1+x)}$ so $\tau_x=\tau_{\rm MB}/b$ in the non viscous case ($x=0$) and $\tau_x=\left( \frac{\tau_{\rm MB}}{b}\right)^{2/3} \left(\frac{9}{4}t_0 \right)^{1/3}$ otherwise ($x=1/2$).

It seems easy to invert this equation to derive the exposure age $t$ for an observed dust fraction $f_m$. In particular, we notice that taking into account the viscous evolution ($x=\frac12$ instead of $0$) slows the growth of $f_m$ since $\frac{4t}{9t_0}<1$, which yields a longer exposure age.
However, if $t \gg \tau_x$, i.e. $t\gg\tau_{\rm MB}/b$, $f_m\approx \frac{c}{b}$ and the solution becomes degenerate. If the time since the rings have been exposed to MB is too large compared to the characteristic time $\tau_{\rm MB}/b$, then we only measure the limit $c/b$ (that is $\eta$ in absence of vaporisation of the target) and \emph{not an exposure age}.

\subsection{Introducing space weathering}

In the previous subsections, dust is assumed to stay for ever in the rings, and remain detectable as pollutant for ever. However, one may question this hypothesis under the effect of space weathering \citep{esposito_space_2024}. Until now, studies on the dust evolution only took into account MB as a source of darkening in Saturn's rings, but they do face solar irradiation and can be altered in their composition. 
Such deterioration process can be studied as it has been for organics trapped in icy environment \citep{gudipati_-situ_2012}. 
One can thus imagine an equilibrium between deposit and degradation of pollutant. 
SW for ice could also be taken into account but this process mainly changes the structure of H$_2$O, going from crystalline to amorphous and vice-versa \citep{fama_radiation-induced_2010,dalle_ore_impact_2015}. Henceforth, water will still be detected as water while silicates might not be detected as 'dust' or 'pollutant' once altered. 

Not only SW and MB are source terms for the evolution of dust, but physically can play together on another aspect. SW is thought to act over a certain depth from the surface. Without renewal of this layer, dust could still be trapped inside the rings particles. In this case, MB could excavate this trapped dust and bring it close to the surface where SW occurs. We could also think of the collisions between the rings particles that would also scatter the blocks and expose the dust to irradiations. These different processes therefore could maintain a constant cleaning process.

Note that SW is known to make asteroids spectrum redder or bluer depending on their composition \citep{dellagiustina_variations_2020} and could happen on Saturn's rings. However, it is not clear if SW could really reduce the fraction of pollutant through time \citep{brisset_laboratory_2025}. Hence, we study this reduction from a theoretical point of view rather than a realistic process occurring in Saturn's rings.

From these different assumptions, we suppose that SW constantly alters pollutant through time and is proportional to the fraction of dust $f$ under the form $-\dot{\sigma}_{\rm sw}f$ in $\partial_t\Sigma_{\rm p}$ where $\dot{\sigma}_{\rm sw}$ is the alteration rate of pollutant per unit surface exposed. The new temporal evolution of $\partial_t f_m$ reads
\begin{equation}
    \partial_tf_m=(-bf_m+c)/t_{\rm MB} - f_m/t_{\rm sw} = (-\tilde{b}f_m+c)/t_{\rm MB}
\end{equation}
where like for MB we have $t_{\rm sw}=\tau_{\rm sw}(t/t_0)^{-x}$ with $\tau_{\rm sw}=\Sigma_0/\dot{\sigma}_{\rm sw}$, $x\in\{0,\frac12\}$ and $\tilde{b}=b+\tau_{\rm MB}/\tau_{\rm sw}$ (we recover $b$ if the space weathering is inefficient and $\tau_{\rm sw}\to\infty$). Thus, the solution is simply given by Eq.~\ref{eq:fm global} with $\tilde{b}$ instead of $b$. An efficient space weathering makes $\tilde{b}>b$ thus decreases the limit value, and accelerates the convergence towards it. 

Having scenarios where rings start with an initial fraction of dust $f_i$ leads to a new term such that
\begin{equation}\label{eq: fm w/ sw + IF}
        f_m(t) = \left(f_i - f_\infty\right)\cdot \text{exp}\left(- \left(\frac{t}{T_x}\right)^{1+x}\right) + f_\infty
\end{equation}
where
\begin{eqnarray}
    {T_x}^{1+x} &=& \left(\tau_{\rm MB}/\tilde{b}\right)\left(9t_0/4\right)^{x}\\
    f_\infty  &=&  \frac{c}{\tilde{b}} = \frac{\eta\ \zeta}{\mu_p(1-\eta\zeta)-\mu_i(1-\zeta_i)+\zeta + \frac{\dot{\sigma}_{\rm sw}}{\simdot}}
\end{eqnarray}
With this formula, one sees that the fraction of dust converges to a final fraction $f_\infty$ independent of the initial condition $f_i$ and of the final surface density $\Sigma_0$.

\section{Results}\label{sec:results}

In this section, the formulas found above are used to derive the exposure age of Saturn's rings, for two scenarios:
\begin{itemize}
    \item \emph{Static rings ($x=0$)\,:} The viscous evolution of the surface density is neglected, and $\Sigma$ only evolve through mass loading due to the bombardment. This scenario is appropriate for young rings or for the C ring which has a very low density hence viscosity. In this case, we compute the value of $A(\tau)$ for the B and C rings, using the data from \cite{durisen_large_2023}. We have then $A(\tau_{\rm C})=0.18$ and $A(\tau_{\rm B})=0.74$ for the C and B ring respectively. The corresponding value of $\tau_B$ ($0.68$) may be inconsistent with those found in the literature \citep{miller_composition_2024}. We use this data for comparative purposes only.
    \item \emph{Viscous rings ($x=\frac12$)\,:} The rings evolve under the effect of viscosity on long timescales. This scenario is appropriate for the main rings (B and A rings), assuming they formed together with Saturn. To simplify the computation as $\Sigma$ evolves in time, we make the assumption that $A(\tau) = 1$.
\end{itemize}
Remember that $x$ is the negative of the power index of $\Sigma(t)/\dot\sigma_{\rm im}$, so $x=0$ (here called \emph{static rings}) could also be seen as viscously evolving rings under a bombardment rate decreasing as $\dot\sigma_{\rm im}\propto t^{-1/2}$.

In what follows, the probably exogenic flux measured by \emph{Cassini} $F_\infty = 2.7\cdot 10^{-16}$ \mbdim \citep{kempf_micrometeoroid_2023} is adopted. Using the certainly exogenic flux, which is 4 times smaller, would yield exposure ages 4 times larger in the static rings case and $4^{2/3}\approx2.5$ times larger in the viscous case.
Section~\ref{sec:4.1} studies the most basic case of rings starting as pure water ice, with no vaporisation of material upon impact ($\mu_i=\mu_p=0$). This latter hypothesis is released in section~\ref{sec:4.2} and the possibility of an initial pollution ($f_i>0$) is also introduced, while space weathering is added in section~\ref{sec:4.3 SW}.

\subsection{Exposure ages with no loss from the rings due to MB}\label{sec:4.1}
In Section \ref{sec:analytical evolution f}, we found the analytical expressions of $f_m(t)$. Inverting Eq.(21) yields the exposure age to reach a given pollutant fraction $f_m$
\begin{equation}\label{eq:t_exp no visco}
    t_{\text{exp}}= \tau_x\cdot\left(-{\rm ln}\left( 1-f_m\frac{b}{c}   \right)    \right)^{1/(1+x)}.
\end{equation}
In this subsection, since we have no vaporisation and loss of the target material, $\mu_p=\mu_i=0$ which gives $b=\zeta$ and hence $\frac{b}{c}=\frac{1}{\eta}$. Furthermore, if $f_m\frac{b}{c} \ll 1$ (i.e. $f_m\ll\eta$), the exposure time can be approximated by 
\begin{equation}
    t_{\rm exp}\approx \tau_x\left(\frac{f_m}{\eta}\right)^{1/(1+x)}.
\end{equation}
In the particular case $x=0$, $t_{\rm exp}\approx (\tau_{\rm MB}/b)\cdot (f_mb/c)=(\tau_{\rm MB}f_m)/(\eta\zeta)$. This linear approximation is similar to what \cite{durisen_large_2023} uses. They define the exposure age for the B ring at $r=r_B$ with
\begin{equation}\label{eq:t_exp DE23}
\begin{split}
    t_{\text{pol}}^B \approx &\ 120\ \text{Myr}\ \left(\frac{\Sigma_B}{520\ \text{kg.m$^{-2}$}}\right)\left(\frac{2.7\cdot 10^{-16}\ \text{[S.I]}}{F_{\infty}}\right) \\ & \left(\frac{30}{F_g(r_B)}\right) \left(\frac{0.1}{\eta}\right)\left(\frac{f_B}{0.93\%}\right)
\end{split}
\end{equation}
Assuming the current mean dust fraction in the B ring equal to $f_B = 0.93\%$ \citep{zhang_exposure_2017}, $\Sigma_B=520\ \text{kg.m$^{-2}$}, F_g(r_B)=6.7$ (new value of $F_g^*(r=r_B)$ with Eq.~\ref{eq:fg_generalised rings}) and introducing $\zeta$, Eq.~\ref{eq:t_exp DE23} becomes
\begin{equation}\label{eq:t_exp simplified DE23}
        t_{\text{pol}}^B \approx 5.40\ \text{Gyr}\ \left(\frac{2.7\ 10^{-16}\ \text{\mbdim}}{F_{\infty}}\right)\left(\frac{0.1}{\eta}\right)\left(\frac{0.1}{\zeta}\right)
\end{equation}
for the B ring. Furthermore, they found a similar exposure age for the C ring, so that Eq.\ref{eq:t_exp simplified DE23} applies also to the C ring. We see that our new value of the gravitational focus and the introduction of $\zeta$ naturally make the default exposure age five billion years, thereby killing the main argument in favour of young rings.

\begin{figure}
    \centering
    \includegraphics[width=\columnwidth]{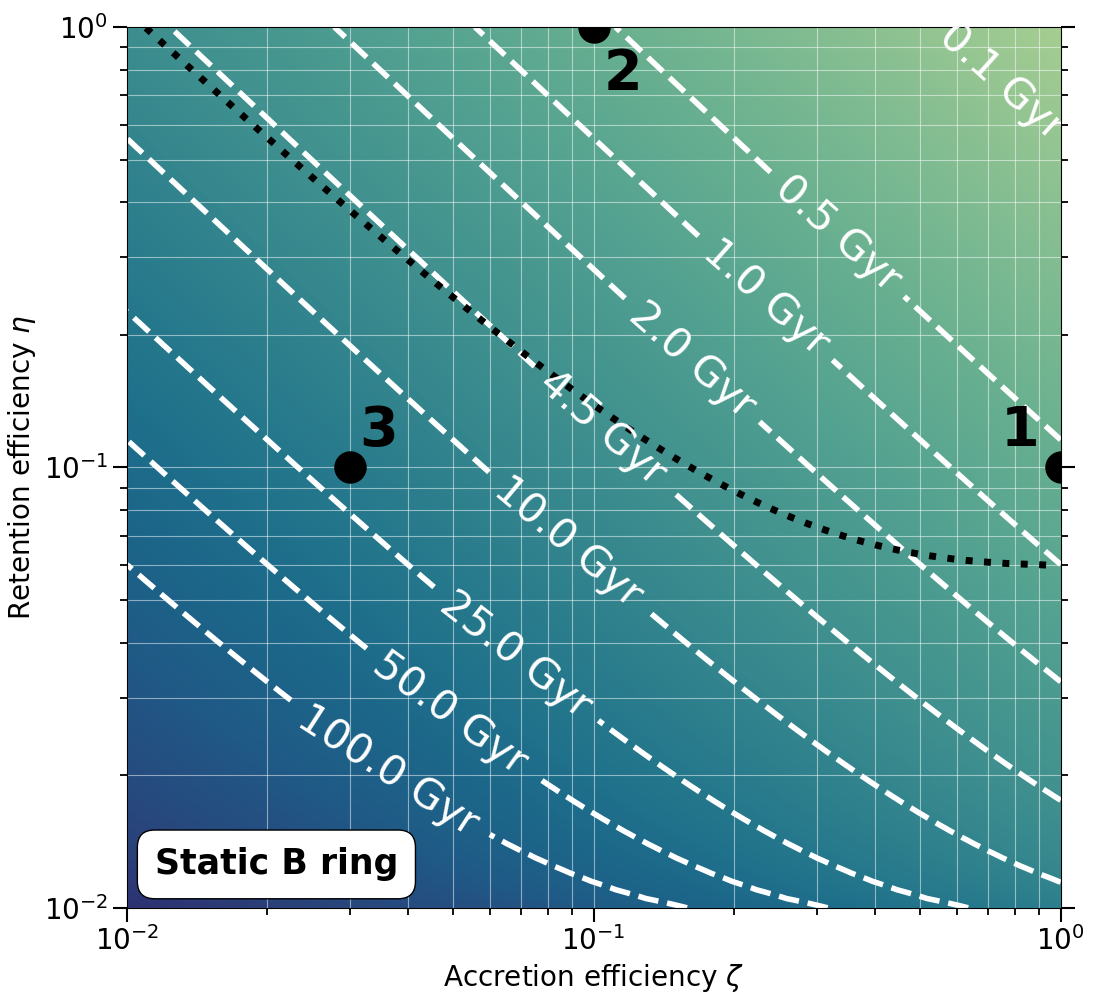}    
    \includegraphics[width=\columnwidth]{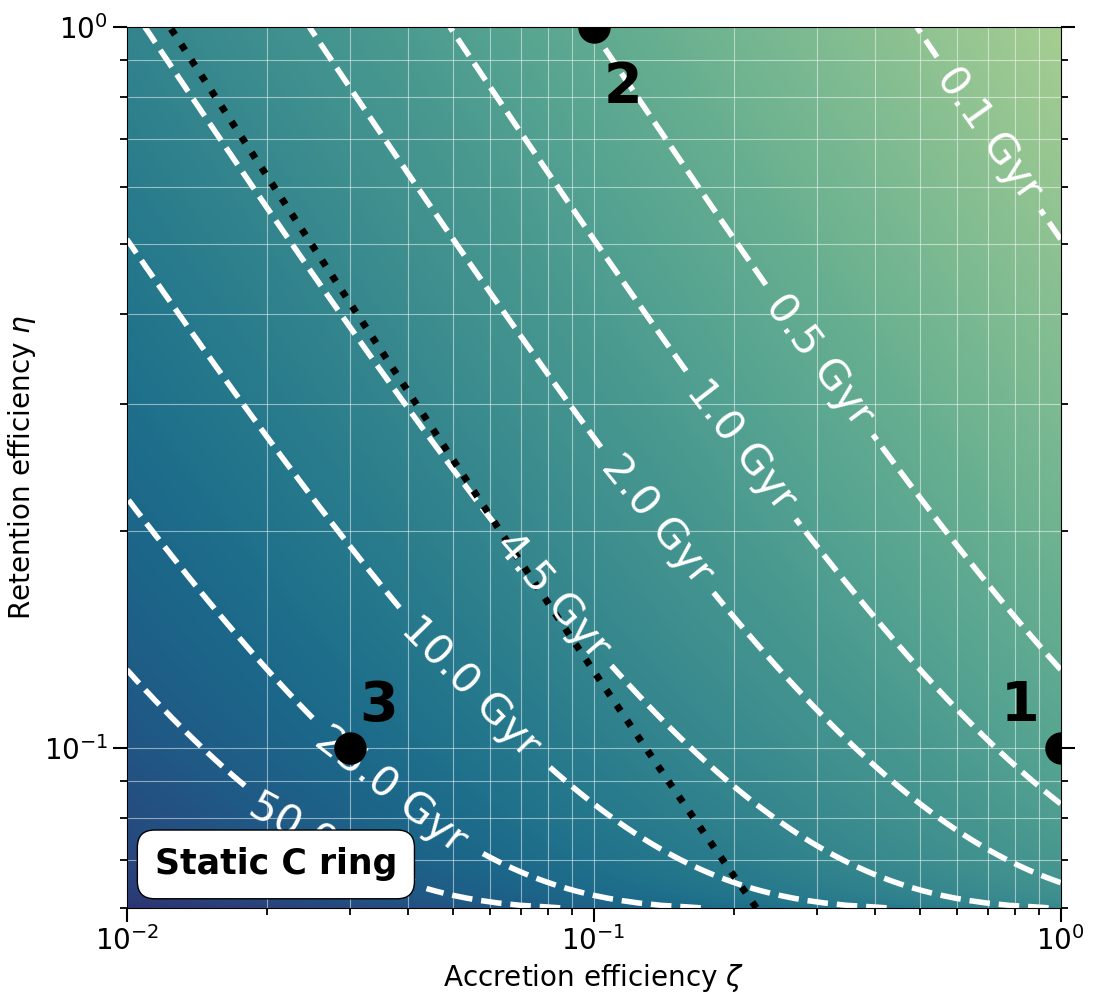}    
    \caption{Colormaps of the exposure age (in Gyr) to reach a fraction of dust $f_m$ depending on $(\eta, \zeta)$. The three dots represent the values of $t_{\rm exp}$ for the three reference cases of $(\eta, \zeta)$: 1. \cite{estrada_constraints_2023}, 2. \cite{doyle_radiative_1989}, 3. \cite{hyodo_pollution_2025} (for $\zeta$ only). Top panel: B ring where $f_m=0.93\%$. Bottom panel: C ring case with $f_m=6\%$. The different ages are 1: 0.58 Gyr, 2: 0.56 Gyr, 3: 19.4 Gyr for the B ring. For the C ring we have 1: 0.73 Gyr, 2: 0.50 Gyr, 3: 24.43 Gyr. Here the rings do not evolve viscously. The black dotted level curve is the one of $4.5$ Gyr for the other ring.}
    \label{fig:texp colormap no visco}
\end{figure}

In the following, we explore the $(\eta,\zeta)$ parameter space to look for all possible exposure ages. Three reference cases are highlighted for their sets of accretion and retention efficiencies: \cite{estrada_constraints_2023}, \cite{doyle_radiative_1989} and \cite{hyodo_pollution_2025}. The first two papers have two different but roughly equivalent assumptions for the impact. The first one stated that the whole impactor is accreted but only $10\%$ keep its polluting property ($\eta=0.1$ and $\zeta=1$, set 1). The second paper suggested that only $10\%$ was accreted and chemically unaltered ($\eta=1$, $\zeta=0.1$, set 2). In the case of \cite{hyodo_pollution_2025}, they found $\zeta=0.03$ for a non porous target. However, they did not estimate a value for the retention efficiency $\eta$ so we assume $\eta=0.1$ to complete set 3.
\begin{figure}
    \centering
    \includegraphics[width=\columnwidth]{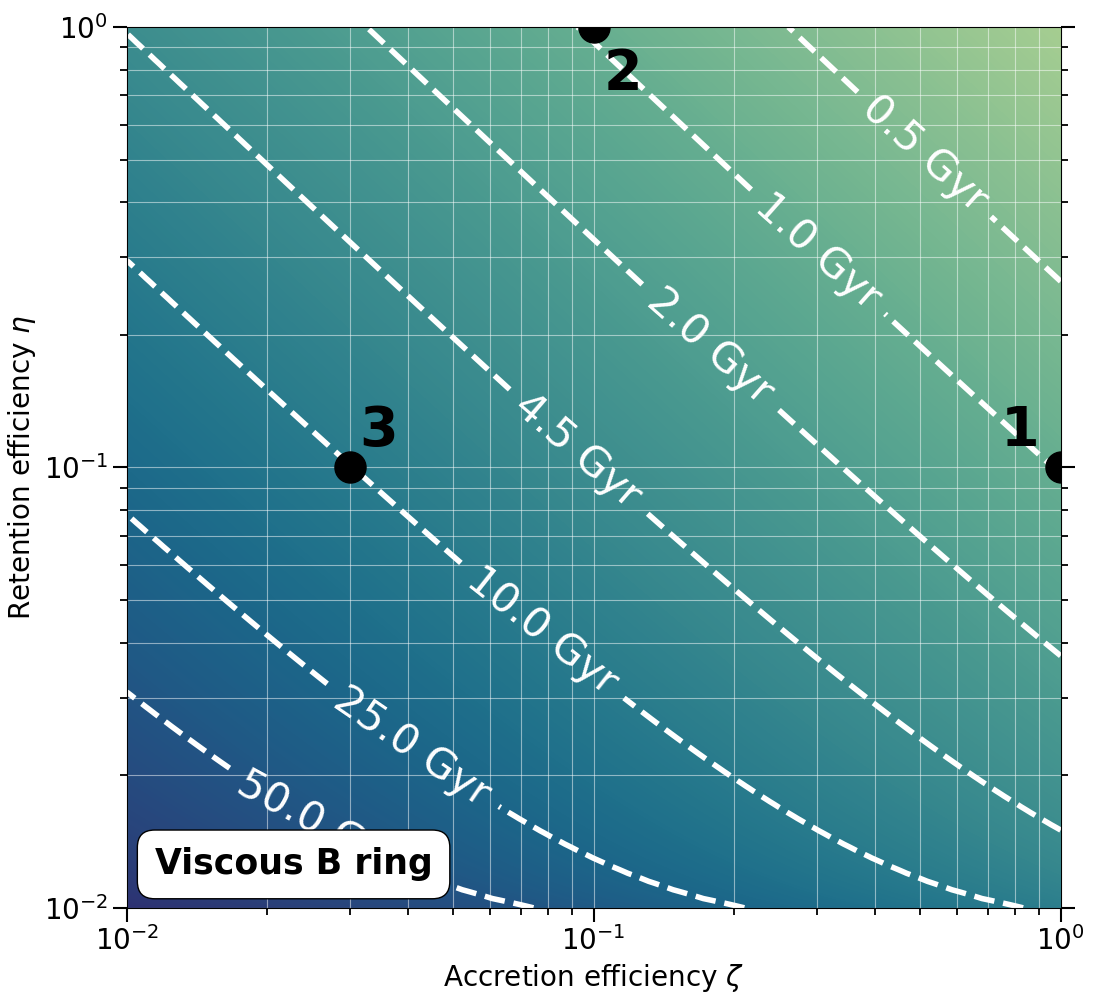}    
    \caption{Same figure as Fig.~\ref{fig:texp colormap no visco} but where the viscosity is taken into account in the evolution model of $\Sigma(t)$. Only the B ring is studied. The exposure ages from the dots $1,2\ \text{and}\ 3$ are 1: 0.98 Gyr, 2: 0.95 Gyr and 3: 10.2 Gyr.}
    \label{fig:texp colormap with visco}
\end{figure}

In Fig.~\ref{fig:texp colormap no visco} are given the colormaps of the exposure age to reach the actual fraction of dust in the B ring (top panel) and the C ring (bottom panel), as a function of $\eta$ and $\zeta$ in the case of static rings. Some level curves are plotted and the three dots marked $1, 2$ and $3$ represent the three sets mentioned before. In the case of the C ring, the range for $\eta$ only goes from $>6\%$ to $100\%$ because Eq.~\ref{eq:t_exp no visco} does not have a solution for $\eta<f_m$. 
In each panel, the level curve for $t_{\rm exp}=4.5$~Gyr in the other ring is overplotted in dotted black. It shows that the conditions $\eta\times\zeta\approx 0.012$ and $\zeta<0.1$ yield an exposure age equal to that of Saturn simultaneously in both the B and the C rings (consistent with Eq.~\ref{eq:t_exp simplified DE23}).
We also notice that with $\zeta=0.03$ \citep{hyodo_pollution_2025}, the exposure age is always larger than $2$ Gyr.

On such timescales, the viscous evolution should be taken into account. On Fig.~\ref{fig:texp colormap with visco}, we show the exposure age of the B ring in the viscous case (that is: $x=1/2$). For the sets of parameters 1 and 2, the age is almost doubled to $\sim 1$~Gyr, because the rings being initially more massive, this slows down the darkening process \citep[see also][]{crida_are_2019, estrada_constraints_2023}. Remember that in this simple model, the impact probability $A(\tau)$ is assumed to be always $1$, while as the rings spread and their density decreases, one may expect $A(\tau)<1$, which would yield even longer exposure ages.

From these two figures, it appears very clearly that there exists an infinite number of values of $(\eta,\zeta)$ which would yield any given exposure age for a ring. In addition, Fig.~\ref{fig:texp colormap no visco} shows that the exposure ages of the B and C rings would be very similar for $\eta\geqslant 0.1$. Our knowledge of the physics of the high-velocity impact of a micrometeorite on a more or less porous ice block being rather low, $\eta$ and $\zeta$ are almost unknown from first principles. Previous estimates in the literature are hand-wavy or based on a fit to reproduce particular structures. Instead of using a more or less arbitrary value of $\eta$ and $\zeta$ to estimate a ring exposure age \citep[as is classically done, e.g.][]{iess_measurement_2019}, one could revert the proposition\,: considering the rings to be as old as Saturn (and assuming no vaporisation upon impact), we would suggest $\eta=0.10-0.15$, $\zeta=0.1$ (which seems reasonable).

\subsection{Exposure ages with possible loss from the rings due to MB}\label{sec:4.2}
\begin{figure}
    \centering
    \includegraphics[width=\columnwidth]{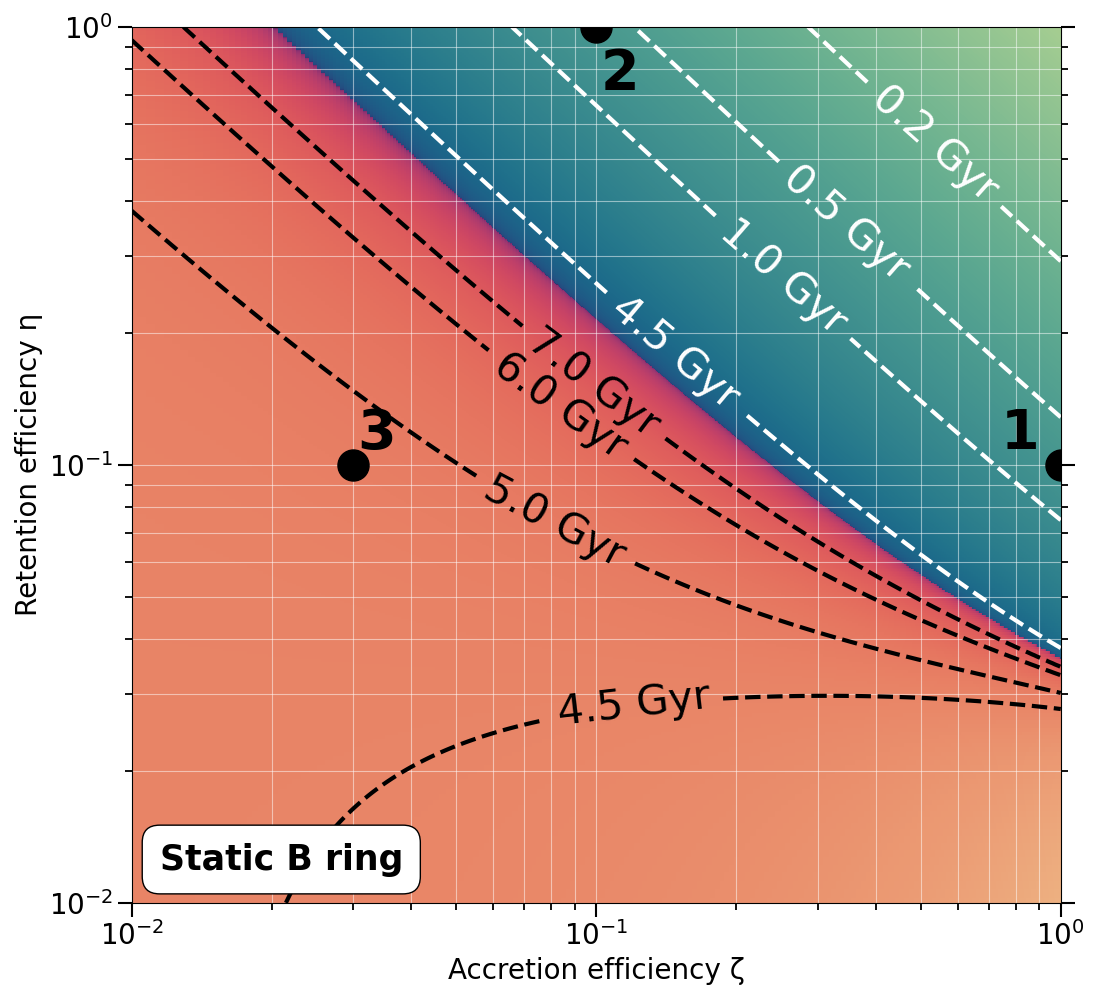}    
    \includegraphics[width=\columnwidth]{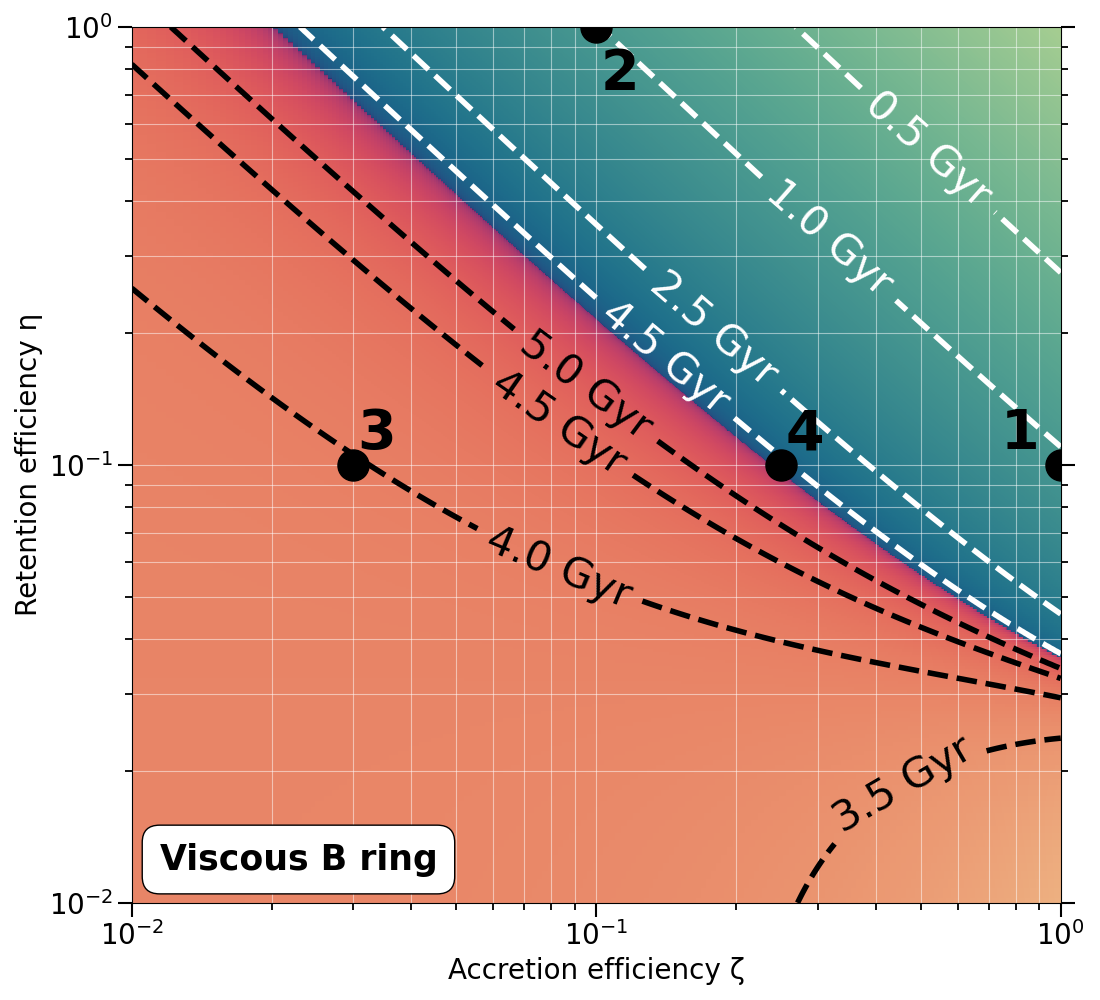}
    \caption{Colormaps of the exposure age for the B ring for both evolution scenarios. Vaporisation and loss from the rings after the impacts are possible in this case ($\mu_p=3$, $\mu_i=0.8$). The blue regions represent the domain where the rings start at pure ice while the red ones are when the initial pollutant fraction is around $5\%$. The dot number 4 corresponds to the choice of parameters adopted in our test simulation ($\eta=0.1, \zeta=0.25$) to see the ring and dust profiles after $4.5$ Gyr in Section \ref{sec:4.2}.}
    \label{fig:texp colormap B ring vaptar}
\end{figure}
\begin{figure}
    \centering
    \includegraphics[width=\columnwidth]{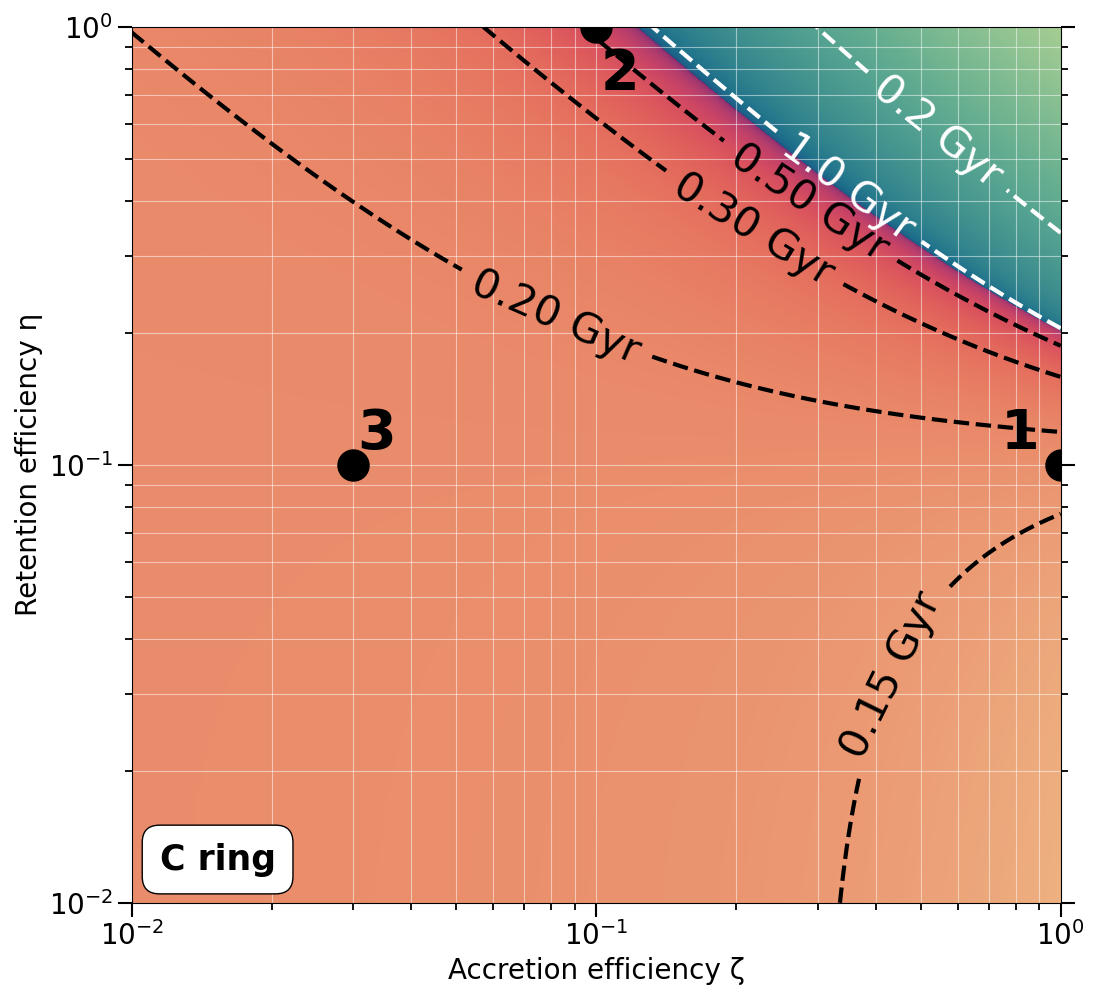}
    \caption{Colormap for the exposure age of the C ring in the static case. The colored regions are the same as in Fig.~\ref{fig:texp colormap B ring vaptar} but with $f_i=10\%$ (red area). The vaporisation parameters are still the same used in for the previous plot.}
    \label{fig:texp C ring fi}
\end{figure}
Allowing vaporisation of the target material upon impact can heavily affect the evolution of the rings dust fraction. To do so, we will use the values provided in Figure S3 of \cite{hyodo_pollution_2025} which gives the amount of target material vaporised in impactor masses depending on the impact velocity. It is important to know that the values are obtained assuming that the target and impactor have the same density. Around Saturn, the mean impact velocity is around $28\ {\rm km.s^{-1}}$. Our impactors can hit two types of material: ice and dust. For the second one, we can read that the amount of vaporised mass for dust ($\rho=3000\ {\rm kg.m^{-3}}$) is around $\mu_p=3\ [m_{\text{imp}}] $. However, to target ice ($\rho=900\ {\rm kg.m^{-3}}$), the value we infer and take is around $\mu_i=0.8\ [m_{\text{imp}}]$. Remember that in this case only, it is like assuming the impactor is made of ice. It is not clear how the value of $\mu_i$ would change if the impactor is denser than the target (Hyodo R., private communication) but we keep the inferred value as an example for the purpose of the work. Note that the vaporised target mass is much larger if the impactor hits a dust grain than water ice. This could be the cause of the larger fraction of dust observed in the ring rain compared to what is present in the rings \citep{hsu_situ_2018, linti_cassinis_2024, linti_dust_2025}, even if this dust rich fraction in the ring rain is not well understood yet. 

As ice and dust can have different accretion efficiencies once emitted as vapour in the system, $\zeta_i$ and $\zeta$ could be different, but for simplification, we assume $\zeta_i=\zeta$. While the limit fraction $c/b$ was $\eta$ in the previous subsection, it has a more complex expression with vaporisation. There are more chances that it becomes smaller than the current fraction in the rings so that Eq.~\ref{eq:fm global} does not have a solution.

On the other hand, the vaporisation process allows to consider an initial dust fraction $f_i>0$ (but still $f_i\ll1$ so that Eq.~\ref{eq: dfm_dt} remains a valid approximation). The exposure age is then defined as 
\begin{equation}
    t_{\rm exp} = \tau_x\cdot\left(-{\rm ln}\left( \frac{f_m-c/b}{f_i- c/b}   \right)    \right)^{1/(1+x)}.
\end{equation}
In Fig.~\ref{fig:texp colormap B ring vaptar} is represented the colour map of $t_{\rm exp}$ for the B ring. The blue areas are for the case with $f_i=0$ while the red ones are for $f_i=5\%$. In the area where the rings start as pure ice rings, the main difference with the previous section ($f_i=0$, no vaporisation) is that the level curves moved towards the top right corner and the bottom left part of the diagram has no solution for $f_i=0$. One can easily imagine that if more dust is vaporised on the target than deposited by the impact (low $\eta$ and $\zeta$), the rings can not be darkened. To have the B ring as old as the planet (bottom panel, viscous evolution case), a reasonable choice for $\eta, \zeta$ vary around $0.15-0.2$ with $f_i=0$ or $0.1$ if $f_i=5\%$. Note that the level curves in the region $f_i>0$ actually depend on $f_i$. Adding the initial fraction of dust in the rings as a free parameter opens a new dimension in our parameter space, and offers more possibilities for $\eta$ and $\zeta$. Seeking all possible solutions is not the scope of this paper, but we stress that there are possibilities to have an exposure age of $4.5$ Gyr.

The problem is different for the C ring. As shown in Fig.~\ref{fig:texp C ring fi}, the large ages, which are not higher than 1 Gyr, are located on a narrow line on the top right corner, at the interface of the red and blue sectors, independently of $f_i$. It leads to larger values of $(\eta,\zeta)$ than for the B ring. 

\begin{figure}
    \centering
    \includegraphics[width=\columnwidth]{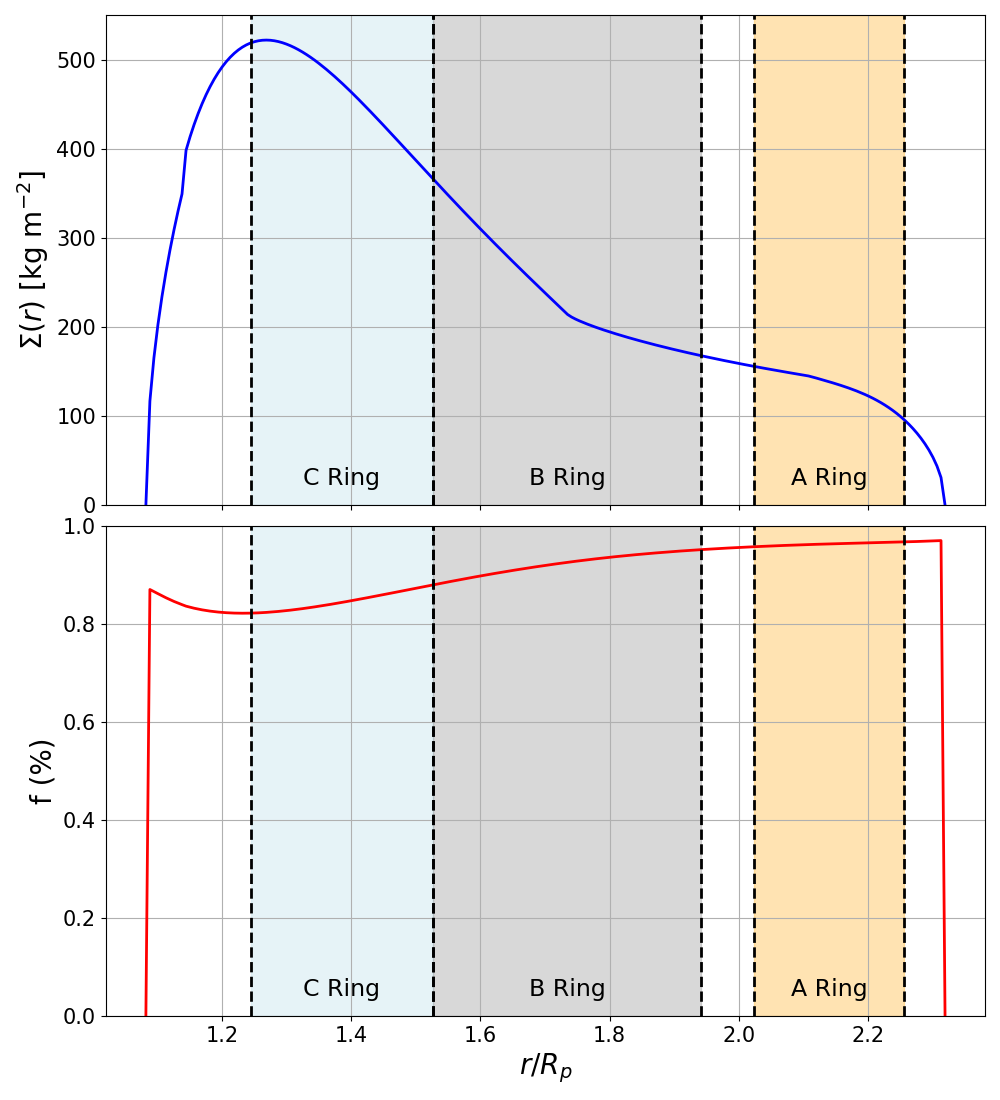}
    \caption{Surface density and dust fraction (in $\%$) profiles after $4.5$~Gyr of evolution under viscosity and bombardment. The location of the main rings (A, B, C) are shown with the colored areas (orange, grey, light blue for the A, B and C ring, from right to left respectively). Here, we adopted $\eta=0.1, \zeta=0.25, \mu_p=3, \mu_i=0.8$.}
    \label{fig: sigma f vap example}
\end{figure}

\paragraph{Numerical simulation}
Using our code, we integrate numerically the evolution of the ring profile described by Eqs.~\ref{eq:MSD eq new} and~\ref{eq:Spol MSD eq new} with vaporisation. Like in \cite{salmon_long-term_2010}, the rings are modelled with an initial Gaussian profile centred at $110\ 000$ km with an initial mass of $150\ M_{\rm Mimas}$ (Mimas masses) and the simulation domain extends from $r_{\rm min}=65\ 000$ km to $r_{\rm Roche}=140\ 000$ km over $201$ cells. Furthermore, we adopt a smooth transition of the viscosity regime around $Q=2$ with a width of $0.1$ to deal with the discontinuity associated with the \citet{daisaka_viscosity_2001} prescription. For the optical depth, we use here an effective radius of $15$~cm for a density of $\rho_p=900$ kg/m$^{3}$, with $\Sigma_B=600\ \rm{kg.m^{-2}}$, $\Sigma_A=150\ \rm{kg.m^{-2}}$ \citep{miller_composition_2024} such that the optical depth in the B ring is roughly $3.3$ and $0.8$ for the A ring \citep{cuzzi_evolving_2010}. Finally, the rings start with $100\%$ ice and, guided by the above calculations, we choose $\eta=0.1, \zeta=0.25$ to possibly reach the current fraction of pollutant in the B ring over the age of the Solar System, according to the bottom panel of figure \ref{fig:texp colormap B ring vaptar} (dot number 4). 

After $4.5$~Gyr, $\Sigma(r)$ and $f(r)$ are given in Fig.~\ref{fig: sigma f vap example}. The rings profile has some similarities with what we observe today in shape, except that the C and D rings are too massive and that the Cassini division is missing. Besides, but consistently, the Cassini Division and the C ring should be much more polluted than what we have in our simulations \citep{hedman_connections_2013}. These features are probably structures that have been carved recently in old rings \citep[e.g.][]{baillie_formation_2019,noyelles_formation_2019} and modelling them is not the scope of this paper which focusses on the global evolution of the mass and composition of the rings. Hence, we focus here on the main rings. We recover a mean fraction of dust $f$ of $0.93\%$ in the B ring and less than $1\%$ in the A ring. This value is slightly higher than the mean fraction of $f_A=0.78\%$ \citep{zhang_exposure_2017} but is the good order of magnitude. In addition, we observed the growth of the mean fraction of pollutant with time to slow down and almost reach a stationary state in the last billion years of the simulation, as expected from Eq.~\ref{eq:fm global}.
\begin{figure}
    \centering
    \includegraphics[width=\columnwidth]{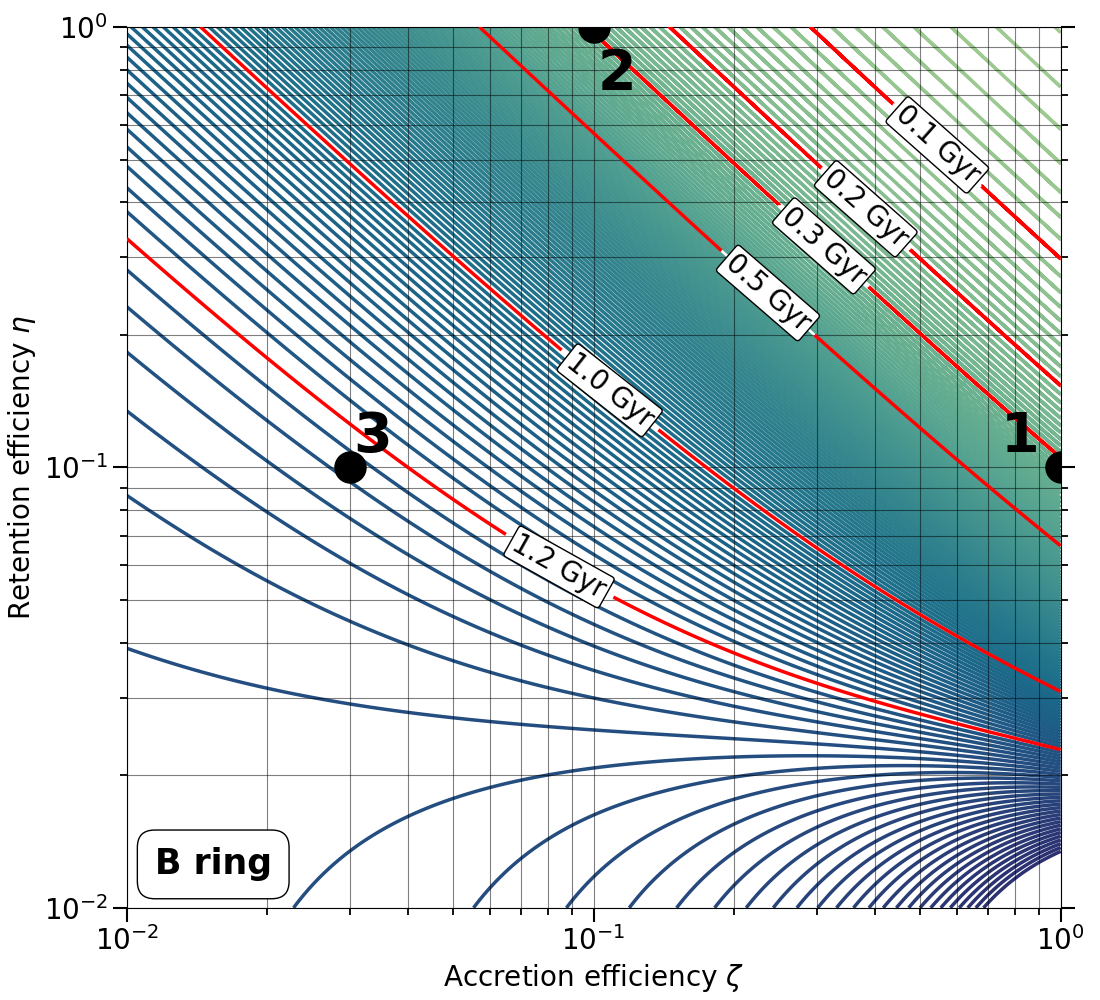}
    \caption{Different level curves of $\tau_{\rm sw}$ (every $10$~Myr) such that the fraction $f_B$ is reached after 4.5 Gyr with an initial fraction of dust equal to $10\%$, and no vaporisation upon impact. Set 1: This scenario is possible if $\tau_{\rm sw} \approx 320$ Myr. The characteristic space weathering times for sets 2 and 3 are $290$ Myr and $1.2$ Gyr respectively.}
    \label{fig:level curves tsw for texp visco + SW + IF}
\end{figure}
\subsection{Consequences of SW on exposure age}\label{sec:4.3 SW}

The other possible mechanism that can remove or alter the dust is the space weathering which also allows to start with an initial fraction $f_i>0$.
Since space weathering plays a role on long timescales, and we look for a solution to have old rings with little pollution, we use $x=1/2$ (viscously evolving rings) in this section. Inverting Eq.~\ref{eq: fm w/ sw + IF}, the exposure age is 
\begin{equation}
    t_{\rm exp}= T_{(x=1/2)}\left( -{\rm ln}\left( \frac{f_m- c/\tilde{b}}{f_i-c/\tilde{b}} \right) \right)^{2/3}.
\end{equation}
Fixing $f_i=0.1$ and assuming no vaporisation ($\mu_i=\mu_p=0$), $t_{\rm exp}$ is a function of 3 parameters: $\eta,\zeta$ and $\tau_{\rm sw}$. For any set ($\eta,\zeta$), one can find the value of $\tau_{\rm sw}$ such that $t_{\rm exp,B}=4.5$ Gyr (Fig.~\ref{fig:level curves tsw for texp visco + SW + IF}). For instance, if the characteristic time $\tau_{\rm sw}$ is roughly $320$ Myr, the rings can be initially more polluted than today and reach through cleaning the actual fraction of dust over the age of the Solar System with the parameters of set 1 ($\zeta=1, \eta=0.1$).

\begin{figure*}
    \centering
    \includegraphics[width=\columnwidth]{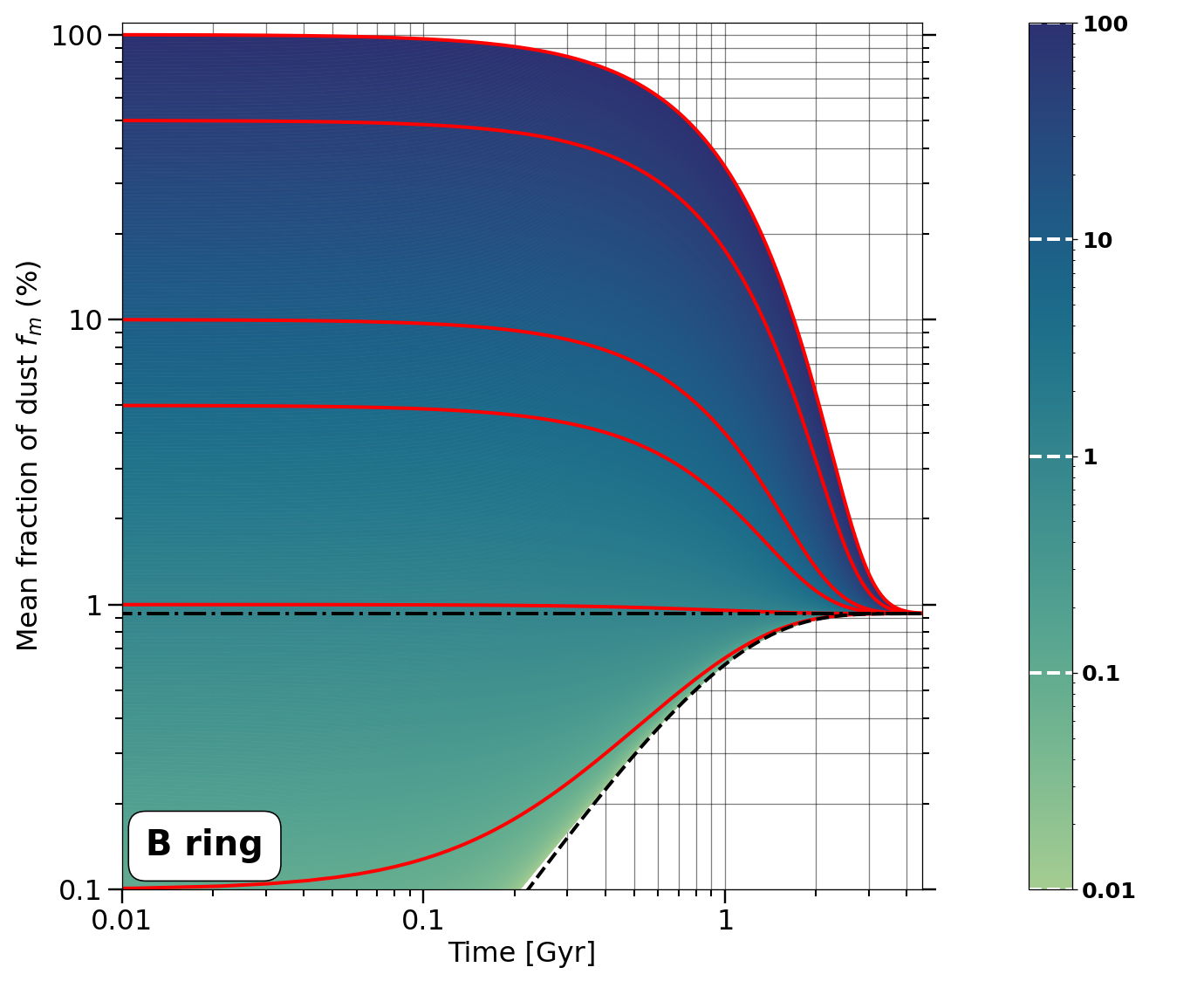}
    \includegraphics[width=\columnwidth]{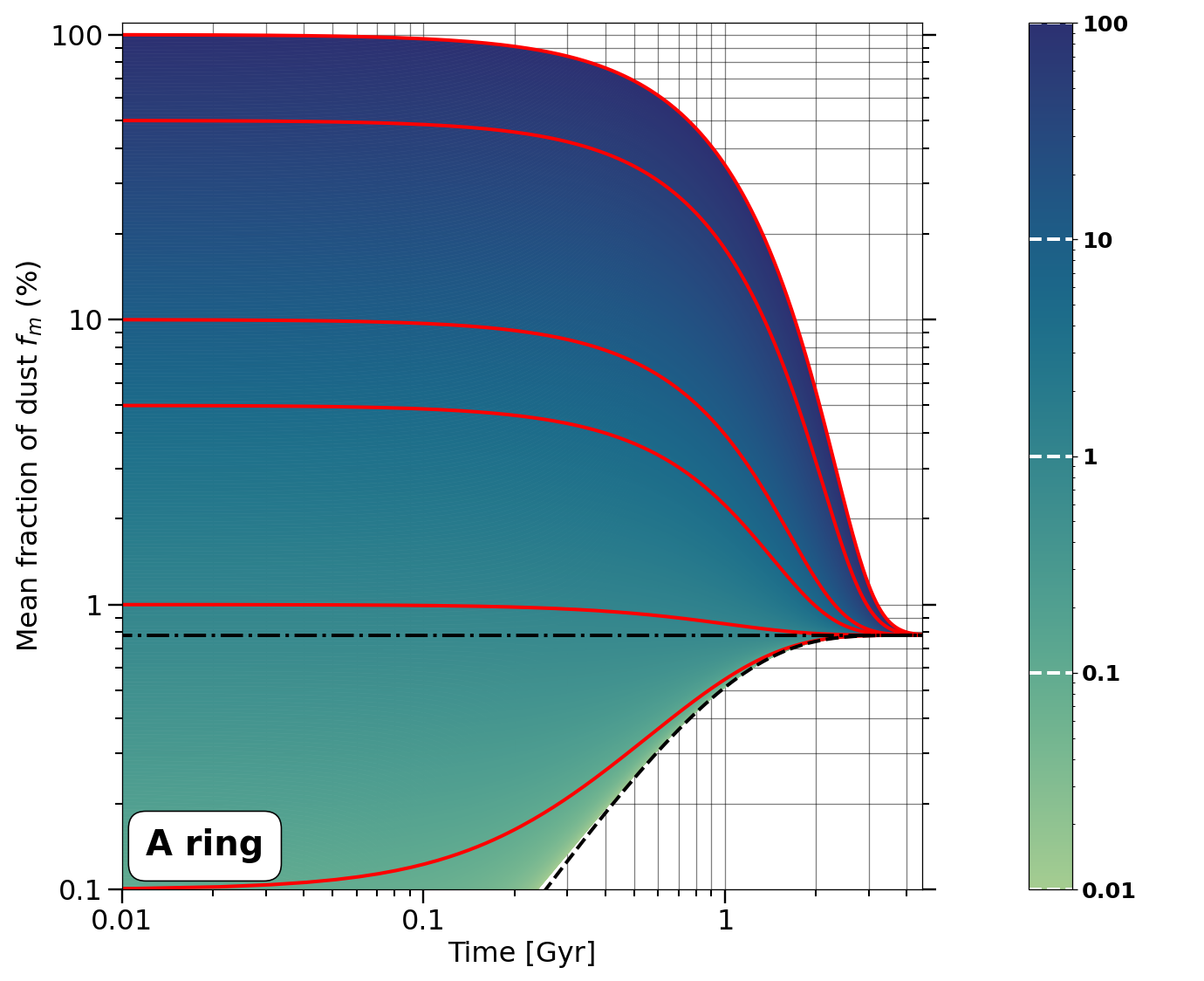}
    \caption{Evolution of the mean fraction of dust in the B and A rings for different initial fractions $f_i$ (colour-coded in $f_i$). All evolutions have the same values for $(\eta,\zeta,\tau_{\rm sw})=(0.1, 1, 320\ {\rm Myr})$. Some initial fractions are highlighted in red. From top to bottom: $f_i=100\%, 50\%, 10\%, 5\%, 1\%, 0.1\%$. Black dashed line: $f_i=0\%$. Black dash-dotted lines: current fraction of dust in the B ring ($0.93\%$) and the A ring ($0.78\%$, derived from the data of \cite{zhang_exposure_2017}).}
    \label{fig:evol fm 1st eta zeta tau in B/A ring}
\end{figure*}

Setting now $\eta=0.1,\zeta=1, \tau_{\rm sw} = 320$ Myr, the time evolution of $f_m$ can be computed for various initial fractions $f_i$. This is shown for the B ring in the left panel of Fig.~\ref{fig:evol fm 1st eta zeta tau in B/A ring}. All curves converge after $\sim 3$~Gyr to $c/\tilde{b}\approx f_B$. In other words, no matter the initial fraction $f_i$, we only measure after $4.5$ Gyr the limit value of $f_m(t)$ as $t\to\infty$; and neither $f_i$ nor $t$ can be determined.

In the case of the A ring, the mean value of $F_g^*$ is $\overline{F_g^*}=5.7$. Keeping the same values of $\eta, \zeta, \tau_{\rm sw}$ as for the B ring case above, we obtain the evolutions of $f_m(t)$ in the right panel of Fig.~\ref{fig:evol fm 1st eta zeta tau in B/A ring}. Interestingly, the curves converge towards a value of $f_m$ very close to $f_A=0.78\%$. This surprising result can be explained: the ratio of $\overline{F_g^*}/f_m$ for the B and A rings are very similar ($7.4$ for the B ring and $7.3$ for the A ring). The different pollution rates of the A and B rings could therefore be explained by the same chemical evolution process, with $\eta=0.1, \zeta=1, \tau_{\rm sw}=320$~Myr. 
\begin{figure}
    \centering
    \includegraphics[width=\columnwidth]{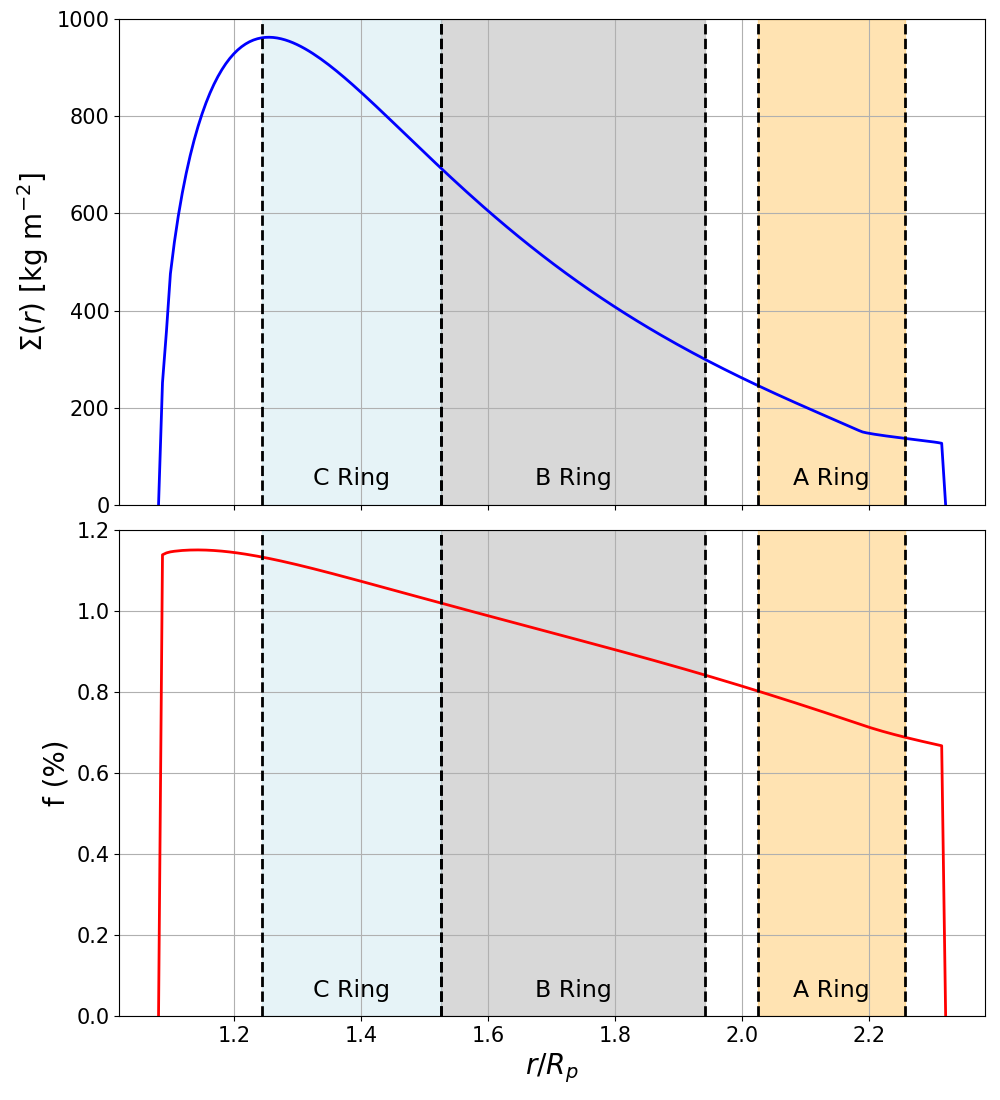}
    \caption{Same figure as Fig.~\ref{fig: sigma f vap example} without vaporisation upon impact but with space weathering. Here we adopt the values of $\eta,\zeta$ from \cite{estrada_constraints_2023} (i.e. $\eta=0.1, \zeta=1$) and an effective $\tau_{\rm sw}\approx320\ \rm{Myr}$.}
    \label{fig: sigma f SW example}
\end{figure}

More generally, if the rings are old enough to have approached the limit value of $f_m(t)$, that is $t> T_{1/2}\sim950$ Myr, we can only measure $c/\tilde{b}$ and not $f_i$ or the time. However, this is very interesting as it constrains $\eta$, $\zeta$ and $\tau_{\rm sw}$.

Combining vaporisation and space weathering automatically leads to new possible solutions for our problem. Taking $\mu_p=3, \mu_i=0.8$, one needs a characteristic time $\tau_{\rm sw}\sim 430$~Myr to reach the current fraction in the B ring. The main difference is that $\tau_{\rm sw}$ has to be slightly smaller for the A ring, around $420$~Myr. Physically, one could think it takes less time to clean the A ring as it spends less time in Saturn's shadow compared to the B ring, therefore having a bigger $\tau_{\rm sw}$. Otherwise, the evolutions of $f_m(t)$ are the same as in Fig.~\ref{fig:evol fm 1st eta zeta tau in B/A ring} with vaporisation of the target material upon impacts.

We point out that $\tau_{\rm sw}$ used in this section is the \emph{effective} space weathering time for silicates in the rings. It should most likely be longer than the space weathering time of a dust plate exposed perpendicularly to the Sun light, because the rings are inclined, and made of blocks which cast a shadow on each other. 
\begin{figure}
    \centering
    \includegraphics[width=\columnwidth]{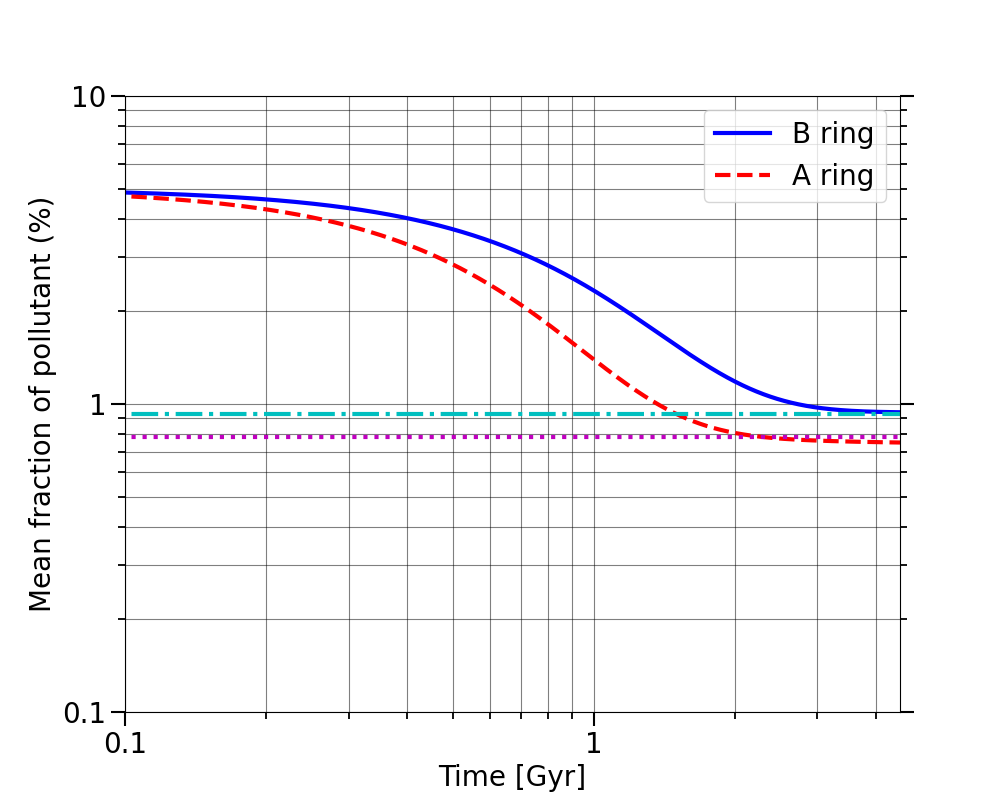}
    \caption{Evolution of the mean fraction of pollutant in the B and A rings, respectively the solid blue and dashed red curves. The dash-dotted cyan and dotted magenta lines represent the mean fraction of dust measured in both rings ($0.93\%$ and $0.78\%$).}
    \label{fig: f mean B A rings SW}
\end{figure}

\paragraph{Numerical simulation}
As in the previous subsection, Fig.~\ref{fig: sigma f SW example} shows the final result of a numerical simulation of $\Sigma(r, t\!=\!4.5\ \rm{Gyr})$ and $f(r, t\!=\!4.5\ \rm{Gyr})$ 
with the same initial condition, except that the rings start with a dust fraction $f_i=5\%$, and that only space weathering is considered here ($\mu_i=\mu_p=0$, $\tau_{\rm sw}=320$~Myr).
To account for the fact that optically thin rings do not capture all the solar flux, we multiplied in the simulation the space weathering term by a factor $\left[1-\exp(-\tau/c_0)\right]$, where $\tau/c_0$ is the optical depth in the direction of the Sun light and $c_0\sim0.28$ is the average of the cosine of the rings obliquity over one revolution of Saturn around the Sun \citep[Eq. 39 of][]{vokrouhlicky_thermal_2007}. The time evolution of the average dust fraction in the A and B ring regions is shown in Fig.~\ref{fig: f mean B A rings SW}. Again, the result is very satisfactory in terms of the density profile and composition of the A and B rings. It is not clear how realistic $\tau_{\rm sw}\approx320$~Myr (which corresponds to a space weathering time of $320c_0=90$~Myr for a surface facing the Sun) is, but it provides a very attractive ring pollution history.

\section{Summary, discussion and perspectives}
\label{sec:sdp}

\subsection{A bombardment five times lower than expected due to gravitational focusing}
A major result from this work is the calculation of the gravitational focusing which applies to the interplanetary dust particles that hit Saturn's rings. We have first shown that in order to compute the bombardment rate at a given radius, one needs the \emph{local} gravitational focusing $F_g^*(r)$, which is the differentiation of the classical, \emph{global} $F_g(r)$ ---\,more precisely $d\left(r^2F_g(r)\right)/dr^2$.
Second, we have noticed that the most commonly used expression for $F_g(r)$ was an approximate fit of the analytical expression for a sphere, while we are dealing with a planar surface. The exact expression of the local gravitational focusing on a disc has thus been derived (Eq.~\ref{eq:fg_generalised rings}). Third, we revised the value of the velocity at infinity of incoming particles, using the distribution which has been measured by \emph{Cassini} around Saturn. Debiasing for the oversampling of slow particles, and taking the average of $v_\infty^{-2}$ (which is the relevant quantity to compute $F_g$), we found an effective $v_\infty$ of $6.8$~km/s, larger than the modal value of the \emph{Cassini} sample ($4.3$ km/s) which is generally used in the literature. This increase is not surprising, as the populations of slow and fast IDPs entering Saturn's system are thought to be relatively similar \citep{altobelli_exogenic_2025}.

All together, these results lead to a flux hitting the rings about 5 times lower than with the previous estimate of the gravitational focusing. We stress that this result does not rely on arbitrary hypotheses about the physics of the impact on the rings or the chemistry of the dust, but solely on geometrical considerations. Therefore, one can robustly assess that the rings suffer a bombardment five times lower than previously considered. Thus, the chemical and dynamical effects of the bombardment are actually five times weaker, which has a huge impact on all the rings evolution models. In particular, all the exposure ages previously derived should be multiplied by five.

\subsection{A countless number of possibilities for any exposure ages}

In sections 3 and 4, we derived and studied an analytical expression of the mean exposure age of the rings, in the case of static or viscously evolving rings. Separating $\eta$ the fraction of impactor material that retains its absorptive properties after impact and $\zeta$ the fraction of impactor material actually accreted, we provide a map of the exposure age of the B ring in the $(\eta,\zeta)$ parameter space in Fig.~5. Our color map goes from 0.1 to 100 Gyr.

Here, one should humbly admit that $\eta$ and $\zeta$ are mostly unknown. Three sets of parameters used in reference papers of the literature are shown in the figures and are widely spread across the parameter space.
One of them is from \cite{hyodo_pollution_2025}, who estimated $\zeta$ based on SPH simulations of impacts at 30 km/s on isolated solid ice blocks; their values lead to an exposure age of 10~Gyr. Because the real rings may correspond to a different set-up (variety of impact speeds, porosity, presence of neighbouring blocks), $\eta$ and mainly $\zeta$ may well be larger in the rings than in this study \citep[see][section 2.2]{crida_age_2025}, yielding a shorter exposure age.

We let the reader choose their favourite values for these key parameters; our main point is that virtually any exposure age is actually possible with an appropriate choice of $(\eta,\zeta)$. We stress that previous studies made more or less implicitly a firm choice on $\eta$ and $\zeta$, often arbitrarily setting one of them to 1.

\subsection{Vaporisation and space weathering}

In our analytical model of the evolution of the pollution rate of the rings, we have added two physical phenomena which were previously discarded\,: vaporisation of the target material, and space weathering of the pollution. Since they decrease the pollution rate of the rings, they both introduce the possibility of an equilibrium between pollution by the bombardment and cleaning. In fact, we find that the dust fraction converges towards this equilibrium value $f_\infty$ as time goes to infinity. Therefore, it opens the possibility for a non zero initial dust fraction in the rings.

Besides, if the typical timescale for convergence is shorter than the age of the rings, then the measure of the dust fraction now only constrains $f_\infty$, and neither the initial composition nor the age of the rings. Interestingly, we find that without vaporisation but with an effective space weathering typical timescale of $320$~Myr in the rings, $f_\infty$ in both the A ring and the B ring is close to the observed value now. This may mean that the pollution rate observed in the rings today does not constrain their age, but shows that detectable dust grains in the rings have an average life-time of only 320~Myr.

\subsection{Towards a global picture}

In the end, our work opens more possibilities than it provides conclusions.
Lifting the constraint that the rings should be only $\sim 100$~Myr old allows a larger variety of scenarios for their origin. In particular primordial rings born at the same time as the planet appear possible for low enough $\eta$ and $\zeta$, or with suitable vaporisation and space weathering rates.
However, a complete full-consistent model for the origin of the rings should address not only the question of their composition, but also that of their mass and density profile. So far, no such scenario exists \citep[see][section 5 for a recent review]{crida_age_2025}.

The crucial importance of the parameters at play --\,namely the fraction of impactor that keeps its absorptive properties $\eta$, the fraction of impactor actually accreted $\zeta$, the ratios of vaporised ice and dust of target mass to impactor mass $\mu_i$ and $\mu_p$, and the effective space weathering timescale $\tau_{\rm sw}$\,-- advocates for theoretical or experimental studies to constrain the values of these parameters. We hope that in the future, the parameter space will be narrower. Meanwhile, any study on the age of the rings should explicitly mention which value of these parameters they consider, in particular when by default $\mu_i=\mu_p={\tau_{\rm sw}}^{-1}=0$. Indeed, this choice is simpler, but not necessarily more realistic than non zero values \citep[see for instance][]{hyodo_pollution_2025,esposito_space_2024}.

The meteoritic bombardment also has a dynamical effect on the rings, which leads to their slow drift towards Saturn. This effect has purposely not been addressed in this work, where we focus on the rings composition. \citet{estrada_constraints_2023} show that the mass loading and mostly the ballistic transport due to the impacts would make the present rings vanish in hundreds of millions of years. This questions their lifetime and the possibility that they survived 4.5~Gyr. We remark that their calculation was performed with the old $F_g$; reducing the bombardment rate by a factor 5 with the new gravitational focusing should reduce five times the dynamical effects too, all other things equal. In a forthcoming paper, we will address the dynamical evolution of the rings (including the effects of the bombardment) and, based on the results of this paper, try to build a self-consistent scenario to reproduce both the density profile (including the inner edge of the B ring and the low density C ring) and the composition of Saturn's rings today.

\section*{Acknowledgments}
We thank L. Esposito for enriching discussions about the space weathering in Saturn's rings.
We also thank the two reviewers (in particular Luke Dones) whose positive and constructive comments helped improve this article.

\appendix
\section{Derivation of the global gravitational focusing for the rings.}\label{appendix: derivation Fgstar}

\begin{figure}
    \centering
    \includegraphics[width=\columnwidth]{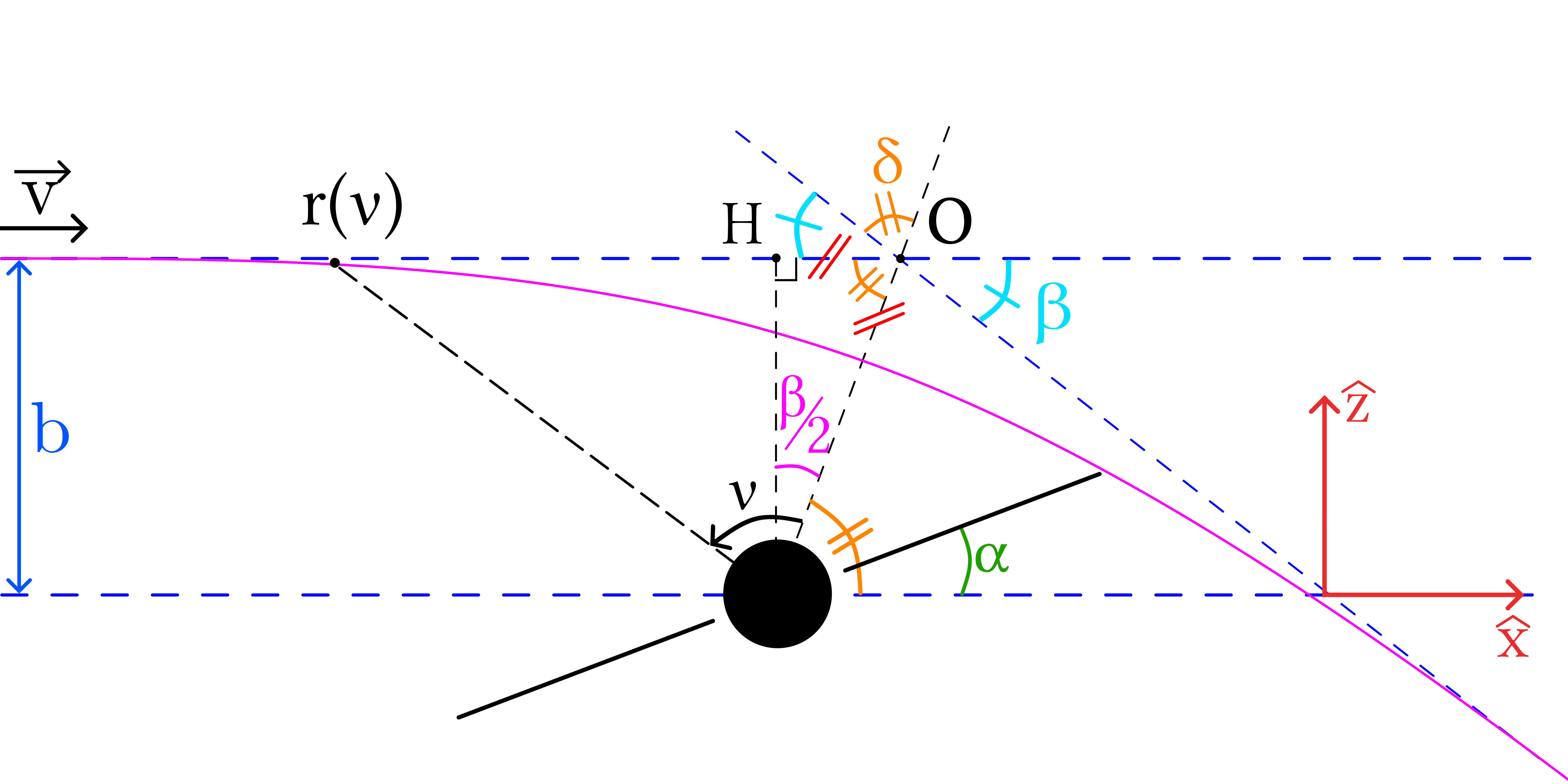}
    \caption{Scheme inside the trajectory plane of a particle deflected by an angle $\beta$ after a close encounter with Saturn (black disc). Here the rings are represented by the solid black line tilted by an angle $\alpha$ from the $x$-axis. A test particle is represented by $r(\nu)$ along its trajectory (solid magenta line).}
    \label{fig:traj_plane}
\end{figure}

The notations we use to prepare the mathematical work on deriving $F_g^*(r)$ are:
\begin{itemize}
    \item We adopt a frame centred on Saturn with the rings spreading from $r_{\text{min}}$ to $r_{\text{Roche}}$ in their own plane. 
    \item We choose the $x$-axis to be colinear to $\overrightarrow{v_{\infty}}$, the velocity vector of the incoming particle at infinity. The bombardment being isotropic, its orientation is not linked to the rings anyhow.
    \item We define $\hat{y}$ orthogonal to $\hat{x}$ in the rings plane and $\hat{z}$ completes the orthonormal frame. The angle between the rings plane and $\pxy$ is noted $\theta$.
    \item In the trajectory plane $\ptraj$ (which contains the $x$-axis), the particles follow hyperbolic trajectories (see Fig.~\ref{fig:traj_plane}), i.e. we neglect the influence of the Sun. The separation between the asymptote of the trajectory at infinity and the $x$-axis is called $b$ the impact parameter.
    \item In $\ptraj$, the axis perpendicular to $x$ is noted $z'$, and makes an angle $\gamma$ with the $z$-axis.
    \item In $\ptraj$, the rings are inclined by an angle $\alpha$ with respect to the $x$-axis (see Fig.~\ref{fig:traj_plane}). Note that $\alpha=\theta$ if and only if $\ptraj=\pxz$ (that is: $\gamma=0$). Otherwise, $\alpha$ is a function of $\theta$ and $\gamma$.
\end{itemize}

In Section~\ref{sec:no_inc ptraj}, the goal is to find in $\ptraj$ the exact value of $b(\alpha, r)$ to fall at the distance $r$ on the rings plane inclined by an angle $\alpha$ with respect to the $x$-axis. The expression of $\alpha(\theta,\gamma)$ is given in Section~\ref{sec:inclined ptraj} and the gravitational cross-section for the rings $\sigma_{\text{grav}}$ in Section~\ref{sec:cross}.
\subsection{In the trajectory plane}\label{sec:no_inc ptraj}
We place ourselves in the trajectory plane shown in Fig.~\ref{fig:traj_plane} where the particle is described along the trajectory by $r(\nu)$. It is well-known that the trajectory is deflected by an angle $\beta$ that is linked to $b$ through the relation
\begin{equation}
    \beta = 2\text{arctan}\left( \frac{GM_p}{v_{\infty}^2b} \right)
\end{equation}
where $v_{\infty}$ is the velocity of the particle at 'infinity' (we recall that the infinity is set to be at Saturn's Hill sphere). 

In the trajectory plane, we have the position of the particle given by 
\begin{equation}
    r(\nu) = \frac{a(1-e^2)}{1+e\ \text{cos}(\nu)}
\end{equation}
where $a=-GM_p/v_{\infty}^2$ is the semi-major axis of the hyperbola with $G$ the gravitational constant, $M_p$ Saturn's mass and $e>1$. Combining properties from the hyperbola and trigonometric functions, we can find that $e=\text{sin}^{-1}(\beta/2)$, and that the particle intercepts the ring plane at $\nu_{\rm rings} = \alpha - \frac{\pi - \beta}{2}$.

As $e$ and $\nu_{\rm rings}$ are function of $\beta$ thus $b$, falling at distance $r$ in the rings plane tilted by $\alpha$ with respect to $\overrightarrow{v_\infty}$ yields to the \emph{analytical solution of the impact parameter $b(\alpha, r)$} 
\begin{equation}\label{eq:b_alpha_r}
    b(\alpha, r) = \frac{2\Theta r\ \text{cos}(\alpha/2)}{\sqrt{\text{sin}^2(\alpha/2) + 2\Theta} - \text{sin}(\alpha/2)},
\end{equation}
where $\Theta$ is the Safronov number defined as $(1/2)(v_{\rm esc}/v_\infty)^2$, with $v_{\rm esc}$ the escape velocity around Saturn.
We found the analytical form of $b(\alpha, r)$. What is $\alpha$ as a function of $\theta$ and $\gamma$ ?

\begin{figure}
    \centering
    \includegraphics[width=\columnwidth]{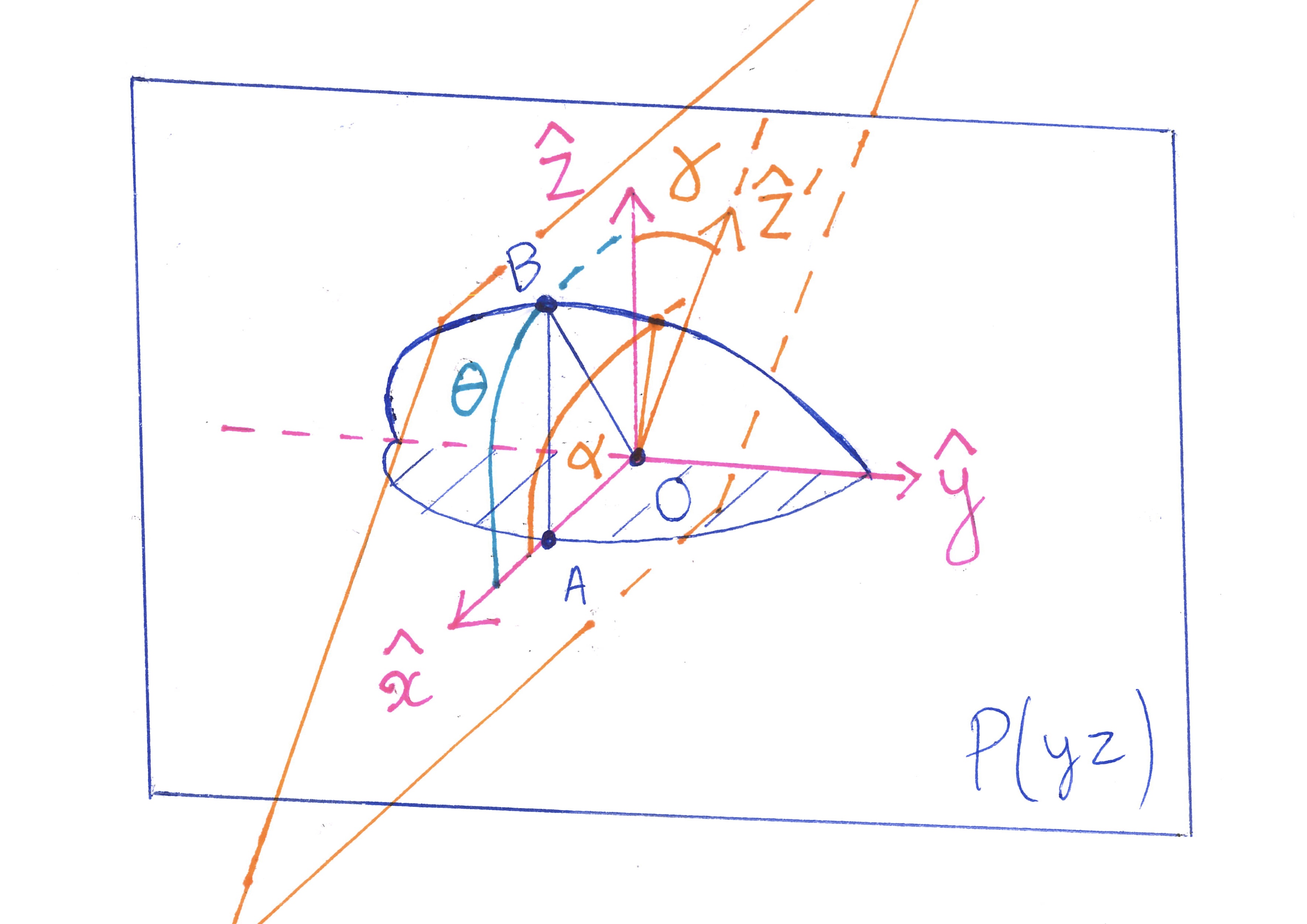}
    \caption{Scheme of the three-dimensional problem when $\ptraj$ (orange plane) is inclined by an angle $\gamma$ from $\pxz$. The intersection of the rings with $\ptraj$ is the orange line going from the centre of the frame to the border of the rings (blue solid line). $\alpha$ is the angle seen by the particle in its trajectory plane. The rings make an angle $\theta$ with the plane $\pxy$ where their shadow is projected (blue hatched area). The angle $\alpha$ depends on the two other angles $\theta$ and $\gamma$.}
    \label{fig:ptraj_3d}
\end{figure}
\begin{figure}
    \centering
    \includegraphics[width=0.8\columnwidth]{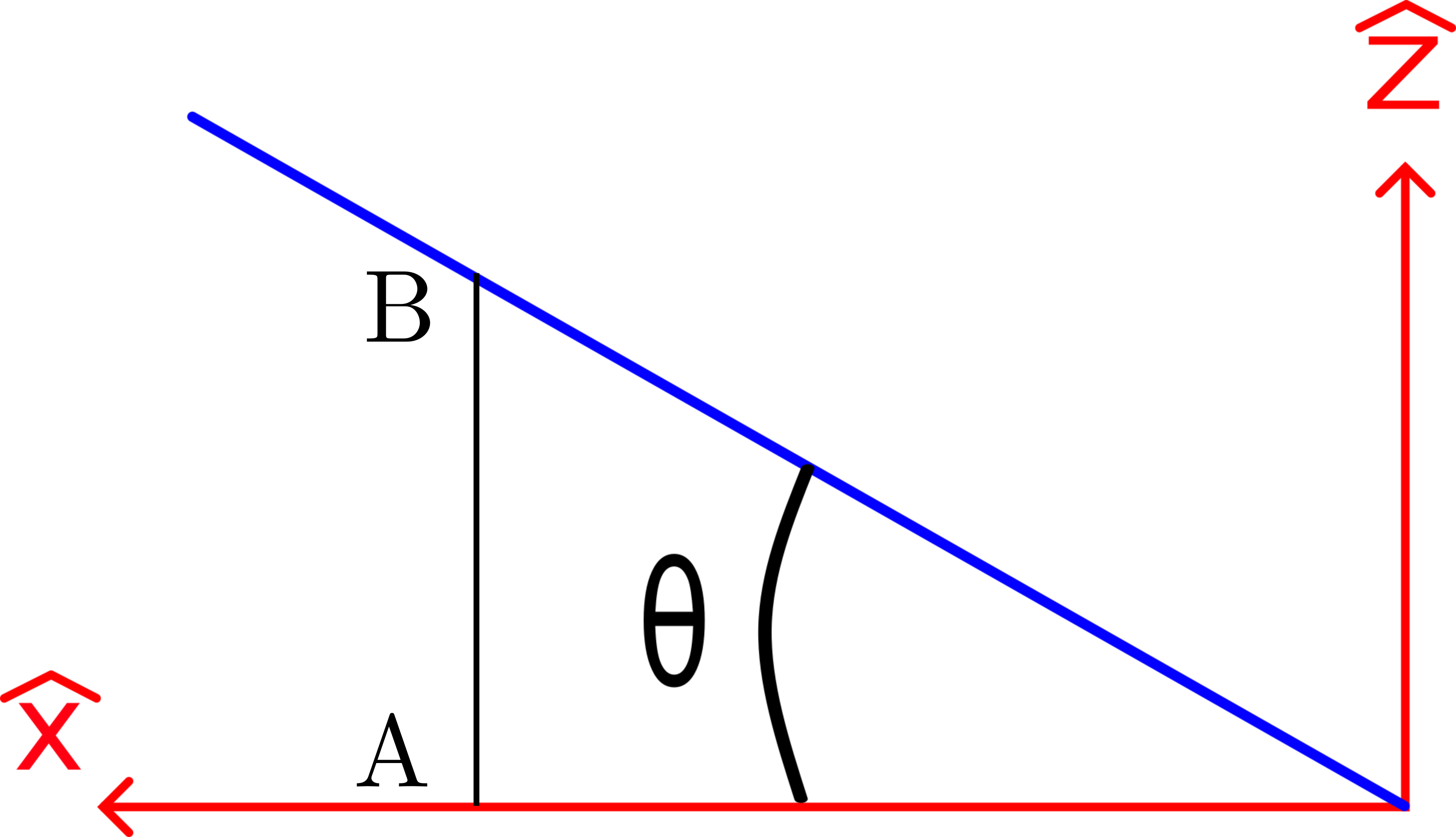}
    \caption{Scheme of the inclinations of the rings (blue line) in $\pxz$. The segment AB is located at $OA = x=1$.}
    \label{fig:pxz_angle}
\end{figure}
\subsection{Trajectory plane inclined by $\gamma$ from $\pxz$}\label{sec:inclined ptraj}
There are two ways to find the relation between $\alpha$, $\theta$ and $\gamma$. We are going to use the geometrical method in this work. The problem is illustrated in Fig.~\ref{fig:ptraj_3d}.

\begin{figure}
    \centering
    \includegraphics[width=0.8\columnwidth]{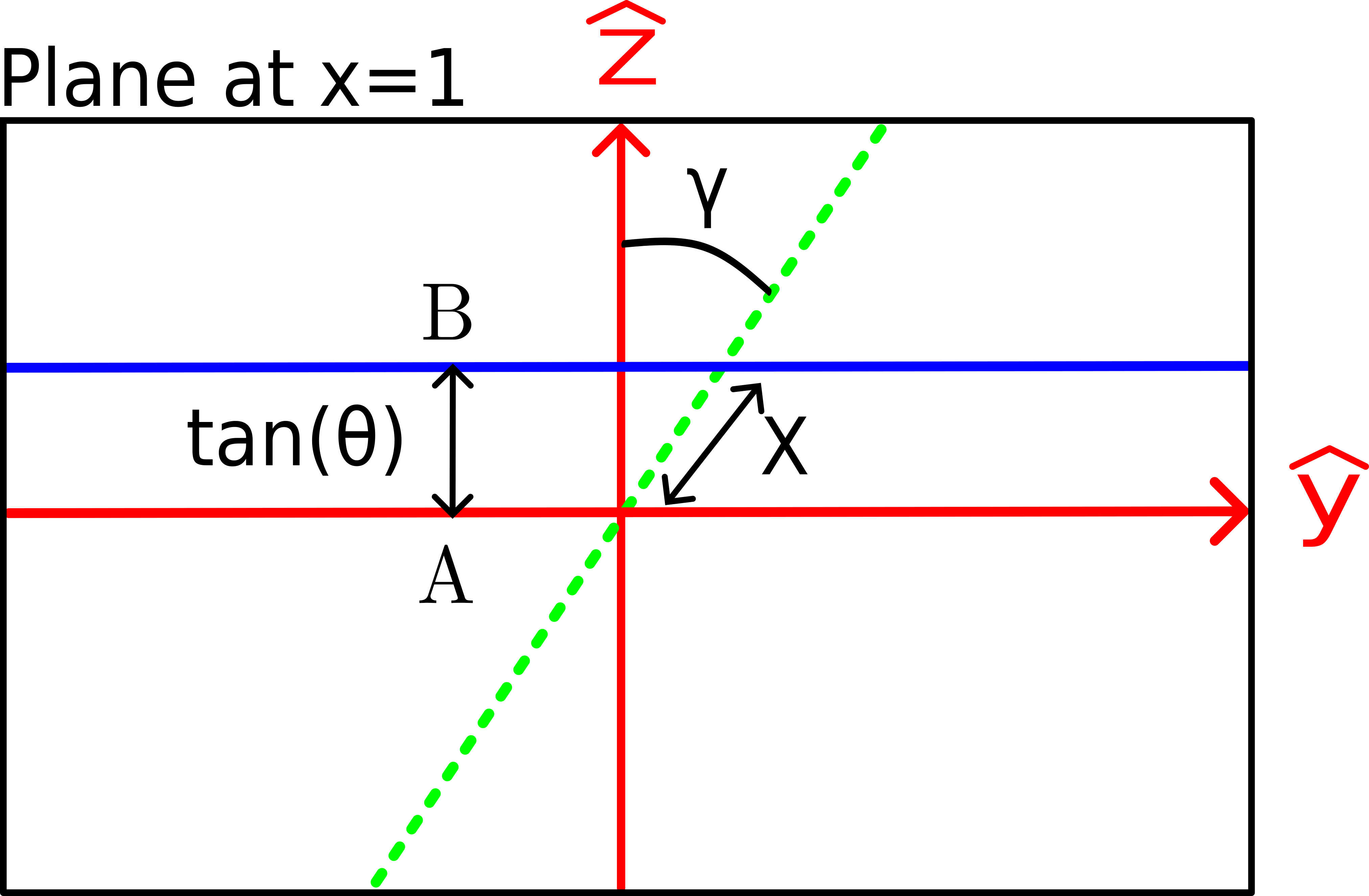}
    \caption{$\pyz$ at x = 1. The tilt of $\ptraj$ (dashed green line) is $\gamma$. $X$ is also defined in $\ptraj$ and is equal to $X = z' = \text{tan}(\alpha)$. }
    \label{fig:pyz_angles}
\end{figure}

\begin{itemize}
    \item In $\pxz$, we have $\text{tan}(\theta) = z/x$ in the rings plane (Fig.~\ref{fig:pxz_angle}). At $x=1$, we have $z=\tan(\theta)$.
    \item In the plane \{x=1\} (Fig.~\ref{fig:pyz_angles}), we have $\text{cos}(\gamma) = \text{tan}(\theta)/X$. However, we know that this distance $X$ in $\ptraj$ is $\text{tan}(\alpha)$ as we have in $\ptraj$ the relation $\text{tan}(\alpha) = z'$ where $z'$ denotes the coordinate perpendicular to $x$ in the trajectory plane. 
\end{itemize}

In the end, the angle made by the rings in $\ptraj$ is
\begin{equation}\label{eq:expr_alpha_3D}
    \alpha = \text{arctan}\left(\frac{\text{tan}(\theta)}{\text{cos}(\gamma)}\right),
\end{equation}
and from this relation\footnote{The \textbf{arctan} function results depend on the signs of the denominator and numerator, but this can be easily handled with the \textbf{arctan2} numerical function.} we can rewrite $b(\alpha, r)$ as $b(\theta, \gamma, r)$.

\subsection{Cross section}
\label{sec:cross}
The effective cross-section $\sigma_{\text{eff}}(\theta,r)$ is
\begin{equation}
    \sigma_{\text{eff}}(\theta,r) = \int_{\gamma=0}^{2\pi} \int_{r'=0}^{b(\theta, \gamma, r)} r'dr'd\gamma
\end{equation}
which gives when integrating over $r'$
\begin{equation}\label{eq:sigma_eff final}
    \sigma_{\text{eff}}(\theta,r) = \int_{\gamma=0}^{2\pi} \frac{b^2(\theta, \gamma, r)}{2}d\gamma.
\end{equation}

Unfortunately, this integral does not have an analytical expression due to trigonometric functions nested within each other but it is easily found doing numerical integrations. 

Once the effective cross-section derived, we can average over all possible directions to find the gravitational cross-section of a planar ring of radius $r$ under an isotropic bombardment\,:
\begin{equation}
    \sigma_{\text{grav}}(r) = \frac{1}{4\pi}\int_{\phi=0}^{2\pi}\int_{\theta=-\pi/2}^{\pi/2} \sigma_{\text{eff}}(\theta, r)\text{cos}(\theta)d\theta d\phi
\end{equation}
and as the problem is invariant with rotation $\phi$, the latter equation becomes
\begin{equation}\label{eq:sigma_grav}
    \sigma_{\text{grav}}(r) = \frac{1}{2}\int_{\theta=-\pi/2}^{\pi/2} \sigma_{\text{eff}}(\theta, r)\text{cos}(\theta)d\theta.
\end{equation}

\subsection{Final expression of the local gravitational focusing for the rings}\label{sec: final exp dFg rings}
\begin{figure}
    \centering
    \includegraphics[width=0.8\columnwidth]{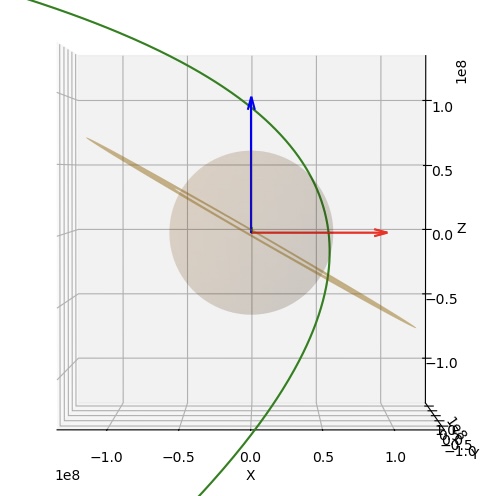}
    \caption{A trajectory of one particle is represented by the green curve. Its plane is not inclined with $\pxz$ ($\gamma = 0$) and the rings are inclined from an angle $\theta=-\pi/6$ (brown plane). Here, $\alpha = \theta < (\pi - \beta)/2$ and $d_s< R_p$, the particle 'enters' the planet before touching the rings. These trajectories are forbidden and justify the need to put $b=0$ in such case.}
    \label{fig:pb_traj}
\end{figure}

\begin{figure}
    \centering
    \includegraphics[width=\columnwidth]{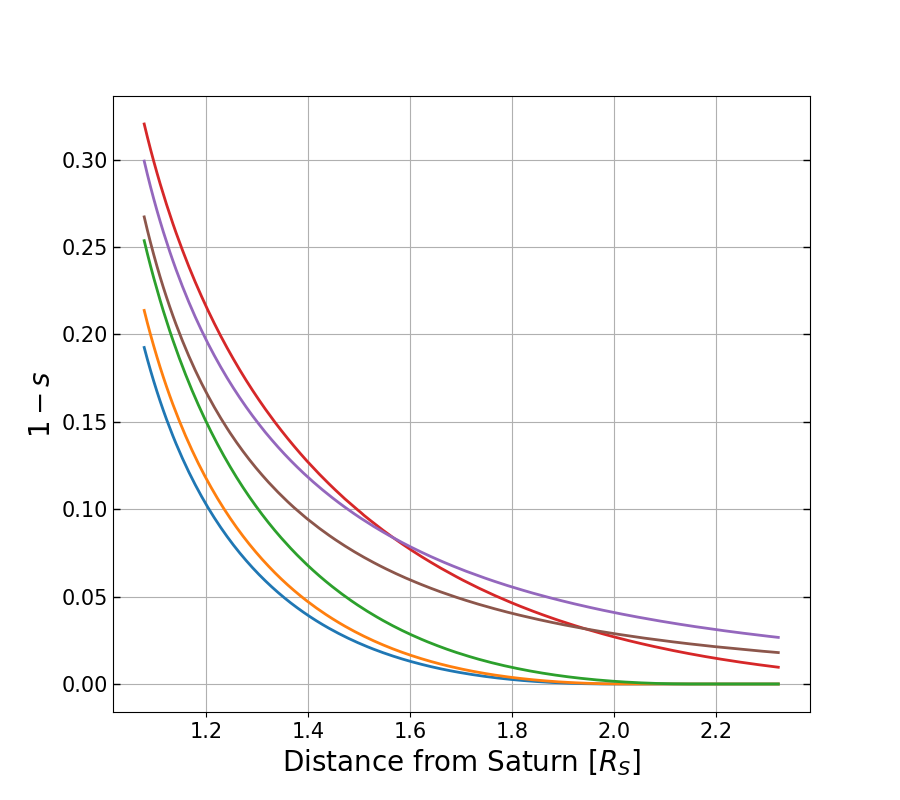}
    \caption{Plot of the fraction of particles lost due to the shielding ($1-s$) as we change the value of $v_{\infty}$. Blue: $v_{\infty}=1$ km/s. Orange: $v_{\infty}=4.3$ km/s. Green: $v_{\infty}=10$ km/s. Red: $v_{\infty}=25$ km/s. Purple: $v_{\infty}=100$ km/s. Brown: $v_{\infty}=5000$ km/s.}
    \label{fig:diff_s_in_vinfty}
\end{figure}

We want to add two constraints on the expression of $b$:
\begin{itemize}
    \item $b(\theta, \gamma, r)$ is the impact parameter needed to fall at the distance $r$ on the rings and on their 'upper' face (i.e. the one facing the incoming particle). We can have however particles hitting the rings at the same distance on the 'lower' face, which is the same problem as setting $\theta$ to $-\theta$ and $\gamma$ to $\pi - \gamma$. So we compute both values of $b$ and keep the max between the two such that we fall at the ring of radius $r$. This is shown in Fig.~\ref{fig:mushroom_cross_sections} where each curve is a polar coordinates $(\rho,\gamma)$ plot of $\rho(\gamma)=b(\theta_0,\gamma, r_{\rm rings})$ with $r_{\rm rings}$ the radius of the rings and $\theta_0$ fixed (4 different values are shown).
    \item Some trajectories can not be kept in our computations, those for which we fall into the planet before touching the rings (Fig.~\ref{fig:pb_traj}). The idea is to compute $b$ for both faces of the rings and check if $\alpha < (\pi - \beta)/2$ (i.e. the angle is smaller than the angle at the closest approach) and if the closest distance to the centre of the planet $d_s$ is smaller than the radius of the planet $R_p$. If both conditions are satisfied then $b$ is set to 0. Like this, we introduce a shielding effect from the planet on the inner part of the rings, which we define a coefficient $s(r)$ equal to the ratio between the cross section with and without taking this second condition into account. It is shown in Fig.~\ref{fig:diff_s_in_vinfty} for various values of $v_\infty$. As $v_\infty$ increases, we see that $s$ decreases when we are around $25$ km/s for $v_\infty$ before reaching an asymptote that is sort of an in-between curve for $v_\infty=4.3$ km/s and $v_\infty\rightarrow \infty$.
\end{itemize}

With these two constraints, we can compute $\sigma_{\text{eff}}(\theta, r)$ and thus $\sigma_{\text{grav}}(r)$ from Eq.~\ref{eq:sigma_grav} and by extension the local gravitational focusing for rings, given in Eq.~\ref{eq:fg_generalised rings}. 

Our curves for the shielding coefficient $s(r)$ shown in Fig.~\ref{fig:diff_s_in_vinfty} differ from the expression found by \citet[][Eq.~(19)\,]{lissauer_bombardment_1988} for a satellite on an orbit of radius $r$. In particular, we find that the shielding vanishes completely beyond $2R_S$ at low velocity, while \citet{lissauer_bombardment_1988}'s formula does not. This is because of the azimuthal extension of the rings. Take the green trajectory on Fig.~\ref{fig:pb_traj}: it is intercepted by Saturn before it hits the rings at $\sim1.1 R_S$. Bending the rings more with respect to the horizontal $z$-axis in the figure (that is: decreasing $\theta$ towards more negative values), the radius at which this trajectory hits the rings increases, and would reach $2R_S$ for $\theta\approx -80^\circ$. However, at this inclination, the trajectory would hit the lower face of the rings on its way to Saturn at $r<2R_S$, before the closest approach (near $x=-0.2, z=1.1$). Therefore, this particle should be accreted. Because the rings extend all around the planet, it is impossible for Saturn to shield them beyond a given radius. In the absence of rings which intercept incoming particles before their closest approach, the calculations of \citet{lissauer_bombardment_1988} for the flux apply.

The above discussion also shows that at low $v_\infty$, there exists trajectories, which intercept the rings plane twice, without necessarily having a closest approach smaller than $R_S$. 
A particle on such a trajectory has two chances of being accreted by the rings, and its probability of going through is $(1-A(\tau))^2$ instead of $1-A(\tau)$. The probability of capture should thus be $1-(1-A(\tau))^2$ instead of $A(\tau)$. For simplicity, we always use $A(\tau)$, but keep in mind that it is in some rare cases a lower bound.

Note that in Fig.~\ref{fig:pb_traj}, the trajectory intercepts the $x$-axis at $\sim 0.6\cdot10^8$~m from the centre (along the red arrow), while it intercepts the $z$-axis at $\sim10^8$~m (along the blue arrow). This shows that hitting rings in the $x-y$ plane (seen edge-on) is easier at low velocity than in the $y-z$ plane (face-on), and explains why the circle corresponding to $\theta=0$ is larger than the one for $\theta=\pi/2$ in Fig.~\ref{fig:mushroom_cross_sections}.

\section{Global or local: Discussions on the expression of $F_g$ to use}\label{app:discuss Fg}
Appendix B of \cite{cuzzi_compositional_1998} raises a debate on the expression of $F_g(r)$ to use between the one from Eq.~\ref{eq:fg_sphere full} \citep{shoemaker_cratering_1982, lissauer_bombardment_1988, cuzzi_bombardment_1990}, and the one given by Eq.~\ref{eq:fg_sphere local} \citep{morfill_consequences_1983, colwell_origins_1992, colwell_disruption_1994}. According to \cite{cuzzi_compositional_1998}, using the local expression is valid only when hitting concentric annuli and if the impact is normal to the rings plane. In addition, they explained it should be averaged over all incident angles. Hence, the formula to use would be the one for the global gravitational focusing. However, the global $F_g(r)$ from Eq.~\ref{eq:fg_sphere full} is already averaged over all incident angles as one can take $b^2(\theta, \gamma, r) = r^2 F_g(r)$ in Eq.~\ref{eq:sigma_eff final} and use it in Eq.~\ref{eq:sigma_grav} to find the fundamental result of $\sigma_{\text{grav}} = \pi r^2 F_g(r)$. Besides, this expression is to fall \emph{inside} the sphere of radius $r$ by definition. Thus, it is necessary to take the derivative in order not to overestimate what is falling \emph{at} the distance $r$ when computing the local bombardment rate.

\bibliographystyle{cas-model2-names}

\bibliography{Saturn_rings_icarus}

\begin{thebibliography}{59}
\expandafter\ifx\csname natexlab\endcsname\relax\def\natexlab#1{#1}\fi
\providecommand{\url}[1]{\texttt{#1}}
\providecommand{\href}[2]{#2}
\providecommand{\path}[1]{#1}
\providecommand{\DOIprefix}{doi:}
\providecommand{\ArXivprefix}{arXiv:}
\providecommand{\URLprefix}{URL: }
\providecommand{\Pubmedprefix}{pmid:}
\providecommand{\doi}[1]{\href{http://dx.doi.org/#1}{\path{#1}}}
\providecommand{\Pubmed}[1]{\href{pmid:#1}{\path{#1}}}
\providecommand{\bibinfo}[2]{#2}
\ifx\xfnm\relax \def\xfnm[#1]{\unskip,\space#1}\fi
%Type = Article
\bibitem[{Altobelli et~al.(2025)Altobelli, Postberg, Kempf, Poppe, Schmidt,
  Fischer, Moragas-Klostermeyer and Srama}]{altobelli_exogenic_2025}
\bibinfo{author}{Altobelli, N.}, \bibinfo{author}{Postberg, F.},
  \bibinfo{author}{Kempf, S.}, \bibinfo{author}{Poppe, A.R.},
  \bibinfo{author}{Schmidt, J.}, \bibinfo{author}{Fischer, C.},
  \bibinfo{author}{Moragas-Klostermeyer, G.}, \bibinfo{author}{Srama, R.},
  \bibinfo{year}{2025}.
\newblock \bibinfo{title}{Exogenic dust inventory in the {Saturn} system: the
  {Cassini} {Cosmic} {Dust} {Analyzer} perspective}.
\newblock \bibinfo{journal}{Monthly Notices of the Royal Astronomical Society}
  \bibinfo{volume}{539}, \bibinfo{pages}{3551--3567}.
\newblock \URLprefix
  \url{https://ui.adsabs.harvard.edu/abs/2025MNRAS.539.3551A},
  \DOIprefix\doi{10.1093/mnras/staf656}.
%Type = Article
\bibitem[{Baillié et~al.(2019)Baillié, Noyelles, Lainey, Charnoz and
  Tobie}]{baillie_formation_2019}
\bibinfo{author}{Baillié, K.}, \bibinfo{author}{Noyelles, B.},
  \bibinfo{author}{Lainey, V.}, \bibinfo{author}{Charnoz, S.},
  \bibinfo{author}{Tobie, G.}, \bibinfo{year}{2019}.
\newblock \bibinfo{title}{Formation of the {Cassini} {Division} - {I}.
  {Shaping} the rings by {Mimas} inward migration}.
\newblock \bibinfo{journal}{Monthly Notices of the Royal Astronomical Society}
  \bibinfo{volume}{486}, \bibinfo{pages}{2933--2946}.
\newblock \URLprefix
  \url{https://ui.adsabs.harvard.edu/abs/2019MNRAS.486.2933B},
  \DOIprefix\doi{10.1093/mnras/stz548}.
%Type = Incollection
\bibitem[{Brisset et~al.(2025)Brisset, Blum, Hénault, Burchell, Grundy,
  Gudipati and Brunetto}]{brisset_laboratory_2025}
\bibinfo{author}{Brisset, J.}, \bibinfo{author}{Blum, J.},
  \bibinfo{author}{Hénault, E.}, \bibinfo{author}{Burchell, M.},
  \bibinfo{author}{Grundy, W.}, \bibinfo{author}{Gudipati, M.S.},
  \bibinfo{author}{Brunetto, R.}, \bibinfo{year}{2025}.
\newblock \bibinfo{title}{Laboratory {Studies} {Applicable} to {Centaurs}}, in:
  \bibinfo{booktitle}{Centaurs}, pp. \bibinfo{pages}{13--1}.
\newblock \URLprefix
  \url{https://ui.adsabs.harvard.edu/abs/2025cent.book...13B},
  \DOIprefix\doi{10.1088/10.1088/2514-3433/ada267ch13}. \bibinfo{note}{aDS
  Bibcode: 2025cent.book...13B}.
%Type = Article
\bibitem[{Canup(2010)}]{canup_origin_2010}
\bibinfo{author}{Canup, R.M.}, \bibinfo{year}{2010}.
\newblock \bibinfo{title}{Origin of {Saturn}’s rings and inner moons by mass
  removal from a lost {Titan}-sized satellite}.
\newblock \bibinfo{journal}{Nature} \bibinfo{volume}{468},
  \bibinfo{pages}{943--946}.
\newblock \URLprefix \url{https://www.nature.com/articles/nature09661},
  \DOIprefix\doi{10.1038/nature09661}.
%Type = Article
\bibitem[{Charnoz et~al.(2009)Charnoz, Morbidelli, Dones and
  Salmon}]{charnoz_did_2009}
\bibinfo{author}{Charnoz, S.}, \bibinfo{author}{Morbidelli, A.},
  \bibinfo{author}{Dones, L.}, \bibinfo{author}{Salmon, J.},
  \bibinfo{year}{2009}.
\newblock \bibinfo{title}{Did {Saturn}'s rings form during the {Late} {Heavy}
  {Bombardment}?}
\newblock \bibinfo{journal}{Icarus} \bibinfo{volume}{199},
  \bibinfo{pages}{413--428}.
\newblock \URLprefix
  \url{https://ui.adsabs.harvard.edu/abs/2009Icar..199..413C},
  \DOIprefix\doi{10.1016/j.icarus.2008.10.019}.
%Type = Article
\bibitem[{Colombo et~al.(1966)Colombo, Lautman and
  Shapiro}]{colombo_earths_1966}
\bibinfo{author}{Colombo, G.}, \bibinfo{author}{Lautman, D.A.},
  \bibinfo{author}{Shapiro, I.I.}, \bibinfo{year}{1966}.
\newblock \bibinfo{title}{The {Earth}'s {Dust} {Belt}: {Fact} or fiction?: 2.
  {Gravitational} focusing and {Jacobi} capture}.
\newblock \bibinfo{journal}{Journal of Geophysical Research}
  \bibinfo{volume}{71}, \bibinfo{pages}{5705--5717}.
\newblock \URLprefix
  \url{https://ui.adsabs.harvard.edu/abs/1966JGR....71.5705C},
  \DOIprefix\doi{10.1029/JZ071i023p05705}.
%Type = Article
\bibitem[{Colwell(1994)}]{colwell_disruption_1994}
\bibinfo{author}{Colwell, J.E.}, \bibinfo{year}{1994}.
\newblock \bibinfo{title}{The disruption of planetary satellites and the
  creation of planetary rings}.
\newblock \bibinfo{journal}{Planetary and Space Science} \bibinfo{volume}{42},
  \bibinfo{pages}{1139--1149}.
\newblock \URLprefix
  \url{https://ui.adsabs.harvard.edu/abs/1994P&SS...42.1139C},
  \DOIprefix\doi{10.1016/0032-0633(94)90013-2}.
%Type = Article
\bibitem[{Colwell and Esposito(1992)}]{colwell_origins_1992}
\bibinfo{author}{Colwell, J.E.}, \bibinfo{author}{Esposito, L.W.},
  \bibinfo{year}{1992}.
\newblock \bibinfo{title}{Origins of the rings of {Uranus} and {Neptune} 1.
  {Statistics} of satellite disruptions}.
\newblock \bibinfo{journal}{Journal of Geophysical Research}
  \bibinfo{volume}{97}, \bibinfo{pages}{10227--10241}.
\newblock \URLprefix
  \url{https://ui.adsabs.harvard.edu/abs/1992JGR....9710227C},
  \DOIprefix\doi{10.1029/92JE00788}.
%Type = Inproceedings
\bibitem[{Crida and Charnoz(2014)}]{crida_complex_2014}
\bibinfo{author}{Crida, A.}, \bibinfo{author}{Charnoz, S.},
  \bibinfo{year}{2014}.
\newblock \bibinfo{title}{Complex satellite systems: a general model of
  formation from rings}, in: \bibinfo{booktitle}{Complex {Planetary} {Systems},
  {Proceedings} of the {International} {Astronomical} {Union}}, pp.
  \bibinfo{pages}{182--189}.
\newblock \URLprefix
  \url{https://ui.adsabs.harvard.edu/abs/2014IAUS..310..182C/abstract},
  \DOIprefix\doi{10.1017/S1743921314008229}.
%Type = Article
\bibitem[{Crida et~al.(2019)Crida, Charnoz, Hsu and Dones}]{crida_are_2019}
\bibinfo{author}{Crida, A.}, \bibinfo{author}{Charnoz, S.},
  \bibinfo{author}{Hsu, H.W.}, \bibinfo{author}{Dones, L.},
  \bibinfo{year}{2019}.
\newblock \bibinfo{title}{Are {Saturn}'s rings actually young?}
\newblock \bibinfo{journal}{Nature Astronomy} \bibinfo{volume}{3},
  \bibinfo{pages}{967--970}.
\newblock \URLprefix
  \url{https://ui.adsabs.harvard.edu/abs/2019NatAs...3..967C/abstract},
  \DOIprefix\doi{10.1038/s41550-019-0876-y}.
%Type = Article
\bibitem[{Crida et~al.(2025)Crida, Estrada, Nicholson and
  Murray}]{crida_age_2025}
\bibinfo{author}{Crida, A.}, \bibinfo{author}{Estrada, P.R.},
  \bibinfo{author}{Nicholson, P.D.}, \bibinfo{author}{Murray, C.D.},
  \bibinfo{year}{2025}.
\newblock \bibinfo{title}{The {Age} and {Origin} of {Saturn}’s {Rings}}.
\newblock \bibinfo{journal}{Space Science Reviews} \bibinfo{volume}{221},
  \bibinfo{pages}{66}.
\newblock \URLprefix \url{https://doi.org/10.1007/s11214-025-01189-z},
  \DOIprefix\doi{10.1007/s11214-025-01189-z}.
%Type = Article
\bibitem[{Cuzzi et~al.(2010)Cuzzi, Burns, Charnoz, Clark, Colwell, Dones,
  Esposito, Filacchione, French, Hedman, Kempf, Marouf, Murray, Nicholson,
  Porco, Schmidt, Showalter, Spilker, Spitale, Srama, Sremčević, Tiscareno
  and Weiss}]{cuzzi_evolving_2010}
\bibinfo{author}{Cuzzi, J.N.}, \bibinfo{author}{Burns, J.A.},
  \bibinfo{author}{Charnoz, S.}, \bibinfo{author}{Clark, R.N.},
  \bibinfo{author}{Colwell, J.E.}, \bibinfo{author}{Dones, L.},
  \bibinfo{author}{Esposito, L.W.}, \bibinfo{author}{Filacchione, G.},
  \bibinfo{author}{French, R.G.}, \bibinfo{author}{Hedman, M.M.},
  \bibinfo{author}{Kempf, S.}, \bibinfo{author}{Marouf, E.A.},
  \bibinfo{author}{Murray, C.D.}, \bibinfo{author}{Nicholson, P.D.},
  \bibinfo{author}{Porco, C.C.}, \bibinfo{author}{Schmidt, J.},
  \bibinfo{author}{Showalter, M.R.}, \bibinfo{author}{Spilker, L.J.},
  \bibinfo{author}{Spitale, J.N.}, \bibinfo{author}{Srama, R.},
  \bibinfo{author}{Sremčević, M.}, \bibinfo{author}{Tiscareno, M.S.},
  \bibinfo{author}{Weiss, J.}, \bibinfo{year}{2010}.
\newblock \bibinfo{title}{An {Evolving} {View} of {Saturn}’s {Dynamic}
  {Rings}}.
\newblock \bibinfo{journal}{Science} \bibinfo{volume}{327},
  \bibinfo{pages}{1470--1475}.
\newblock \URLprefix
  \url{https://ui.adsabs.harvard.edu/abs/2010Sci...327.1470C},
  \DOIprefix\doi{10.1126/science.1179118}.
%Type = Article
\bibitem[{Cuzzi and Durisen(1990)}]{cuzzi_bombardment_1990}
\bibinfo{author}{Cuzzi, J.N.}, \bibinfo{author}{Durisen, R.H.},
  \bibinfo{year}{1990}.
\newblock \bibinfo{title}{Bombardment of planetary rings by meteoroids:
  {General} formulation and effects of {Oort} cloud projectiles}.
\newblock \bibinfo{journal}{Icarus} \bibinfo{volume}{84},
  \bibinfo{pages}{467--501}.
\newblock \URLprefix
  \url{https://ui.adsabs.harvard.edu/abs/1990Icar...84..467C},
  \DOIprefix\doi{10.1016/0019-1035(90)90049-F}.
%Type = Article
\bibitem[{Cuzzi and Estrada(1998)}]{cuzzi_compositional_1998}
\bibinfo{author}{Cuzzi, J.N.}, \bibinfo{author}{Estrada, P.R.},
  \bibinfo{year}{1998}.
\newblock \bibinfo{title}{Compositional {Evolution} of {Saturn}'s {Rings} {Due}
  to {Meteoroid} {Bombardment}}.
\newblock \bibinfo{journal}{Icarus} \bibinfo{volume}{132},
  \bibinfo{pages}{1--35}.
\newblock \URLprefix
  \url{https://ui.adsabs.harvard.edu/abs/1998Icar..132....1C},
  \DOIprefix\doi{10.1006/icar.1997.5863}.
%Type = Article
\bibitem[{Daisaka et~al.(2001)Daisaka, Tanaka and Ida}]{daisaka_viscosity_2001}
\bibinfo{author}{Daisaka, H.}, \bibinfo{author}{Tanaka, H.},
  \bibinfo{author}{Ida, S.}, \bibinfo{year}{2001}.
\newblock \bibinfo{title}{Viscosity in a {Dense} {Planetary} {Ring} with
  {Self}-{Gravitating} {Particles}}.
\newblock \bibinfo{journal}{Icarus} \bibinfo{volume}{154},
  \bibinfo{pages}{296--312}.
\newblock \URLprefix
  \url{https://ui.adsabs.harvard.edu/abs/2001Icar..154..296D},
  \DOIprefix\doi{10.1006/icar.2001.6716}.
%Type = Article
\bibitem[{Dalle~Ore et~al.(2015)Dalle~Ore, Cruikshank, Mastrapa, Lewis and
  White}]{dalle_ore_impact_2015}
\bibinfo{author}{Dalle~Ore, C.M.}, \bibinfo{author}{Cruikshank, D.P.},
  \bibinfo{author}{Mastrapa, R.M.E.}, \bibinfo{author}{Lewis, E.},
  \bibinfo{author}{White, O.L.}, \bibinfo{year}{2015}.
\newblock \bibinfo{title}{Impact craters: {An} ice study on {Rhea}}.
\newblock \bibinfo{journal}{Icarus} \bibinfo{volume}{261},
  \bibinfo{pages}{80--90}.
\newblock \URLprefix
  \url{https://ui.adsabs.harvard.edu/abs/2015Icar..261...80D},
  \DOIprefix\doi{10.1016/j.icarus.2015.08.008}.
%Type = Incollection
\bibitem[{De~Pater et~al.(2018)De~Pater, Hamilton, Showalter, Throop and
  Burns}]{de_pater_rings_2018}
\bibinfo{author}{De~Pater, I.}, \bibinfo{author}{Hamilton, D.P.},
  \bibinfo{author}{Showalter, M.R.}, \bibinfo{author}{Throop, H.B.},
  \bibinfo{author}{Burns, J.A.}, \bibinfo{year}{2018}.
\newblock \bibinfo{title}{The {Rings} of {Jupiter}}, in:
  \bibinfo{editor}{Tiscareno, M.S.}, \bibinfo{editor}{Murray, C.D.} (Eds.),
  \bibinfo{booktitle}{Planetary {Ring} {Systems}. {Properties}, {Structure},
  and {Evolution}}, pp. \bibinfo{pages}{125--134}.
\newblock \URLprefix
  \url{https://ui.adsabs.harvard.edu/abs/2018prs..book..125D/abstract},
  \DOIprefix\doi{10.1017/9781316286791.006}.
%Type = Article
\bibitem[{DellaGiustina et~al.(2020)DellaGiustina, Burke, Walsh, Smith, Golish,
  Bierhaus, Ballouz, Becker, Campins, Tatsumi, Yumoto, Sugita, Deshapriya,
  Cloutis, Clark, Hendrix, Sen, Al~Asad, Daly, Applin, Avdellidou, Barucci,
  Becker, Bennett, Bottke, Brodbeck, Connolly, Delbo, de~Leon,
  Drouet~d'Aubigny, Edmundson, Fornasier, Hamilton, Hasselmann, Hergenrother,
  Howell, Jawin, Kaplan, Le~Corre, Lim, Li, Michel, Molaro, Nolan, Nolau,
  Pajola, Parkinson, Popescu, Porter, Rizk, Rizos, Ryan, Rozitis, Shultz,
  Simon, Trang, Van~Auken, Wolner and Lauretta}]{dellagiustina_variations_2020}
\bibinfo{author}{DellaGiustina, D.N.}, \bibinfo{author}{Burke, K.N.},
  \bibinfo{author}{Walsh, K.J.}, \bibinfo{author}{Smith, P.H.},
  \bibinfo{author}{Golish, D.R.}, \bibinfo{author}{Bierhaus, E.B.},
  \bibinfo{author}{Ballouz, R.L.}, \bibinfo{author}{Becker, T.L.},
  \bibinfo{author}{Campins, H.}, \bibinfo{author}{Tatsumi, E.},
  \bibinfo{author}{Yumoto, K.}, \bibinfo{author}{Sugita, S.},
  \bibinfo{author}{Deshapriya, J.D.P.}, \bibinfo{author}{Cloutis, E.A.},
  \bibinfo{author}{Clark, B.E.}, \bibinfo{author}{Hendrix, A.R.},
  \bibinfo{author}{Sen, A.}, \bibinfo{author}{Al~Asad, M.M.},
  \bibinfo{author}{Daly, M.G.}, \bibinfo{author}{Applin, D.M.},
  \bibinfo{author}{Avdellidou, C.}, \bibinfo{author}{Barucci, M.A.},
  \bibinfo{author}{Becker, K.J.}, \bibinfo{author}{Bennett, C.A.},
  \bibinfo{author}{Bottke, W.F.}, \bibinfo{author}{Brodbeck, J.I.},
  \bibinfo{author}{Connolly, H.C.}, \bibinfo{author}{Delbo, M.},
  \bibinfo{author}{de~Leon, J.}, \bibinfo{author}{Drouet~d'Aubigny, C.Y.},
  \bibinfo{author}{Edmundson, K.L.}, \bibinfo{author}{Fornasier, S.},
  \bibinfo{author}{Hamilton, V.E.}, \bibinfo{author}{Hasselmann, P.H.},
  \bibinfo{author}{Hergenrother, C.W.}, \bibinfo{author}{Howell, E.S.},
  \bibinfo{author}{Jawin, E.R.}, \bibinfo{author}{Kaplan, H.H.},
  \bibinfo{author}{Le~Corre, L.}, \bibinfo{author}{Lim, L.F.},
  \bibinfo{author}{Li, J.Y.}, \bibinfo{author}{Michel, P.},
  \bibinfo{author}{Molaro, J.L.}, \bibinfo{author}{Nolan, M.C.},
  \bibinfo{author}{Nolau, J.}, \bibinfo{author}{Pajola, M.},
  \bibinfo{author}{Parkinson, A.}, \bibinfo{author}{Popescu, M.},
  \bibinfo{author}{Porter, N.A.}, \bibinfo{author}{Rizk, B.},
  \bibinfo{author}{Rizos, J.L.}, \bibinfo{author}{Ryan, A.J.},
  \bibinfo{author}{Rozitis, B.}, \bibinfo{author}{Shultz, N.K.},
  \bibinfo{author}{Simon, A.A.}, \bibinfo{author}{Trang, D.},
  \bibinfo{author}{Van~Auken, R.B.}, \bibinfo{author}{Wolner, C.W.V.},
  \bibinfo{author}{Lauretta, D.S.}, \bibinfo{year}{2020}.
\newblock \bibinfo{title}{Variations in color and reflectance on the surface of
  asteroid (101955) {Bennu}}.
\newblock \bibinfo{journal}{Science} \bibinfo{volume}{370},
  \bibinfo{pages}{eabc3660}.
\newblock \URLprefix
  \url{https://ui.adsabs.harvard.edu/abs/2020Sci...370.3660D},
  \DOIprefix\doi{10.1126/science.abc3660}. \bibinfo{note}{aDS Bibcode:
  2020Sci...370.3660D}.
%Type = Article
\bibitem[{Dones(1991)}]{dones_recent_1991}
\bibinfo{author}{Dones, L.}, \bibinfo{year}{1991}.
\newblock \bibinfo{title}{A recent cometary origin for {Saturn}'s rings?}
\newblock \bibinfo{journal}{Icarus} \bibinfo{volume}{92},
  \bibinfo{pages}{194--203}.
\newblock \URLprefix
  \url{https://ui.adsabs.harvard.edu/abs/1991Icar...92..194D},
  \DOIprefix\doi{10.1016/0019-1035(91)90045-U}.
%Type = Article
\bibitem[{Doyle et~al.(1989)Doyle, Dones and Cuzzi}]{doyle_radiative_1989}
\bibinfo{author}{Doyle, L.R.}, \bibinfo{author}{Dones, L.},
  \bibinfo{author}{Cuzzi, J.N.}, \bibinfo{year}{1989}.
\newblock \bibinfo{title}{Radiative transfer modeling of {Saturn}'s {Outer} {B}
  ring}.
\newblock \bibinfo{journal}{Icarus} \bibinfo{volume}{80},
  \bibinfo{pages}{104--135}.
\newblock \URLprefix
  \url{https://ui.adsabs.harvard.edu/abs/1989Icar...80..104D},
  \DOIprefix\doi{10.1016/0019-1035(89)90163-2}.
%Type = Article
\bibitem[{Durisen et~al.(1992)Durisen, Bode, Cuzzi, Cederbloom and
  Murphy}]{durisen_ballistic_1992}
\bibinfo{author}{Durisen, R.H.}, \bibinfo{author}{Bode, P.W.},
  \bibinfo{author}{Cuzzi, J.N.}, \bibinfo{author}{Cederbloom, S.E.},
  \bibinfo{author}{Murphy, B.W.}, \bibinfo{year}{1992}.
\newblock \bibinfo{title}{Ballistic transport in planetary ring systems due to
  particle erosion mechanisms {II}. {Theoretical} models for {Saturn}'s {A}-
  and {B}-ring inner edges}.
\newblock \bibinfo{journal}{Icarus} \bibinfo{volume}{100},
  \bibinfo{pages}{364--393}.
\newblock \URLprefix
  \url{https://ui.adsabs.harvard.edu/abs/1992Icar..100..364D},
  \DOIprefix\doi{10.1016/0019-1035(92)90106-H}.
%Type = Article
\bibitem[{Durisen et~al.(1989)Durisen, Cramer, Murphy, Cuzzi, Mullikin and
  Cederbloom}]{durisen_ballistic_1989}
\bibinfo{author}{Durisen, R.H.}, \bibinfo{author}{Cramer, N.L.},
  \bibinfo{author}{Murphy, B.W.}, \bibinfo{author}{Cuzzi, J.N.},
  \bibinfo{author}{Mullikin, T.L.}, \bibinfo{author}{Cederbloom, S.E.},
  \bibinfo{year}{1989}.
\newblock \bibinfo{title}{Ballistic transport in planetary ring systems due to
  particle erosion mechanisms {I}. {Theory}, numerical methods, and
  illustrative examples}.
\newblock \bibinfo{journal}{Icarus} \bibinfo{volume}{80},
  \bibinfo{pages}{136--166}.
\newblock \URLprefix
  \url{https://ui.adsabs.harvard.edu/abs/1989Icar...80..136D},
  \DOIprefix\doi{10.1016/0019-1035(89)90164-4}.
%Type = Article
\bibitem[{Durisen and Estrada(2023)}]{durisen_large_2023}
\bibinfo{author}{Durisen, R.H.}, \bibinfo{author}{Estrada, P.R.},
  \bibinfo{year}{2023}.
\newblock \bibinfo{title}{Large mass inflow rates in {Saturn}'s rings due to
  ballistic transport and mass loading}.
\newblock \bibinfo{journal}{Icarus} \bibinfo{volume}{400},
  \bibinfo{pages}{115221}.
\newblock \URLprefix
  \url{https://ui.adsabs.harvard.edu/abs/2023Icar..40015221D},
  \DOIprefix\doi{10.1016/j.icarus.2022.115221}.
%Type = Inproceedings
\bibitem[{{Esposito} et~al.(2024){Esposito}, {Elliott} and
  {Bradley}}]{esposito_space_2024}
\bibinfo{author}{{Esposito}, L.W.}, \bibinfo{author}{{Elliott}, J.P.},
  \bibinfo{author}{{Bradley}, E.T.}, \bibinfo{year}{2024}.
\newblock \bibinfo{title}{{Space Weathering Provides a Lower Limit on the Age
  of Saturn's Rings}}, in: \bibinfo{booktitle}{European Planetary Science
  Congress}, pp. \bibinfo{pages}{EPSC2024--562}.
\newblock \DOIprefix\doi{10.5194/epsc2024-562}.
%Type = Article
\bibitem[{Estrada and Durisen(2023)}]{estrada_constraints_2023}
\bibinfo{author}{Estrada, P.R.}, \bibinfo{author}{Durisen, R.H.},
  \bibinfo{year}{2023}.
\newblock \bibinfo{title}{Constraints on the initial mass, age and lifetime of
  {Saturn}'s rings from viscous evolutions that include pollution and transport
  due to micrometeoroid bombardment}.
\newblock \bibinfo{journal}{Icarus} \bibinfo{volume}{400},
  \bibinfo{pages}{115296}.
\newblock \URLprefix
  \url{https://ui.adsabs.harvard.edu/abs/2023Icar..40015296E},
  \DOIprefix\doi{10.1016/j.icarus.2022.115296}.
%Type = Book
\bibitem[{Estrada et~al.(2018)Estrada, Durisen and
  Latter}]{estrada_meteoroid_2018}
\bibinfo{author}{Estrada, P.R.}, \bibinfo{author}{Durisen, R.H.},
  \bibinfo{author}{Latter, H.N.}, \bibinfo{year}{2018}.
\newblock \bibinfo{title}{Meteoroid {Bombardment} and {Ballistic} {Transport}
  in {Planetary} {Rings}}.
\newblock \URLprefix
  \url{https://ui.adsabs.harvard.edu/abs/2018prs..book..198E},
  \DOIprefix\doi{10.1017/9781316286791.009}.
%Type = Article
\bibitem[{Famá et~al.(2010)Famá, Loeffler, Raut and
  Baragiola}]{fama_radiation-induced_2010}
\bibinfo{author}{Famá, M.}, \bibinfo{author}{Loeffler, M.J.},
  \bibinfo{author}{Raut, U.}, \bibinfo{author}{Baragiola, R.A.},
  \bibinfo{year}{2010}.
\newblock \bibinfo{title}{Radiation-induced amorphization of crystalline ice}.
\newblock \bibinfo{journal}{Icarus} \bibinfo{volume}{207},
  \bibinfo{pages}{314--319}.
\newblock \URLprefix
  \url{https://ui.adsabs.harvard.edu/abs/2010Icar..207..314F},
  \DOIprefix\doi{10.1016/j.icarus.2009.11.001}.
%Type = Article
\bibitem[{Gomes et~al.(2005)Gomes, Levison, Tsiganis and
  Morbidelli}]{gomes_origin_2005}
\bibinfo{author}{Gomes, R.}, \bibinfo{author}{Levison, H.F.},
  \bibinfo{author}{Tsiganis, K.}, \bibinfo{author}{Morbidelli, A.},
  \bibinfo{year}{2005}.
\newblock \bibinfo{title}{Origin of the cataclysmic {Late} {Heavy}
  {Bombardment} period of the terrestrial planets}.
\newblock \bibinfo{journal}{Nature} \bibinfo{volume}{435},
  \bibinfo{pages}{466--469}.
\newblock \URLprefix
  \url{https://ui.adsabs.harvard.edu/abs/2005Natur.435..466G},
  \DOIprefix\doi{10.1038/nature03676}.
%Type = Phdthesis
\bibitem[{Grossman(1990)}]{grossman_microwave_1990}
\bibinfo{author}{Grossman, A.W.}, \bibinfo{year}{1990}.
\newblock \bibinfo{title}{Microwave {Imaging} of {Saturn}'s {Deep} {Atmosphere}
  and {Rings}.}
\newblock \bibinfo{type}{{PhD} {Thesis}}. California Institute of Technology.
\newblock \URLprefix
  \url{https://ui.adsabs.harvard.edu/abs/1990PhDT.........2G/abstract}.
%Type = Article
\bibitem[{Gudipati and Yang(2012)}]{gudipati_-situ_2012}
\bibinfo{author}{Gudipati, M.S.}, \bibinfo{author}{Yang, R.},
  \bibinfo{year}{2012}.
\newblock \bibinfo{title}{In-situ {Probing} of {Radiation}-induced {Processing}
  of {Organics} in {Astrophysical} {Ice} {Analogs}—{Novel} {Laser}
  {Desorption} {Laser} {Ionization} {Time}-of-flight {Mass} {Spectroscopic}
  {Studies}}.
\newblock \bibinfo{journal}{The Astrophysical Journal} \bibinfo{volume}{756},
  \bibinfo{pages}{L24}.
\newblock \URLprefix
  \url{https://ui.adsabs.harvard.edu/abs/2012ApJ...756L..24G},
  \DOIprefix\doi{10.1088/2041-8205/756/1/L24}.
%Type = Article
\bibitem[{Hedman et~al.(2013)Hedman, Nicholson, Cuzzi, Clark, Filacchione,
  Capaccioni and Ciarniello}]{hedman_connections_2013}
\bibinfo{author}{Hedman, M.M.}, \bibinfo{author}{Nicholson, P.D.},
  \bibinfo{author}{Cuzzi, J.N.}, \bibinfo{author}{Clark, R.N.},
  \bibinfo{author}{Filacchione, G.}, \bibinfo{author}{Capaccioni, F.},
  \bibinfo{author}{Ciarniello, M.}, \bibinfo{year}{2013}.
\newblock \bibinfo{title}{Connections between spectra and structure in
  {Saturn}’s main rings based on {Cassini} {VIMS} data}.
\newblock \bibinfo{journal}{Icarus} \bibinfo{volume}{223},
  \bibinfo{pages}{105--130}.
\newblock \URLprefix
  \url{https://ui.adsabs.harvard.edu/abs/2013Icar..223..105H},
  \DOIprefix\doi{10.1016/j.icarus.2012.10.014}. \bibinfo{note}{publisher:
  Elsevier ADS Bibcode: 2013Icar..223..105H}.
%Type = Article
\bibitem[{Hsu et~al.(2018)Hsu, Schmidt, Kempf, Postberg, Moragas-Klostermeyer,
  Seiß, Hoffmann, Burton, Ye, Kurth, Horányi, Khawaja, Spahn, Schirdewahn,
  O’Donoghue, Moore, Cuzzi, Jones and Srama}]{hsu_situ_2018}
\bibinfo{author}{Hsu, H.W.}, \bibinfo{author}{Schmidt, J.},
  \bibinfo{author}{Kempf, S.}, \bibinfo{author}{Postberg, F.},
  \bibinfo{author}{Moragas-Klostermeyer, G.}, \bibinfo{author}{Seiß, M.},
  \bibinfo{author}{Hoffmann, H.}, \bibinfo{author}{Burton, M.},
  \bibinfo{author}{Ye, S.}, \bibinfo{author}{Kurth, W.S.},
  \bibinfo{author}{Horányi, M.}, \bibinfo{author}{Khawaja, N.},
  \bibinfo{author}{Spahn, F.}, \bibinfo{author}{Schirdewahn, D.},
  \bibinfo{author}{O’Donoghue, J.}, \bibinfo{author}{Moore, L.},
  \bibinfo{author}{Cuzzi, J.}, \bibinfo{author}{Jones, G.H.},
  \bibinfo{author}{Srama, R.}, \bibinfo{year}{2018}.
\newblock \bibinfo{title}{In situ collection of dust grains falling from
  {Saturn}’s rings into its atmosphere}.
\newblock \bibinfo{journal}{Science} \bibinfo{volume}{362},
  \bibinfo{pages}{eaat3185}.
\newblock \URLprefix \url{https://www.science.org/doi/10.1126/science.aat3185},
  \DOIprefix\doi{10.1126/science.aat3185}.
%Type = Article
\bibitem[{Humes(1980)}]{humes_results_1980}
\bibinfo{author}{Humes, D.H.}, \bibinfo{year}{1980}.
\newblock \bibinfo{title}{Results of {Pioneer} 10 and 11 {Meteoroid}
  {Experiments}: {Interplanetary} and {Near}-{Saturn}}.
\newblock \bibinfo{journal}{Journal of Geophysical Research}
  \bibinfo{volume}{85}, \bibinfo{pages}{5841--5852}.
\newblock \URLprefix
  \url{https://ui.adsabs.harvard.edu/abs/1980JGR....85.5841H},
  \DOIprefix\doi{10.1029/JA085iA11p05841}.
%Type = Article
\bibitem[{Hyodo et~al.(2025)Hyodo, Genda and Madeira}]{hyodo_pollution_2025}
\bibinfo{author}{Hyodo, R.}, \bibinfo{author}{Genda, H.},
  \bibinfo{author}{Madeira, G.}, \bibinfo{year}{2025}.
\newblock \bibinfo{title}{Pollution resistance of {Saturn}’s ring particles
  during micrometeoroid impact}.
\newblock \bibinfo{journal}{Nature Geoscience} \bibinfo{volume}{18},
  \bibinfo{pages}{44--49}.
\newblock \URLprefix \url{https://www.nature.com/articles/s41561-024-01598-9},
  \DOIprefix\doi{10.1038/s41561-024-01598-9}.
%Type = Article
\bibitem[{Iess et~al.(2019)Iess, Militzer, Kaspi, Nicholson, Durante, Racioppa,
  Anabtawi, Galanti, Hubbard, Mariani, Tortora, Wahl and
  Zannoni}]{iess_measurement_2019}
\bibinfo{author}{Iess, L.}, \bibinfo{author}{Militzer, B.},
  \bibinfo{author}{Kaspi, Y.}, \bibinfo{author}{Nicholson, P.},
  \bibinfo{author}{Durante, D.}, \bibinfo{author}{Racioppa, P.},
  \bibinfo{author}{Anabtawi, A.}, \bibinfo{author}{Galanti, E.},
  \bibinfo{author}{Hubbard, W.}, \bibinfo{author}{Mariani, M.J.},
  \bibinfo{author}{Tortora, P.}, \bibinfo{author}{Wahl, S.},
  \bibinfo{author}{Zannoni, M.}, \bibinfo{year}{2019}.
\newblock \bibinfo{title}{Measurement and implications of {Saturn}'s gravity
  field and ring mass}.
\newblock \bibinfo{journal}{Science} \bibinfo{volume}{364},
  \bibinfo{pages}{aat2965}.
\newblock \URLprefix
  \url{https://ui.adsabs.harvard.edu/abs/1985Icar...62..244G/abstract},
  \DOIprefix\doi{10.1126/science.aat2965}.
%Type = Article
\bibitem[{Ip(1984)}]{ip_ring_1984}
\bibinfo{author}{Ip, W.H.}, \bibinfo{year}{1984}.
\newblock \bibinfo{title}{Ring torque of {Saturn} from interplanetary meteoroid
  impact}.
\newblock \bibinfo{journal}{Icarus} \bibinfo{volume}{60},
  \bibinfo{pages}{547--552}.
\newblock \URLprefix
  \url{https://ui.adsabs.harvard.edu/abs/1984Icar...60..547I},
  \DOIprefix\doi{10.1016/0019-1035(84)90163-5}.
%Type = Article
\bibitem[{Kempf et~al.(2023)Kempf, Altobelli, Schmidt, Cuzzi, Estrada and
  Srama}]{kempf_micrometeoroid_2023}
\bibinfo{author}{Kempf, S.}, \bibinfo{author}{Altobelli, N.},
  \bibinfo{author}{Schmidt, J.}, \bibinfo{author}{Cuzzi, J.N.},
  \bibinfo{author}{Estrada, P.R.}, \bibinfo{author}{Srama, R.},
  \bibinfo{year}{2023}.
\newblock \bibinfo{title}{Micrometeoroid infall onto {Saturn}'s rings
  constrains their age to no more than a few hundred million years}.
\newblock \bibinfo{journal}{Science Advances} \bibinfo{volume}{9},
  \bibinfo{pages}{eadf8537}.
\newblock \URLprefix
  \url{https://ui.adsabs.harvard.edu/abs/2023SciA....9F8537K/abstract},
  \DOIprefix\doi{10.1126/sciadv.adf8537}.
%Type = Article
\bibitem[{Linti et~al.(2024)Linti, Khawaja, Hillier, Nölle, Fischer, Hsu,
  Srama and Postberg}]{linti_cassinis_2024}
\bibinfo{author}{Linti, S.}, \bibinfo{author}{Khawaja, N.},
  \bibinfo{author}{Hillier, J.K.}, \bibinfo{author}{Nölle, L.},
  \bibinfo{author}{Fischer, C.}, \bibinfo{author}{Hsu, H.W.},
  \bibinfo{author}{Srama, R.}, \bibinfo{author}{Postberg, F.},
  \bibinfo{year}{2024}.
\newblock \bibinfo{title}{Cassini's {CDA} observes a variety of dust
  populations just outside {Saturn}'s main rings}.
\newblock \bibinfo{journal}{Monthly Notices of the Royal Astronomical Society}
  \bibinfo{volume}{529}, \bibinfo{pages}{3121--3139}.
\newblock \URLprefix
  \url{https://ui.adsabs.harvard.edu/abs/2024MNRAS.529.3121L},
  \DOIprefix\doi{10.1093/mnras/stae238}. \bibinfo{note}{publisher: OUP ADS
  Bibcode: 2024MNRAS.529.3121L}.
%Type = Article
\bibitem[{Linti et~al.(2025)Linti, Postberg, Hsu, Schmidt, Hillier, Nölle,
  Fischer and Srama}]{linti_dust_2025}
\bibinfo{author}{Linti, S.}, \bibinfo{author}{Postberg, F.},
  \bibinfo{author}{Hsu, H.W.}, \bibinfo{author}{Schmidt, J.},
  \bibinfo{author}{Hillier, J.K.}, \bibinfo{author}{Nölle, L.},
  \bibinfo{author}{Fischer, C.}, \bibinfo{author}{Srama, R.},
  \bibinfo{year}{2025}.
\newblock \bibinfo{title}{A {Dust} {Halo} from {Saturn}'s {Main} {Rings}
  {Extending} {Several} {Saturnian} {Radii} above the {Ring} {Plane}}.
\newblock \bibinfo{journal}{The Planetary Science Journal} \bibinfo{volume}{6},
  \bibinfo{pages}{273}.
\newblock \URLprefix
  \url{https://ui.adsabs.harvard.edu/abs/2025PSJ.....6..273L},
  \DOIprefix\doi{10.3847/PSJ/ae18c1}. \bibinfo{note}{publisher: IOP ADS
  Bibcode: 2025PSJ.....6..273L}.
%Type = Article
\bibitem[{Lissauer(1984)}]{lissauer_ballistic_1984}
\bibinfo{author}{Lissauer, J.J.}, \bibinfo{year}{1984}.
\newblock \bibinfo{title}{Ballistic transport in {Saturn}'s rings: {An}
  analytic theory}.
\newblock \bibinfo{journal}{Icarus} \bibinfo{volume}{57},
  \bibinfo{pages}{63--71}.
\newblock \URLprefix
  \url{https://ui.adsabs.harvard.edu/abs/1984Icar...57...63L},
  \DOIprefix\doi{10.1016/0019-1035(84)90008-3}.
%Type = Article
\bibitem[{Lissauer et~al.(1988)Lissauer, Squyres and
  Hartmann}]{lissauer_bombardment_1988}
\bibinfo{author}{Lissauer, J.J.}, \bibinfo{author}{Squyres, S.W.},
  \bibinfo{author}{Hartmann, W.K.}, \bibinfo{year}{1988}.
\newblock \bibinfo{title}{Bombardment history of the {Saturn} system.}
\newblock \bibinfo{journal}{Journal of Geophysical Research}
  \bibinfo{volume}{93}, \bibinfo{pages}{13776--13804}.
\newblock \URLprefix
  \url{https://ui.adsabs.harvard.edu/abs/1988JGR....9313776L},
  \DOIprefix\doi{10.1029/JB093iB11p13776}.
%Type = Article
\bibitem[{Miller et~al.(2024)Miller, Filacchione, Cuzzi, Nicholson, Hedman,
  Baillié, Johnson, Tseng, Estrada, Waite, Ciarniello, Ferrari, Zhang,
  Hendrix, Moses and Hsu}]{miller_composition_2024}
\bibinfo{author}{Miller, K.E.}, \bibinfo{author}{Filacchione, G.},
  \bibinfo{author}{Cuzzi, J.N.}, \bibinfo{author}{Nicholson, P.D.},
  \bibinfo{author}{Hedman, M.M.}, \bibinfo{author}{Baillié, K.},
  \bibinfo{author}{Johnson, R.E.}, \bibinfo{author}{Tseng, W.L.},
  \bibinfo{author}{Estrada, P.R.}, \bibinfo{author}{Waite, J.H.},
  \bibinfo{author}{Ciarniello, M.}, \bibinfo{author}{Ferrari, C.},
  \bibinfo{author}{Zhang, Z.}, \bibinfo{author}{Hendrix, A.},
  \bibinfo{author}{Moses, J.I.}, \bibinfo{author}{Hsu, H.W.},
  \bibinfo{year}{2024}.
\newblock \bibinfo{title}{The {Composition} of {Saturn}’s {Rings}}.
\newblock \bibinfo{journal}{Space Science Reviews} \bibinfo{volume}{220},
  \bibinfo{pages}{70}.
\newblock \URLprefix \url{https://doi.org/10.1007/s11214-024-01104-y},
  \DOIprefix\doi{10.1007/s11214-024-01104-y}.
%Type = Article
\bibitem[{Morbidelli et~al.(2005)Morbidelli, Levison, Tsiganis and
  Gomes}]{morbidelli_chaotic_2005}
\bibinfo{author}{Morbidelli, A.}, \bibinfo{author}{Levison, H.F.},
  \bibinfo{author}{Tsiganis, K.}, \bibinfo{author}{Gomes, R.},
  \bibinfo{year}{2005}.
\newblock \bibinfo{title}{Chaotic capture of {Jupiter}'s {Trojan} asteroids in
  the early {Solar} {System}}.
\newblock \bibinfo{journal}{Nature} \bibinfo{volume}{435},
  \bibinfo{pages}{462--465}.
\newblock \URLprefix
  \url{https://ui.adsabs.harvard.edu/abs/2005Natur.435..462M},
  \DOIprefix\doi{10.1038/nature03540}.
%Type = Article
\bibitem[{Morfill et~al.(1983)Morfill, Fechtig, Gruen and
  Goertz}]{morfill_consequences_1983}
\bibinfo{author}{Morfill, G.E.}, \bibinfo{author}{Fechtig, H.},
  \bibinfo{author}{Gruen, E.}, \bibinfo{author}{Goertz, C.K.},
  \bibinfo{year}{1983}.
\newblock \bibinfo{title}{Some consequences of meteoroid impacts on {Saturn}'s
  rings}.
\newblock \bibinfo{journal}{Icarus} \bibinfo{volume}{55},
  \bibinfo{pages}{439--447}.
\newblock \URLprefix
  \url{https://ui.adsabs.harvard.edu/abs/1983Icar...55..439M},
  \DOIprefix\doi{10.1016/0019-1035(83)90114-8}.
%Type = Inproceedings
\bibitem[{Nicholson et~al.(1984)Nicholson, Jones and
  Matthews}]{nicholson_infrared_1984}
\bibinfo{author}{Nicholson, P.D.}, \bibinfo{author}{Jones, T.J.},
  \bibinfo{author}{Matthews, K.}, \bibinfo{year}{1984}.
\newblock \bibinfo{title}{Infrared {Observations} of the {Uranian} {Rings}},
  in: \bibinfo{editor}{Brahic, A.} (Ed.), \bibinfo{booktitle}{Planetary
  {Rings}}, p. \bibinfo{pages}{169}.
\newblock \URLprefix
  \url{https://ui.adsabs.harvard.edu/abs/1984plri.coll..169N/abstract}.
%Type = Article
\bibitem[{Noyelles et~al.(2019)Noyelles, Baillié, Charnoz, Lainey and
  Tobie}]{noyelles_formation_2019}
\bibinfo{author}{Noyelles, B.}, \bibinfo{author}{Baillié, K.},
  \bibinfo{author}{Charnoz, S.}, \bibinfo{author}{Lainey, V.},
  \bibinfo{author}{Tobie, G.}, \bibinfo{year}{2019}.
\newblock \bibinfo{title}{Formation of the {Cassini} {Division} - {II}.
  {Possible} histories of {Mimas} and {Enceladus}}.
\newblock \bibinfo{journal}{Monthly Notices of the Royal Astronomical Society}
  \bibinfo{volume}{486}, \bibinfo{pages}{2947--2963}.
\newblock \URLprefix
  \url{https://ui.adsabs.harvard.edu/abs/2019MNRAS.486.2947N},
  \DOIprefix\doi{10.1093/mnras/stz445}.
%Type = Article
\bibitem[{Pollack et~al.(1976)Pollack, Grossman, Moore and
  Graboske}]{pollack_formation_1976}
\bibinfo{author}{Pollack, J.B.}, \bibinfo{author}{Grossman, A.S.},
  \bibinfo{author}{Moore, R.}, \bibinfo{author}{Graboske, Jr., H.C.},
  \bibinfo{year}{1976}.
\newblock \bibinfo{title}{The {Formation} of {Saturn}'s {Satellites} and
  {Rings}, as {Influenced} by {Saturn}'s {Contraction} {History}}.
\newblock \bibinfo{journal}{Icarus} \bibinfo{volume}{29},
  \bibinfo{pages}{35--48}.
\newblock \URLprefix
  \url{https://ui.adsabs.harvard.edu/abs/1976Icar...29...35P},
  \DOIprefix\doi{10.1016/0019-1035(76)90100-7}.
%Type = Article
\bibitem[{Porco et~al.(1987)Porco, Cuzzi, Ockert and
  Terrile}]{porco_color_1987}
\bibinfo{author}{Porco, C.C.}, \bibinfo{author}{Cuzzi, J.N.},
  \bibinfo{author}{Ockert, M.E.}, \bibinfo{author}{Terrile, R.J.},
  \bibinfo{year}{1987}.
\newblock \bibinfo{title}{The color of the {Uranian} rings}.
\newblock \bibinfo{journal}{Icarus} \bibinfo{volume}{72},
  \bibinfo{pages}{69--78}.
\newblock \URLprefix
  \url{https://ui.adsabs.harvard.edu/abs/1987Icar...72...69P},
  \DOIprefix\doi{10.1016/0019-1035(87)90120-5}.
%Type = Article
\bibitem[{Pringle(1981)}]{pringle_accretion_1981}
\bibinfo{author}{Pringle, J.E.}, \bibinfo{year}{1981}.
\newblock \bibinfo{title}{Accretion discs in astrophysics}.
\newblock \bibinfo{journal}{Annual Review of Astronomy and Astrophysics}
  \bibinfo{volume}{19}, \bibinfo{pages}{137--162}.
\newblock \URLprefix
  \url{https://ui.adsabs.harvard.edu/abs/1981ARA&A..19..137P},
  \DOIprefix\doi{10.1146/annurev.aa.19.090181.001033}.
%Type = Book
\bibitem[{Safronov(1972)}]{safronov_evolution_1972}
\bibinfo{author}{Safronov, V.S.}, \bibinfo{year}{1972}.
\newblock \bibinfo{title}{Evolution of the protoplanetary cloud and formation
  of the earth and planets.}
\newblock \URLprefix
  \url{https://ui.adsabs.harvard.edu/abs/1972epcf.book.....S}.
%Type = Article
\bibitem[{Salmon et~al.(2010)Salmon, Charnoz, Crida and
  Brahic}]{salmon_long-term_2010}
\bibinfo{author}{Salmon, J.}, \bibinfo{author}{Charnoz, S.},
  \bibinfo{author}{Crida, A.}, \bibinfo{author}{Brahic, A.},
  \bibinfo{year}{2010}.
\newblock \bibinfo{title}{Long-term and large-scale viscous evolution of dense
  planetary rings}.
\newblock \bibinfo{journal}{Icarus} \bibinfo{volume}{209},
  \bibinfo{pages}{771--785}.
\newblock \URLprefix
  \url{https://ui.adsabs.harvard.edu/abs/2010Icar..209..771S},
  \DOIprefix\doi{10.1016/j.icarus.2010.05.030}.
%Type = Book
\bibitem[{Shoemaker and Wolfe(1982)}]{shoemaker_cratering_1982}
\bibinfo{author}{Shoemaker, E.M.}, \bibinfo{author}{Wolfe, R.F.},
  \bibinfo{year}{1982}.
\newblock \bibinfo{title}{Cratering time scales for the {Galilean} satellites}.
\newblock \URLprefix
  \url{https://ui.adsabs.harvard.edu/abs/1982stjp.conf..277S}.
%Type = Article
\bibitem[{Spahn et~al.(2006)Spahn, Albers, Hörning, Kempf, Krivov, Makuch,
  Schmidt, Seiß and {Miodrag Sremčević}}]{spahn_e_2006}
\bibinfo{author}{Spahn, F.}, \bibinfo{author}{Albers, N.},
  \bibinfo{author}{Hörning, M.}, \bibinfo{author}{Kempf, S.},
  \bibinfo{author}{Krivov, A.V.}, \bibinfo{author}{Makuch, M.},
  \bibinfo{author}{Schmidt, J.}, \bibinfo{author}{Seiß, M.},
  \bibinfo{author}{{Miodrag Sremčević}}, \bibinfo{year}{2006}.
\newblock \bibinfo{title}{E ring dust sources: {Implications} from {Cassini}'s
  dust measurements}.
\newblock \bibinfo{journal}{Planetary and Space Science} \bibinfo{volume}{54},
  \bibinfo{pages}{1024--1032}.
\newblock \URLprefix
  \url{https://ui.adsabs.harvard.edu/abs/2006P&SS...54.1024S},
  \DOIprefix\doi{10.1016/j.pss.2006.05.022}.
%Type = Article
\bibitem[{Teodoro et~al.(2023)Teodoro, Kegerreis, Estrada, Ćuk, Eke, Cuzzi,
  Massey and Sandnes}]{teodoro_recent_2023}
\bibinfo{author}{Teodoro, L.F.A.}, \bibinfo{author}{Kegerreis, J.A.},
  \bibinfo{author}{Estrada, P.R.}, \bibinfo{author}{Ćuk, M.},
  \bibinfo{author}{Eke, V.R.}, \bibinfo{author}{Cuzzi, J.N.},
  \bibinfo{author}{Massey, R.J.}, \bibinfo{author}{Sandnes, T.D.},
  \bibinfo{year}{2023}.
\newblock \bibinfo{title}{A {Recent} {Impact} {Origin} of {Saturn}'s {Rings}
  and {Mid}-sized {Moons}}.
\newblock \bibinfo{journal}{The Astrophysical Journal} \bibinfo{volume}{955},
  \bibinfo{pages}{137}.
\newblock \URLprefix
  \url{https://ui.adsabs.harvard.edu/abs/2023ApJ...955..137T/abstract},
  \DOIprefix\doi{10.3847/1538-4357/acf4ed}.
%Type = Article
\bibitem[{Tsiganis et~al.(2005)Tsiganis, Gomes, Morbidelli and
  Levison}]{tsiganis_origin_2005}
\bibinfo{author}{Tsiganis, K.}, \bibinfo{author}{Gomes, R.},
  \bibinfo{author}{Morbidelli, A.}, \bibinfo{author}{Levison, H.F.},
  \bibinfo{year}{2005}.
\newblock \bibinfo{title}{Origin of the orbital architecture of the giant
  planets of the {Solar} {System}}.
\newblock \bibinfo{journal}{Nature} \bibinfo{volume}{435},
  \bibinfo{pages}{459--461}.
\newblock \URLprefix
  \url{https://ui.adsabs.harvard.edu/abs/2005Natur.435..459T},
  \DOIprefix\doi{10.1038/nature03539}.
%Type = Article
\bibitem[{Vokrouhlický et~al.(2007)Vokrouhlický, Nesvorný, Dones and
  Bottke}]{vokrouhlicky_thermal_2007}
\bibinfo{author}{Vokrouhlický, D.}, \bibinfo{author}{Nesvorný, D.},
  \bibinfo{author}{Dones, L.}, \bibinfo{author}{Bottke, W.F.},
  \bibinfo{year}{2007}.
\newblock \bibinfo{title}{Thermal forces on planetary ring particles:
  application to the main system of {Saturn}}.
\newblock \bibinfo{journal}{Astronomy and Astrophysics} \bibinfo{volume}{471},
  \bibinfo{pages}{717--730}.
\newblock \URLprefix
  \url{https://ui.adsabs.harvard.edu/abs/2007A&A...471..717V},
  \DOIprefix\doi{10.1051/0004-6361:20067029}.
%Type = Article
\bibitem[{Wisdom et~al.(2022)Wisdom, Dbouk, Militzer, Hubbard, Nimmo, Downey
  and French}]{wisdom_loss_2022}
\bibinfo{author}{Wisdom, J.}, \bibinfo{author}{Dbouk, R.},
  \bibinfo{author}{Militzer, B.}, \bibinfo{author}{Hubbard, W.B.},
  \bibinfo{author}{Nimmo, F.}, \bibinfo{author}{Downey, B.G.},
  \bibinfo{author}{French, R.G.}, \bibinfo{year}{2022}.
\newblock \bibinfo{title}{Loss of a satellite could explain {Saturn}’s
  obliquity and young rings}.
\newblock \bibinfo{journal}{Science} \bibinfo{volume}{377},
  \bibinfo{pages}{1285--1289}.
\newblock \URLprefix
  \url{https://ui.adsabs.harvard.edu/abs/2022Sci...377.1285W/abstract},
  \DOIprefix\doi{10.1126/science.abn1234}.
%Type = Article
\bibitem[{Zhang et~al.(2017a)Zhang, Hayes, Janssen, Nicholson, Cuzzi, de~Pater
  and Dunn}]{zhang_exposure_2017}
\bibinfo{author}{Zhang, Z.}, \bibinfo{author}{Hayes, A.G.},
  \bibinfo{author}{Janssen, M.A.}, \bibinfo{author}{Nicholson, P.D.},
  \bibinfo{author}{Cuzzi, J.N.}, \bibinfo{author}{de~Pater, I.},
  \bibinfo{author}{Dunn, D.E.}, \bibinfo{year}{2017}a.
\newblock \bibinfo{title}{Exposure age of {Saturn}'s {A} and {B} rings, and the
  {Cassini} {Division} as suggested by their non-icy material content}.
\newblock \bibinfo{journal}{Icarus} \bibinfo{volume}{294},
  \bibinfo{pages}{14--42}.
\newblock \URLprefix
  \url{https://www.sciencedirect.com/science/article/pii/S0019103516305760},
  \DOIprefix\doi{10.1016/j.icarus.2017.04.008}.
%Type = Article
\bibitem[{Zhang et~al.(2017b)Zhang, Hayes, Janssen, Nicholson, Cuzzi, de~Pater,
  Dunn, Estrada and Hedman}]{zhang_cassini_2017}
\bibinfo{author}{Zhang, Z.}, \bibinfo{author}{Hayes, A.G.},
  \bibinfo{author}{Janssen, M.A.}, \bibinfo{author}{Nicholson, P.D.},
  \bibinfo{author}{Cuzzi, J.N.}, \bibinfo{author}{de~Pater, I.},
  \bibinfo{author}{Dunn, D.E.}, \bibinfo{author}{Estrada, P.R.},
  \bibinfo{author}{Hedman, M.M.}, \bibinfo{year}{2017}b.
\newblock \bibinfo{title}{Cassini microwave observations provide clues to the
  origin of {Saturn}'s {C} ring}.
\newblock \bibinfo{journal}{Icarus} \bibinfo{volume}{281},
  \bibinfo{pages}{297--321}.
\newblock \URLprefix
  \url{https://www.sciencedirect.com/science/article/pii/S0019103516304316},
  \DOIprefix\doi{10.1016/j.icarus.2016.07.020}.

\end{thebibliography}

\end{document}